\documentclass[pra,letterpaper,aps,10pt,superscriptaddress,twocolumn,floatfix,showpacs]{revtex4-1}
\usepackage{graphicx}
\usepackage{amsmath}
\usepackage{amsfonts}
\usepackage{amssymb}
\usepackage{epsfig}
\usepackage{epstopdf}
\DeclareGraphicsExtensions{.pdf,.eps,.png,.jpg,.mps}
\usepackage[pdftex]{color}
\usepackage{amsmath,graphicx,amssymb,braket,xcolor,subfigure,upgreek}
\usepackage[colorlinks, linkcolor=blue, citecolor=blue, urlcolor=blue, breaklinks=true]{hyperref}
\usepackage{microtype}
\usepackage{bbm}
\usepackage{color}
\usepackage{braket}
\usepackage{amsmath,amssymb,amsfonts,latexsym,color,dcolumn,bm,amsbsy}
\usepackage[english]{babel}
\usepackage{graphicx}
\usepackage[utf8]{inputenc}
\usepackage{textcomp}
\usepackage{braket}

\newcommand{\omfb}{\omega_{\text{fb}}}

\begin{document}
\title{Multimode cold-damping optomechanics with delayed feedback}
\author{Christian Sommer}
\affiliation{Max Planck Institute for the Science of Light, Staudtstra{\ss}e 2,
D-91058 Erlangen, Germany}
\author{Alekhya Ghosh}
\affiliation{Department of Physics, University of Erlangen-Nuremberg, Staudtstra{\ss}e 2,
D-91058 Erlangen, Germany}
\author{Claudiu Genes}
\affiliation{Max Planck Institute for the Science of Light, Staudtstra{\ss}e 2,
D-91058 Erlangen, Germany}
\affiliation{Department of Physics, University of Erlangen-Nuremberg, Staudtstra{\ss}e 2,
D-91058 Erlangen, Germany}
\date{\today}
\begin{abstract}
We investigate the role of time delay in cold-damping optomechanics with multiple mechanical resonances. For instantaneous electronic response, it was recently shown in \textit{Phys. Rev. Lett. \textbf{123}, 203605 (2019)}, that a single feedback loop is sufficient to simultaneously remove thermal noise from many mechanical modes. While the intrinsic delayed response of the electronics can induce single mode and mutual heating between adjacent modes, we propose to counteract such detrimental effects by introducing an additional time delay to the feedback loop. For lossy cavities and broadband feedback, we derive analytical results for the final occupancies of the mechanical modes within the formalism of quantum Langevin equations. For modes that are frequency degenerate collective effects dominate, mimicking behavior similar to Dicke super- and subradiance. These analytical results, corroborated with numerical simulations of both transient and steady state dynamics, allow to find suitable conditions and strategies for efficient single or multimode feedback optomechanics.
\end{abstract}

\maketitle

\section{Introduction}

A widespread technique for the removal of thermal noise from a given mechanical degree of freedom involves electronic feedback loops. The procedure is based on the continuous monitoring of a system's observable, followed by the application of an adequate cooling action via the feedback device. For example, in optomechanics~\cite{Aspelmeyer2014cavity,Mancini1998Optomechanical,Cohadon1999Cooling,Vitali2002Mirror,steixner2005quantum,Kleckner2006Sub,Arcizet2006High,Bushev2006Feedback,Poggio2007Feedback,Wilson2015Measurement,Rossi2017Enhancing,Kralj2017,rossi2018measurement,conangla2019optimal,tebbenjohanns2019cold,Guo2019Feedback}, a cavity field quadrature is detected and the result is applied either optically (as a radiation pressure force) or electrically to the thermally activated mechanical resonator. While generally one aims for the isolation and cooling of a specific vibrational mode, it has been recently shown that efficient simultaneous cooling of a few independent modes is also possible either in the case of using sideband cooling~\cite{Massel2012Multimode}, via machine learning~\cite{Sommer2020Prospects} or cold damping~\cite{Sommer2019Partial}. Alternatively, cooling and strong light-matter couplings can also be achieved in an approach dubbed pulsed optomechanics~\cite{Wang2011ultraefficient,Vanner2011pulsed,Liao2011cooling,Stefanos2013optimal} or in multi-element optomechanical setups involving a few optical and mechanical modes~\cite{Bhattacharya2008multiple,Xuereb2012strong,Ludwig2013quantum,Xuereb2014reconfigurable,Peano2015topological,Raeisi2020quench,Newsom2020optimal,Manjeshwar2020suspended,Deng2020nonreciprocal}.\\
\indent We provide here a more in-depth analytical treatment of simultaneous cold-damping of many mechanical resonances~\cite{Nielsen2017Multimode, Piergentili2018Two, Wei2019Contollable} and address a crucial aspect extremely relevant in experiments, i.e. the inherent time delay $\tau_\text{inh}$ that characterizes any electronic feedback loop. It is generally agreed that the delayed action of the feedback loop can lead to unwanted heating eventually leading to an instability~\cite{Rosinberg2015Stochastic,Rosinberg2017Stochastic,zippilli2018cavity,Loos2019Heat}. Extending the analytical approach that we have previously introduced in Ref.~\cite{Sommer2019Partial} to include a delay time $\tau$, we show that analytical solutions are still possible to some degree in the fast-feedback-lossy-cavity (FFLC) regime. In this regime, lossy cavities allow for quick read-out and broadband feedback allows for quick cooling. Most importantly, we suggest that in order to counteract unwanted heating effects, the feedback loop delay time could be \textit{further delayed} by introducing an additional delay time $\tau_\text{add}$. By fitting the total delay $\tau=\tau_\text{inh}+\tau_\text{add}$ to the characteristics of the system, cooling efficiency close to the level of the $\tau=0$ can in some cases be achieved.\\
\begin{figure}[t]
	\centering
	\includegraphics[width=0.85\columnwidth]{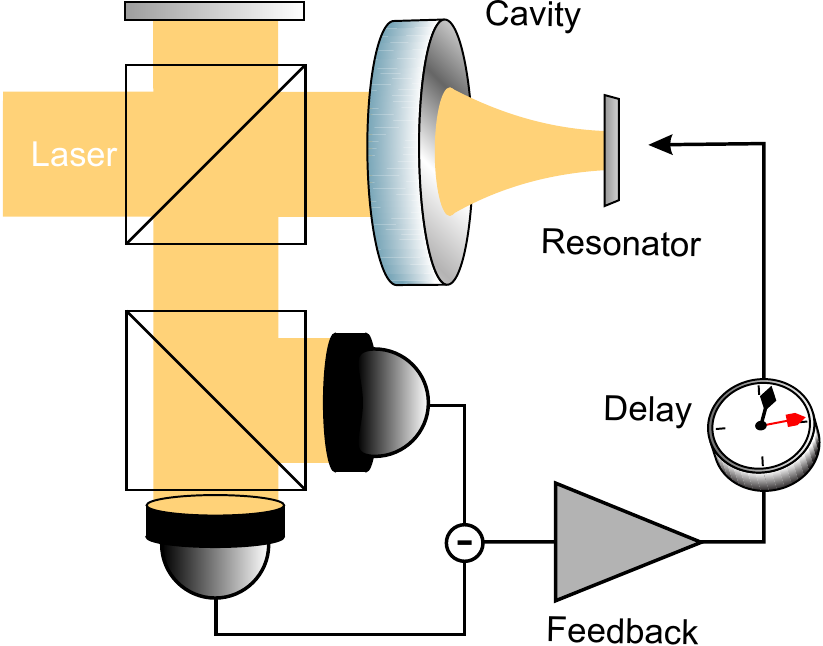}
	\caption{\textit{Schematics.} Cavity optomechanics setup consisting of a vibrating mirror/membrane exhibiting many mechanical resonances and a feedback loop. A cavity output quadrature is monitored and fed into the electronic device which provides the proper cooling action onto the mirror/membrane. The action of the feedback loop is time delayed with $\tau=\tau_\text{inh}+\tau_\text{add}$ consisting of an inherent delay plus an externally controlled additional time $\tau_\text{add}$.}
	\label{fig:single}
\end{figure}
\indent Simultaneous cooling of $N$ mechanical resonances can provide either a wider bandwidth for sensing applications or a stronger optomechanical coupling to a collective mode. We show that a single feedback loop can very efficiently couple to a bright collective mode which, in the near degeneracy case where all mechanical modes lie within a very narrow frequency window, can be up to $N$ times faster damped to a $N$ times lower occupancy than a single mode. This is reminiscent of the superradiance effect as in an increase in the collective radiative rate for a system of $N$ quantum emitters coupled to a single bosonic mode, as in the Dicke model in quantum optics~\cite{Dicke1954coherence,Gross1982Superradiance}. The corresponding effect of subradiance, i.e. strong suppression of radiative rate,  is mimicked by the decoupling of the other $N-1$ collective dark states from the feedback loop. For efficient cooling of many resonances in a wide frequency window, one then has to instead engineer a linear dispersion relation such that all adjacent modes are separated by more than the damping rate introduced by the feedback loop. In terms of collective modes, the spread of many mechanical resonances over a large frequency interval insures strong bright-dark couplings which in turn leads to sympathetic cooling of all dark modes.\\
\indent The analytical treatment followed here is based on solving a system of coupled quantum Langevin equations for $N$ mechanical modes coupled to a single cavity optical mode and subjected to a feedback force. In Sec.~\ref{Model} we introduce the model that includes the feedback force with time delay and detail the procedure that allows for the linearization of the radiation pressure interaction in the high-amplitude field limit. In Sec.~\ref{simplified} we derive the simplified equations of motion for $N$ coupled mechanical degrees of freedom based on the FFLC approximation. The dynamics is generally non-Markovian as the task is to solve a system of coupled integro-differential equations; however, for relatively small time delay, we introduce the weak (wFFLC) and the strong (sFFLC) Markovian approximations which simplify the task by turning the dynamics into a set of coupled linear differential equations. In Sec.~\ref{steady} we provide analytical results for the steady state of the system by using either a time domain analysis, particularly useful under the Markovian approximation but also in the Fourier domain with a generality extending into the non-Markovian regime as well. We benchmark important results in Sec.~\ref{single} for a single resonance, highlighting the role of time delay and showing that an additional delay in the feedback signal with a conveniently chosen $\tau_\text{add}$ can improve cooling efficiency. In Sec.~\ref{two} we extend these results to the two modes case, elucidating the interplay between collective damping and the time delay effects. The different levels of approximations are then tested against exact numerical simulations based on solving time dynamics of a set of stochastic differential equations. Finally, in Sec.~\ref{multi} we present analytical and numerical results for many resonances in particular following a linear dispersion relation and in the case of degenerate modes and illustrate strategies for efficient cooling with adjustable feedback time delay.\\

\section{Model}
\label{Model}

We follow the evolution of an optomechanical system at the level of operators subject to both unitary evolution as well as to dissipation (included as optical and thermal quantum fluctuation input noises, i.e. the standard quantum Langevin approach in optomechanics~\cite{genes2008ground}). The system is comprised of an optical cavity mode coupled via the radiation pressure Hamiltonian to $N$ mechanical resonances of a single vibrating end-mirror. The $N$ independent modes of vibrations have effective mass $m_j$ and frequency $\omega_j$. The quantum Langevin equations of motion~\cite{genes2008ground} for the $N+1$ degrees of freedom read
\begin{subequations}
\begin{align}
\label{App.ModEq.1}
\dot{Q}_j &= \omega_j P_j, \\
\dot{P}_j &= -\omega_j Q_j  -\gamma_j P_j + g_{\text{OM}}^{(j)} A^{\dagger}A + \xi_j,\\
\dot{A} &= -(\kappa + i\Delta_0)A + i\textstyle \sum_{j=1}^{N} g_{\text{OM}}^{(j)} A Q_j + \epsilon + \sqrt{2\kappa}a^{\text{in}}.
\end{align}
\end{subequations}
\indent We have introduced dimensionless position and momentum quadratures $Q_{j}$ and $P_{j}$ for each of the $N$ independent membrane oscillation modes with standard commutations $[Q_j,P_{j'}]=i\delta_{jj'}$. The term $\Delta_0 = \omega_c-\omega_\ell$ describes the detuning of the cavity resonance frequency $\omega_c$ from the laser frequency $\omega_\ell$ and $\kappa$ its decay rate. The input laser power is given by $\epsilon = \sqrt{2\mathcal{P}\kappa/\hbar \omega_\ell}$. The optomechanical coupling is described by the radiation pressure Hamiltonian $\sum_j\hbar g_\text{OM}^{(j)} A^\dagger A Q_j$ where $g_\text{OM}^{(j)}$ is the single-photon-single-phonon coupling rate for the $j$-th mode. The single cavity mode at frequency $\omega$ and loss rate $\kappa$ is described by the bosonic operator $A$ with $[A,A^\dagger]=1$.  The zero-average noise terms are delta-correlated in the time domain $\braket{a^\text{in}(t)a^{\text{in}\dagger}(t')}=\delta(t-t')$. \\
\indent The parameter $\gamma_{j}$ describes the damping of the $j$-th resonator mode and together with the associated zero-averaged Gaussian stochastic noise term $\xi_j$ fulfill the fluctuation-dissipation relation resulting in thermalization with the environment. The noise term can be fully described by the two-time correlation function:
\begin{eqnarray}
\label{App.ModEq.2}
\langle \xi_j(t) \xi_{j'}(t')\rangle &=& \frac{\gamma_{j}}{\omega_j}\int_{-\Omega}^{\Omega} \frac{d\omega}{2\pi}e^{-i\omega(t-t')} S_{\text{th}}(\omega)\delta_{jj'},
\end{eqnarray}
where $\Omega$ is the frequency cutoff of the reservoir and $S_{\text{th}}(\omega)=\omega[\coth\left(\hbar \omega/2k_B T\right) + 1]$ is the thermal noise spectrum. A standard white noise input with delta correlations both in frequency and time is obtained for sufficiently high temperatures $k_B T \gg \hbar \omega_j$ from the correlation function resulting in the approximate form $\langle \xi_j(t) \xi_{j'}(t')\rangle \approx (2\bar{n}_{j}+1)\gamma_{j}\delta(t-t')\delta_{jj'}$, where $\bar{n}_{j} = (\exp(\hbar \omega_j/k_B T)-1)^{-1} \approx k_B T/\hbar \omega_j$ describes the average occupancy of each vibrational mode.\\

\subsection{Linearization}
Let us rewrite all operators $A = \langle A \rangle + a$, $Q_j = \langle Q_j \rangle + q_j$ and $P_j = \langle P_j \rangle + p_j$ as a sum of their expectation value and zero-averaged fluctuations. When the cavity field amplitude is large with respect to the fluctuations, one can simplify the equations of motion by neglecting terms such as $\braket{a^\dagger a}$ as being small compared to $|\braket{A}|^2$. Under this approximation, the classical averages satisfy the following equations of motion
\begin{subequations}
\begin{align}
\label{App.ModEq.3}
\langle \dot{Q}_j \rangle &= \omega_{j} \langle P_{j}\rangle,\\
\langle \dot{P}_j \rangle &= -\omega_{j} \langle Q_{j} \rangle - \gamma_j \langle P_j \rangle + g_{\text{OM}}^{(j)} |\langle A \rangle|^2,\\
\langle \dot{A}\rangle &= -(\kappa + i\Delta_0)\langle A \rangle + i\sum_{j=1}^{N} g_{\text{OM}}^{(j)} \langle A \rangle \langle Q_j \rangle + \epsilon.
\end{align}
\end{subequations}
\indent In steady state (obtained by setting $\langle \dot{Q}_j \rangle = \langle \dot{P}_j \rangle = \langle \dot{A} \rangle = 0$) the cavity field amplitude can be shown to satisfy the following non-linear equation
\begin{eqnarray}
\label{App.ModEq.4}
\langle A \rangle &=& \frac{\epsilon}{\left[\kappa + i\left(\Delta_0 - \sum_{j} \left( g_{\text{OM}}^{(j)} \right)^2|\langle A \rangle|^2 /\omega_j\right) \right]}.
\end{eqnarray}
The non-linearity stems from the intensity dependent cavity detuning $\Delta = \Delta_0 - \sum_{j} \left( g_{\text{OM}}^{(j)}\right)^2|\langle A \rangle|^2/\omega_j$ owing to the radiation pressure induced displacement from equilibrium. The steady state value for the displacements are $\langle Q_j \rangle = (g_{\text{OM}}^{(j)}/\omega_j)|\langle A \rangle|^2$. \\
\indent With the omission of the small nonlinear terms $a^{\dagger}a$ and $a q_j$, we obtain the linearized equations of motion
\begin{subequations} \label{App.ModEq.5}
\begin{align}
\dot{q}_j &= \omega_j p_j,\\
\dot{p}_j &= -\omega_j q_j - \gamma_j p_j + \xi_j + G_j x,\\
\dot{x} &= -\kappa x + \Delta y + \sqrt{2\kappa} x^{\text{in}}, \\
\dot{y} &= -\kappa y - \Delta x + \textstyle \sum_{j=1}^{N} G_jq_j + \sqrt{2\kappa}y^{\text{in}},
\end{align}
\end{subequations}
where $x = (1/\sqrt{2})(a+ a^{\dagger})$ and $y = (i/\sqrt{2})(a^{\dagger}-a)$ are the fluctuations of the quadratures of the cavity field and $x^{\text{in}}$ and $y^{\text{in}}$ are similarly defined in terms of field noise operators. The effective optomechanical coupling terms are given by $G_j=\sqrt{2}g_\text{OM}^{(j)}\braket{A}$ and are enhanced by the large cavity field amplitude. We set the condition that the effective cavity detuning $\Delta=\omega_c-\omega_\ell-\textstyle \sum_j g_{\text{OM}}^{(j)} \braket{Q_j}$, containing a collective mechanically-induced frequency shift is kept at zero value.\\

\subsection{Time-delayed feedback}

The application of the feedback requires the readout of a cavity field quadrature followed by the appropriate action onto the mechanical resonator. Generally, one can express the applied force as
\begin{equation}
F_j = - g^{(\tau)}_j*y^{\text{est}},
\end{equation}
where the convolution term is defined as $(g^{(\tau)}_j \ast y)(t)=\int_{-\infty}^{\infty}ds g^{(\tau)}_j(t-s)y(s)$ and depends on the past of the detected quadrature $y$ that is driven by the weighted sum of the oscillator fluctuations $q_j$. Here, we focus on a particular form of negative derivative feedback also known as cold damping. This form of feedback applies a correcting cooling viscous force proportional to the resonator's velocity and has been experimentally employed to cooling of mirrors~\cite{Cohadon1999Cooling}, microtoroids~\cite{Wilson2015Measurement} and levitated nano-particles~\cite{tebbenjohanns2019cold}. The correcting force can either be applied optically or electrically. In the case of optically-based feedback one has a choice to either apply a second laser beam or simply modulate the cavity input field~\cite{Rossi2017Enhancing,zippilli2018cavity}. The causal kernel for negative derivative feedback can be modelled by the following function
\begin{align}
\label{gcd}
g^{(\tau)}_j(t) =g_\text{cd}^{(j)} \partial_t \left[ \theta(t-\tau)\omfb e^{-\omfb (t-\tau)}\right]
\end{align}
and contains the feedback gain terms $g_\text{cd}^{(j)}$ and feedback bandwidth $\omfb$.
The Fourier transform of the feedback kernel is given by
\begin{eqnarray}
\label{DCDQKernel}
g^{(\tau)}_j(\omega) = \frac{i\omega g_\text{cd}^{(j)} e^{-i\omega\tau}}{1+i(\omega/\omega_{\text{fb}})}= g^{(0)}_j(\omega) e^{-i\omega\tau}
\end{eqnarray}
which resembles a standard derivative high-pass filter which here additionally contains a delay dependent phase term (similar to the term expressed in~\cite{zippilli2018cavity}) in contrast to the previous case \cite{genes2008ground, Sommer2019Partial}. In addition, the parameter $\tau$ (neglected previously in Ref.~\cite{Sommer2019Partial}) is the joined feedback delay originating from the measurement signal processing. Within the convolution with $\theta(t-\tau-s)$ this parameter guarantees that only information up to $t-\tau$ can influence the dynamics of the resonator modes. Notice that in the limit $\omfb \rightarrow \infty$ the feedback becomes $g^{(\tau)}_j(t) = g_\text{cd}^{(j)}\partial_t \delta(t-\tau)$.\\
\indent The quadrature component that is injected into the feedback mechanism $y^{\text{est}}$ is the estimated intra-cavity phase quadrature given by
\begin{equation}
\label{App.Full2}
y^{\text{est}}(t) = y(t) - \frac{y^{\text{in}}(t) + \sqrt{\eta^{-1} -1}y^v(t)}{\sqrt{2\kappa}}.
\end{equation}
which results from a measurement of the output quadrature $y^{\text{out}} = \sqrt{2\kappa} y(t) - y^{\text{in}}(t)$. This follows from the description of a detector with quantum efficiency $\eta$ which is modeled by an ideal detector preceded by a beam splitter with transmissivity $\sqrt{\eta}$, which mixes the input field with an uncorrelated vacuum field $y^v(t)$.\\
\indent Finally, to fully describe the dynamics of an optomechanical system with $N$ mechanical resonances undergoing cold damping with time delay one corrects Eqs.~\eqref{App.ModEq.5} with the following equation
\begin{equation}
\label{DCDEq.1}
\dot{p}_j = -\omega_j q_j - \gamma_j  p_j + G_j x- g^{(\tau)}_j*y^{\text{est}}+ \xi_j.
\end{equation}
\indent In the next two sections we will describe strategies that can be employed to simplify the equations of motion and deliver Markovian approximations accurately describing the time dynamics of the system for small time delays. The approximations also allow for the derivation of analytical estimates of final occupancies for all modes undergoing cold damping. An extension beyond the Markovian regime will then be obtained by a Fourier analysis of the coupled system of equations in steady state.

\section{Multi-mode cold damping: simplified equations}
\label{simplified}
The aim of this section is to arrive at a set of $2N$ coupled equations describing solely the dynamics of the $N$ mechanical resonator modes. To this end we proceed by formally integrating the equations of motion for the optical degree of freedom and replacing them in the equations for the mechanical modes. We first find a general formulation for the integro-differential non-Markovian collective dynamics where feedback-induced damping occurs as a time convolution involving momentum quadratures evaluated at times in the past. Under the FFLC approximation, Markovian collective dissipative dynamics emerges, allowing one to write a linear set of coupled differential equations analytically solvable in steady state.\\

\subsection{Non-Markovian collective damping}
Let us start by formally integrating the dynamics of the optical degrees of freedom. For all further calculations we use the condition that the effective cavity detuning is set to $\Delta = 0$. For the optical degrees of freedom we obtain
\begin{subequations}
\begin{align}
x(t) &= \sqrt{2\kappa} \int^{t}_{-\infty} ds e^{-\kappa(t-s)}x^{\text{in}}(s),\\
\label{App.Bright}
y(t) &= \int^{t}_{-\infty}ds e^{-\kappa (t-s)}\sum_{j=1}^{N}G_j q_{j}(s) \\\nonumber
& +\sqrt{2\kappa}\int^{t}_{-\infty}ds e^{-\kappa (t-s)} y^{\text{in}}(s).
\end{align}
\end{subequations}
We will aim at computing the estimated quadrature $y^{\text{est}}(s)$ which introduces both terms proportional to $q_j$ as well as noise terms stemming from the cavity input noise $y^{\text{in}}(s)$ and from the vacuum filled port noise $y^v$. This will then give rise to the convoluted force acting on all mechanical momentum quadratures. In a first step we estimate the contribution coming from the intracavity field $y$ as
\begin{widetext}
\begin{align}
\label{App.Eq1}
g^{(\tau)}_j \ast y = \int_{-\infty}^{t-\tau} ds \frac{\kappa e^{-\kappa(t-s-\tau)} - \omfb e^{-\omfb (t-s-\tau)}}{(\kappa - \omfb )}\left[\textstyle \sum_{k=1}^{N} g^{(j)}_{\text{cd}}\omfb G_k q_{k}(s) + \sqrt{2\kappa}g^{(j)}_{\text{cd}}\omfb y^{\text{in}}(s)\right].
\end{align}
\end{widetext}
Notice that the feedback force above contains terms proportional to the mechanical displacement quadratures in addition to an extra feedback induced noise. It is however desired to express the effect of feedback as a damping force proportional to a momentum quadrature. To this end we apply integration by parts, noticing that the convolution contains a derivative of the following function $h_{\tau}(t-s)=\left[e^{-\kappa(t-s-\tau)}-e^{-\omfb (t-s-\tau)}\right]/(\omfb -\kappa)$. In addition, we make use of the relation $\dot{q}_j = \omega_j p_j$ to obtain
\begin{widetext}
\begin{align}
\label{App.Eq2}
g^{(\tau)}_j \ast y = \sum_{k=1}^{N} g^{(j)}_{\text{cd}}\omfb G_k \omega_k \int_{-\infty}^{t-\tau}ds h_{\tau}(t-s) p_k(s)- \sqrt{2\kappa}g^{(j)}_{\text{cd}}\omfb  \int_{-\infty}^{t-\tau} ds \partial_sh_{\tau}(t-s) y^{\text{in}}(s).
\end{align}
\end{widetext}
The feedback force explicitly shows a damping term exhibiting a non-local kernel: this indicates that the action depends on the past behavior of the momentum quadratures on timescales defined by the cavity loss rate $\kappa$ and on the feedback bandwidth $\omfb$. We can list the set of coupled $2N$ equations for all resonator modes quadratures
\begin{widetext}
\begin{subequations}
\begin{align}
\label{nonMarkovian}
\dot{q}_j &= \omega_j p_j, \\\nonumber
\dot{p}_j &= -\omega_j q_j - \int^{\infty}_{-\infty}ds\left[\gamma_j\delta(t-s) + g^{(j)}_{\text{cd}}\omfb G_j\omega_j \theta(t-s-\tau)h_{\tau}(t-s)\right] p_j(s) - \sum_{k \neq j} g^{(j)}_{\text{cd}}\omfb G_k\omega_k\int^{t-\tau}_{-\infty} ds  h_{\tau}(t-s) p_k(s)\\
& + \xi_j+\xi_\text{fb}+ \xi_\text{vac}+ \xi_\text{rp}.
\end{align}
\end{subequations}
\end{widetext}
Here, we have broken up the sum stated in Eq.~\eqref{App.Eq2} into a self term containing $p_j$ and a sum over the cross terms $p_k$.\\
The Markovian damping rate $\gamma$ for each mode is supplemented with a diagonal non-Markovian feedback damping kernel as well as with off-diagonal dissipative couplings to all other modes. The three sources of noise, in addition to the thermal one $\xi_j$, stem from the direct feedback action $\xi_\text{fb}$, from the feedback filtered vacuum action in the loss port $\xi_\text{vac}$ and from the intra-cavity radiation pressure effect $\xi_\text{rp}$. They are expressed as
\begin{subequations}
\begin{align}
\label{App.Full5}
\xi_\text{fb} &= - \frac{g^{(j)}_{\text{cd}}\omfb }{\sqrt{2\kappa}}\int^{\infty}_{-\infty}ds \phi^{(\tau)}_1(t-s)y^{\text{in}}(s), \\
\xi_\text{vac} &= - \frac{g^{(j)}_{\text{cd}}\omfb }{\sqrt{2\kappa}}\sqrt{\eta^{-1}-1}\int^{\infty}_{-\infty} ds \phi^{(\tau)}_2(t-s)y^v(s),\\
\xi_\text{rp} &= \sqrt{2\kappa}G_j\int^{\infty}_{-\infty}ds \phi_3(t-s)x^{\text{in}}(s).
\end{align}
\end{subequations}
The explicit forms of the convolution kernels $\phi^{(\tau)}_{1}(t)$, $\phi^{(\tau)}_{2}(t)$ and $\phi_{3}(t)$ are given in Appendix~\ref{A}.\\

%%%%%%%%%%%%%%%%%%%%%%%%%%%%%%%%%%%%%%%%%%%
\subsection{Markovian collective damping}
%%%%%%%%%%%%%%%%%%%%%%%%%%%%%%%%%%%%%%%%%%%
The set of integro-differential equations obtained above is not easily tractable; however, in regimes favorable to cold-damping, a transformation to a much simpler form can be achieved. Let us assume a lossy cavity and relatively fast feedback such that both rates fulfill $\kappa, \omfb  \gg \omega_j$, for all $j\in \{1,\dots,n\}$. In such a case, integration by parts of the non-Markovian kernel can be performed and one can show that, in leading order, the convolution gives rise to a very simple expression
\begin{equation}
\label{App.Markov1}
\int^{t-\tau}_{-\infty}ds h_{\tau}(t-s) p_j(s) \approx   \frac{p_{j}(t-\tau)}{\omfb \kappa}.
\end{equation}
We denote this approximation as the strong fast-feedback lossy-cavity (sFFLC) assumption. Under less stringent conditions with $\kappa, \omfb  > \omega_j$ (which we will refer to as the weak wFFLC condition) a slightly more complicated expression with a larger validity region can be derived and is detailed in Appendix~\ref{A},~\ref{E}.\\
Using Eq.~\eqref{App.Markov1} we derive the non-Markovian equations of motion
\begin{subequations}
\label{Eq.NonMark1}
\begin{align}
\dot{q}_j(t) &= \omega_j p_j(t),\\\nonumber
\dot{p}_j(t) &= -\omega_j q_j(t) - \gamma_j p_j(t) - \sum_{k} \Gamma_{jk}p_k(t -\tau) + \zeta_j(t),\\
\end{align}
\end{subequations}
that depend on past events at time $t-\tau$ (higher order delay terms up to $p_k(t - 2\tau)$ are derived in Appendix~\ref{E} and indicate a dependence following $p_k(t-n\tau)$ for even higher orders). Here, $\Gamma_{jk} = g_{\text{cd}}^{(j)}G_{k}\omega_{k}/\kappa$ is the feedback induced damping rate in the sFFLC approximation (the analogue equations to Eq.~\eqref{Eq.NonMark1} in the case of the wFFLC are given in Appendix~\ref{A}). Notice that all noise terms have been gathered into a single term $\zeta_j = \xi_j+\xi_\text{fb}+ \xi_\text{vac}+ \xi_\text{rp}$. For $\tau=0$ the above expression can be plugged back in to give rise to a set of $2N$ coupled equations where all momentum quadratures are evaluated at the same time $t$ (as detailed in Ref.~\cite{Sommer2019Partial}).\\
\indent For $\tau>0$, however, the equations are more complicated to solve as momenta at time $t$ are coupled to momenta in the past at $t-\tau$. The next important simplification that we use implies that during the time $\tau$ the motion stays periodic with period $\omega_j$. Then one can roughly approximate $p_j(t-\tau)\approx \left[p_{j}(t)\cos(\omega_j \tau) + q_{j}(t)\sin(\omega_j \tau)\right]$. The approximation is valid as long as damping of oscillations within the interval $\tau$ can be neglected. In the sFFLC, the equations of motion become then a simple set of coupled linear quantum Langevin equations
\begin{widetext}
\begin{subequations}
\label{DCDEq.4}
\begin{align}
\dot{q}_j &= \omega_j p_j,\\
\dot{p}_j &= -\left[\omega_j + \Gamma_{jj}\sin(\omega_j \tau)\right] q_j - \left[\gamma_j + \Gamma_{jj} \cos(\omega_j \tau) \right] p_j- \sum_{k\neq j} \Gamma_{jk}\left\{\sin(\omega_k \tau) q_k + \cos(\omega_k \tau) p_k\right\} + \zeta_j.
\end{align}
\end{subequations}
\end{widetext}
Again, for $\tau = 0$ the result reduces to that previously derived in Ref.~\cite{Sommer2019Partial}. Notice that the main effect of non-zero time delay $\tau>0$ is to modify both the individual mode and mutual damping rates by the cosine factors $\cos(\omega_j \tau)$. This also means that for given time delays the system can exhibit instabilities when the feedback-induced heating rate surpasses the natural decay rate $\gamma$. The extra frequency renormalization terms proportional to $\sin(\omega_j \tau)$ are negligible as long as $\omega_j \gg \Gamma_{jk}$, a regime to which we will restrict ourselves in the following.\\
\begin{figure}[t]
	\centering
	\includegraphics[width=0.75\columnwidth]{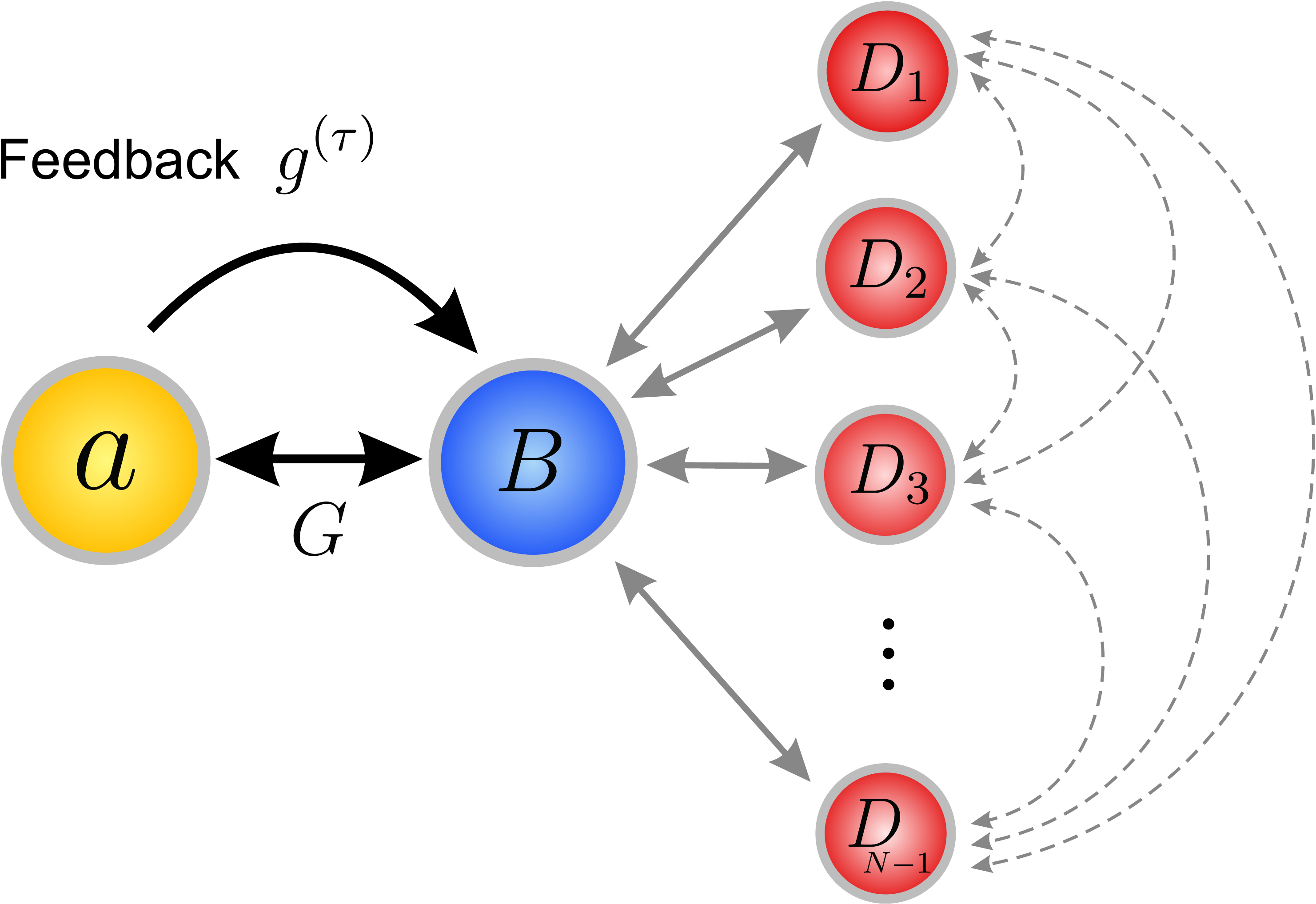}
	\caption{\textit{Bright-dark modes dynamics.} Both the cavity mode and the feedback loop couple to the position quadrature of a collective bright mode $B$. Cooling of the dark state manifold (containing $N-1$ collective dark modes $D_j$) takes place in an indirect, sympathetic way via the dark-bright couplings, strongly dependent on the dispersion relation of the mechanical resonator.}
	\label{fig:Coupling}
\end{figure}
\indent A particularly interesting regime is that of full frequency degeneracy where $\omega_j = \omega$. A simplified picture can be used in this case in terms of collective bright $\mathcal{Q}_1 = \sum_k \alpha_{1k} q_k$ and dark modes $\mathcal{Q}_j = \sum_k \alpha_{jk} q_k$ (See Fig.~\ref{fig:Coupling}). Here, the coefficients from the bright mode $\alpha_{1j} = G_j/\sqrt{\sum_k G^2_k}$ can help to acquire the coefficients for the $N-1$ dark modes via the Gram-Schmidt procedure which satisfies the condition $\sum_j \alpha_{lj}\alpha_{kj} = \delta_{lk}$. The bright mode dynamics is described by
\begin{subequations}
\begin{align}
\dot{\mathcal{Q}}_1 &= \omega \mathcal{P}_1 , \\\nonumber
\dot{\mathcal{P}}_1 &= -\left(\omega + \sum_k \Gamma_{kk}\sin(\omega\tau) \right)\mathcal{Q}_1 \\
&- \left(\gamma + \sum_k \Gamma_{kk}\cos(\omega\tau) \right)\mathcal{P}_1 + \sum_k\alpha_{1k} \zeta_k,
\end{align}
\end{subequations}
while the other $N-1$ orthogonal dark modes satisfy the following equations of motion
\begin{subequations}
\begin{align}
\dot{\mathcal{Q}}_j &= \omega \mathcal{P}_j ,\\\nonumber
\dot{\mathcal{P}}_j &= -\omega \mathcal{Q}_j - \gamma \mathcal{P}_j - \left( \sum_k \frac{\alpha_{jk}}{\alpha_{1k}}\Gamma_{kk}\sin(\omega\tau)\right)\mathcal{Q}_1 \\
&- \left( \sum_k \frac{\alpha_{jk}}{\alpha_{1k}}\Gamma_{kk}\cos(\omega\tau)\right)\mathcal{P}_1 + \sum_k\alpha_{jk} \zeta_k .
\end{align}
\end{subequations}
The dark state manifold dynamics can be further simplified by injecting the solution for the bright mode resulting in
\begin{subequations}
\begin{align}
\dot{\mathcal{Q}}_j &= \omega \mathcal{P}_j , \\
\dot{\mathcal{P}}_j &= -\omega \mathcal{Q}_j - \gamma \mathcal{P}_j + \Xi_j ,
\end{align}
\end{subequations}
where the expression for the compound noise term $\Xi_j$ is detailed in Appendix~\ref{A}. The above dynamics shows that in the fully degenerate case the bright mode is damped at a high rate: for equal coupling this rate is directly proportional to $N$. The dark modes are instead mostly unaffected by the feedback loop (except via the input noise) and decay at the natural decay rates $\gamma$. This is reminiscent of the collective monitoring of a collection of quantum emitters by their electromagnetic environment (either free space or cavity) which leads to collective radiative effects known as super- and subradiance~\cite{Dicke1954coherence,Gross1982Superradiance}. The role of the collective bath is played here by the common electronic feedback loop which provides simultaneous dissipative dynamics for all modes.\\

\section{Multi-mode cold damping: steady state}
\label{steady}

To derive the final achievable occupancies for all modes undergoing cold-damping, one can compute the covariance matrix of the system in steady state. We will follow two different paths: i) a time domain analysis suitable to the Markovian case, where the covariance matrix of the system can be computed from the Lyapunov equation and ii) a Fourier domain analysis which only applies to the steady state but presents the advantage of providing exact solutions in the Fourier domain even for the non-Markovian case.\\

\subsection{Time domain analysis}

The linearized quantum Langevin equations presented in Eq.~\eqref{DCDEq.4} can be rewritten in compact vector form
\begin{equation}
\label{App.Lya1}
\dot{\bold{v}} = M \bold{v} + \bold{n}_{\text{in}},
\end{equation}
with a vector of fluctuations $\bold{v} = (q_1, p_1, \dots q_N, p_N)^{\top}$ and the corresponding input noise vector $ \bold{n}_{\text{in}} = (0, \zeta_1, \dots , 0, \zeta_N)$. Here, the elements of the matrix $M$ are defined by the coefficients for $q_j$ and $p_j$ in Eq.~\eqref{DCDEq.4}. Under the condition that the system is stable, i.e. if all eigenvalues of $M$ have negative real parts, one can find the covariance matrix
\begin{subequations}
\begin{align}
\label{App.Lya2}
V &= \langle \bold{v}(t) \bold{v}^{\top}(t) \rangle\\
&= \int^{t}_{-\infty}ds \int^{t}_{-\infty}ds' e^{M(t-s)}\langle \bold{n}_{\text{in}}(s) \bold{n}_{\text{in}}^{\top}(s')\rangle e^{M^{\top}(t-s')}.
\end{align}
\end{subequations}
This can be greatly simplified as all two-time correlations of noise terms $\langle \bold{n}_{\text{in}}(s) \bold{n}_{\text{in}}^{\top}(s')\rangle$ are delta-like in the strong fast-feedback-lossy-cavity regime with $\omega_{\text{fb}}$, $\kappa \gg \omega_{j}$. One can show that
\begin{eqnarray}
\label{App.Lya4}
\langle n_{\text{in},2i}(s)n_{\text{in},2j}(s') \rangle &=& \langle \zeta_{i}(s)\zeta_{j}(s')\rangle \\\nonumber
&\approx & \left( (2\bar{n}_{i}+1)\gamma_{i}\delta_{ij} + \frac{G_iG_j}{\kappa}\right)\delta(s-s'),
\end{eqnarray}
and zero otherwise. These terms can be gathered into a diffusion matrix with elements $\mathcal{D}_{\text{in},2i,2j} = (2\bar{n}_{i}+1)\gamma_{i}\delta_{ij} + G_iG_j/\kappa$ and zero otherwise. The final step involves solving a Lyapunov equation for the covariance matrix
\begin{eqnarray}
\label{App.Eq3}
MV + VM^{\top} = -\mathcal{D}_{\text{in}}.
\end{eqnarray}
Notice that the diffusion matrix shows no dependence on the time delay.\\
\indent One can proceed to solve the Lyapunov equation by introducing the following notations for momentum correlations, $X_{ij} = \braket{p_i p_j + p_j p_i}$, position correlations $Z_{ij} = \braket{q_i q_j + q_j q_i}$ and cross-terms $Y_{ij} = \braket{q_i p_j + p_j q_i}$. From the diagonal elements $Z_{ii}$ and $X_{ii}$ one can then estimate the final occupancy of each mode. With these notations, analytical results can be obtained by solving the following set of algebraic equations (simplified below under the sFFLC approximation)
\begin{widetext}
\begin{subequations}
\label{DCDEq.4a}
\begin{align}
\label{App.Eq4}
Y_{ii} &= 0, \\
\omega_j Y_{ij} +\omega_i Y_{ji} &= 0,\\
(\gamma_i + \Gamma_{ii}c_i) X_{ii} + \sum_{j\neq i}\Gamma_{ij}s_{j} Y_{ji} + \sum_{j\neq i}\Gamma_{ij}c_j X_{ij} - (2\bar{n}_i + 1)\gamma_i - \frac{G_i^2}{\kappa} &= 0, \\
\omega_i X_{ii} - \left(\omega_i  + \Gamma_{ii}s_{i}\right)Z_{ii} - \sum_{j \neq i} \Gamma_{ij}s_{j} Z_{ij} - \sum_{j \neq i} \Gamma_{ij}c_j Y_{ij} &= 0, \\
\omega_j X_{ij} - \left(\omega_i + \Gamma_{ii}s_i \right)Z_{ij} - \left(\gamma_i + \Gamma_{ii}c_i \right)Y_{ij} - \sum_{k\neq i} \Gamma_{ik}s_k Z_{jk} - \sum_{k\neq i} \Gamma_{ik}c_k Y_{jk}  &= 0, \\
\label{App.Eq4a1}
-\frac{\left(\omega_i^2 - \omega_j^2 \right)}{\omega_i} Y_{ij} - \sum_{k \neq j} \Gamma_{ik}s_k Y_{kj} - \sum_{k \neq i} \Gamma_{jk}s_k Y_{ki}  - \sum_{k} \Gamma_{ik}c_k X_{jk} - \sum_{k} \Gamma_{jk}c_k X_{ik}  + \frac{2G_{i}G_{j}}{\kappa}  &= 0.
\end{align}
\end{subequations}
\end{widetext}
For the wFFLC analogue equations can be obtained in the case $k_{\text{B}}T \gg \hbar \omega_j$ which are presented in Appendix~\ref{A}, where we can ignore the contribution from feedback, measurement and radiation pressure noise terms which do not show delta-like correlations in this regime. The form above is more complex than the $\tau=0$ case studied in Ref.~\cite{Sommer2019Partial} showing the influence of the delay in the weight factors $c_i = \cos(\omega_i \tau)$ and $s_i = \sin(\omega_i \tau)$ multiplying with the rates $\Gamma_{ij}$. Setting $\tau=0$, quasi-exact but cumbersome expressions for the final occupancy of each mode can be obtained (as detailed in Appendix~\ref{A} and presented in \cite{Sommer2019Partial}). The non-zero delay case however is more complicated and analytical expressions are harder to obtain except in the single mode and two adjacent modes case which we detail in the next sections.\\

\subsection{Fourier domain analysis}

The time-domain analysis provided above has a domain of validity restricted by the Markovian assumptions implied in the sFFLC and wFFLC approximations. However, in steady state, one can turn the integro-differential set of coupled equations into a simple set of algebraic equations by transforming to the Fourier domain. This allows for solutions inside the non-Markovian regime. Let us write the equations
\begin{subequations}
\begin{align}
\label{FouEq.1}
i\Omega q_j(\Omega) &= \omega_j p_j(\Omega),\\
\label{App.Four2b}
i\Omega p_j(\Omega) &= -\omega_j q_j(\Omega) - \gamma_j  p_j(\Omega) + G_j x(\Omega)\\\nonumber
& - g^{(\tau)}_j(\Omega) y^{\text{est}}(\Omega) + \xi_j(\Omega),\\
\label{App.Four2c}
i\Omega x(\Omega) &= -\kappa x(\Omega) + \sqrt{2\kappa} x^{\text{in}}(\Omega), \\
\label{App.Four2d}
i\Omega y(\Omega) &= -\kappa y(\Omega) + \textstyle \sum_{j=1}^{N} G_jq_j(\Omega) + \sqrt{2\kappa}y^{\text{in}}(\Omega),
\end{align}
\end{subequations}
and proceed by eliminating the Fourier components of the field quadratures. One then obtains a set of N equations which allows for the derivation of each mode's response to the input noise
\begin{widetext}
\begin{subequations}
\label{FouEq.2}
\begin{align}
\frac{(\omega^2_{j,\text{eff}}(\Omega) - \Omega^2 ) + i\Omega\gamma_{j,\text{eff}}(\Omega)}{\omega_j}q_{j}(\Omega) + \sum_{k\neq j} \left(\frac{g^{(j)}_{\text{cd}}}{g^{(k)}_{\text{cd}}}\right)\frac{(\omega^2_{k,\text{eff}}(\Omega)-\omega^2_k) + i\Omega(\gamma_{k,\text{eff}}(\Omega)-\gamma_k)}{\omega_k} q_k(\Omega) &= \zeta_j(\Omega),\\
\sum_{k=1}^{N}(\chi^{-1})_{jk}(\Omega)q_k(\Omega) &= \zeta_j(\Omega).
\end{align}
\end{subequations}
\end{widetext}
The matrix $\pmb{\chi}$ describes the susceptibility matrix of the system. We have introduced a frequency-dependent effective resonance frequency $\omega^2_{j,\text{eff}}(\Omega) =\omega^2_j + \Omega\delta\tilde{\omega}_j(\Omega)\cos(\Omega\tau) + \Omega\tilde{\Gamma}_j(\Omega)\sin(\Omega\tau)$ and the corresponding frequency-dependent effective decay rate $\gamma_{j,\text{eff}}(\Omega) =\gamma_j + \tilde{\Gamma}_j(\Omega)\cos(\Omega\tau) - \delta \tilde{\omega}_j(\Omega)\sin(\Omega\tau)$. Additionally, we obtain the terms
\begin{subequations}
\begin{align}
\delta \tilde{\omega}_j(\Omega) &= \frac{g^{(j)}_{\text{cd}}G_j\omega_j\omega_{\text{fb}}\Omega \left(\omega_{\text{fb}} + \kappa \right)}{\left(\kappa^2+\Omega^2 \right)\left(\omega_{\text{fb}}^2+\Omega^2 \right)} \\
\tilde{\Gamma}_j(\Omega) &= \frac{g^{(j)}_{\text{cd}}G_j\omega_j \omega_{\text{fb}} \left(\kappa\omega_{\text{fb}}-\Omega^2 \right)}{\left(\kappa^2+\Omega^2 \right)\left(\omega_{\text{fb}}^2+\Omega^2 \right)}.
\end{align}
\end{subequations}
It is interesting to note that one can immediately obtain the wFFLC steady state computed in the time domain, under the approximation $\omega^2_{j,\text{eff}}(\Omega) \approx \omega^2_{j,\text{eff}}(\omega_j)$ and $\gamma_{j,\text{eff}}(\Omega) \approx \gamma_{j,\text{eff}}(\omega_j)$.
\\
\indent To obtain the steady state solution of the resonator mode occupation from the Fourier transform we need to calculate
\begin{eqnarray}
\label{FouEq.3}
\nonumber
\left(n_{\text{eff}}\right)_j &= \frac{1}{2\pi}\int^{\infty}_{\infty}d\Omega \frac{1}{2}\left[ \left(\pmb{\chi}(\Omega)\pmb{S}(\Omega)\pmb{\chi}^{\dagger}(\Omega)\right)_{jj}\left(1 + \frac{\Omega^2}{\omega^{2}_{j}} \right) \right],\\
\end{eqnarray}
where $\pmb{S}(\Omega)$ describes the position-fluctuation spectrum as presented in the Appendix~\ref{C}, which for high temperatures $k_{\text{B}}T \gg \hbar \omega_j$ can be approximated by $S_{jk}(\Omega) \approx \gamma_j(2\bar{n}_j+1)\delta_{jk}$. In the following we will refer to the term $S_{q_j}(\Omega) = \left(\pmb{\chi}(\Omega)\pmb{S}(\Omega)\pmb{\chi}^{\dagger}(\Omega)\right)_{jj}$ as the position spectrum of the $j$-th mode.\\

\section{Single mode cooling}
\label{single}
We provide here an analytical and numerical treatment of the time dynamics and steady state final occupancies for a single mode undergoing cold damping with a variable time delay. In steady state, the solutions to the Lyapunov equations under the weak or strong FFLC approximation provide simple, intuitive results for the final achievable occupancies. We provide a numerical validity check for the steady state solutions and extend the analytical calculations to regions of more general validity by solving the coupled set of non-Markovian equations in the Fourier domain.
%%%%%%%%%%%%%%%%%%%%%%%%%%%%
%%%Figure
%%%%%%%%%%%%%%%%%%%%%%%%%%%%
\begin{figure*}[t]
	\centering
	\includegraphics[width=1.95\columnwidth]{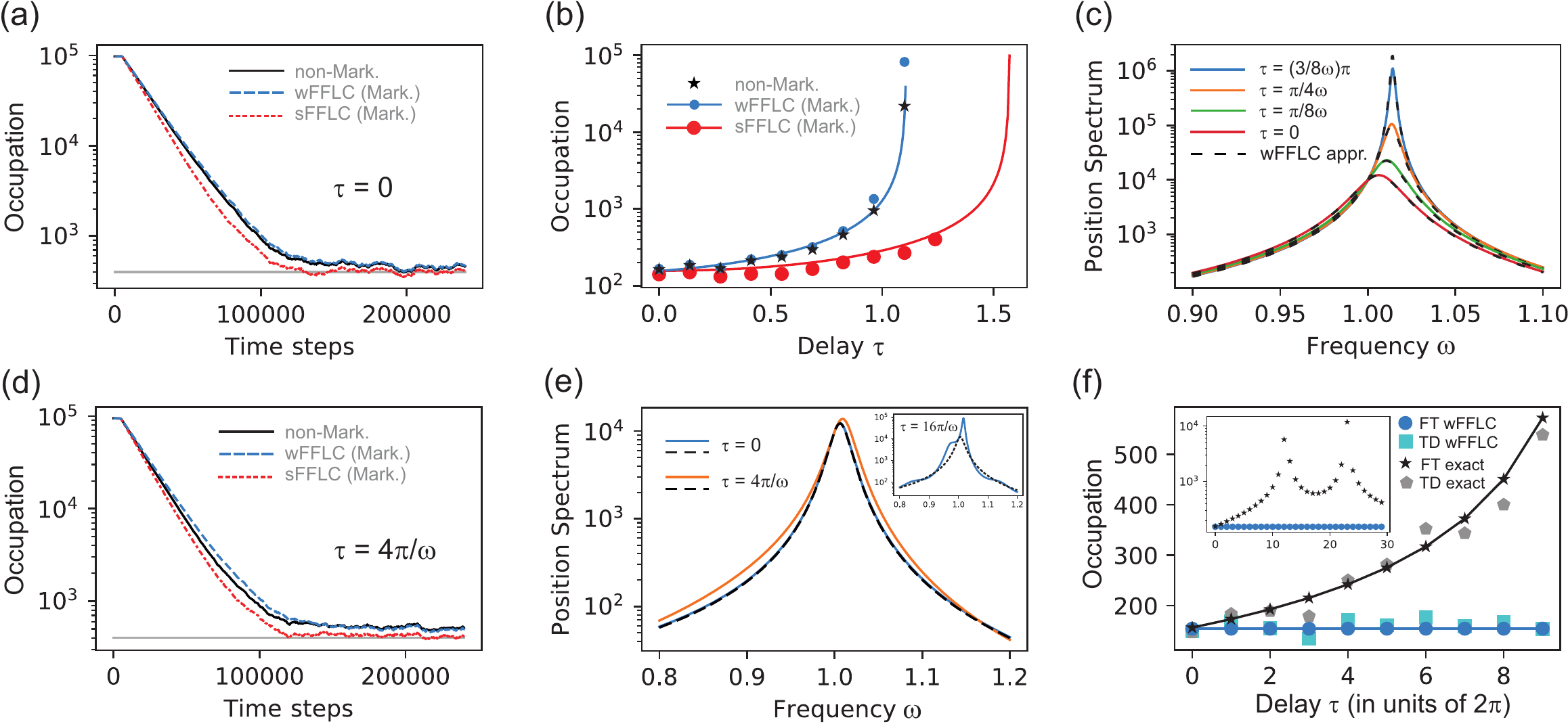}
	\caption{\textit{Single mode cooling with delay.} (a) Effective occupancy (in logarithmic scale) as a function of time with delay $\tau=0$ for a weak (wFFLC) and strong fast-feedback-lossy-cavity (sFFLC) approximation as well as the full non-Markovian convolutional treatment. (b) The dependence of $n_\text{eff}$ as a function of time delay in units of $\omega^{-1}$ varied from 0 close to $\pi/2$ is presented by the solid lines and from numerical trajectory simulations by the stars and dots. Here, the wFFLC approximates the correct behavior much better for delays close to instability regions. Cases approaching $\pi/2$ lead to inefficient damping and eventually an instability as $\omega\tau>\pi/2$. (c) The position spectrum is presented for four values in the range $\tau < \pi/2\omega$ (solid lines) and compared to their wFFLC-approximations (black dashed lines). (d) Effective occupancy (in logarithmic scale) as a function of time delay of $\tau = 4\pi/\omega$. (e) The position spectrum for $\tau = 0$ and $\tau = 4\pi/\omega$ shows that with increasing delay the wFFLC-approximation (dashed lines) deviates more strongly from the exact solution (solid lines), which is especially clear for $\tau = 16\pi/\omega$ shown in the inset. (f) Final occupancies obtained by the temporal evolution (pentagons for the exact and squares for the wFFLC approximation) and via Fourier-transform integration (stars for the exact and circles for the wFFLC approximation) for delays being multiples of $2\pi$. The inset shows an extended calculation for the Fourier-transform case showing repeating structure. We used $\omega = 1$, $g_{\text{cd}} = 0.6$, $G = 0.2$, $\kappa = 4$, $\omega_{\text{fb}} = 4.5$, $\bar{n} = 1\times 10^{5}$ and $\gamma = 4\times 10^{-5}$.}
	\label{fig:singlemode}
\end{figure*}
%%%%%%%%%%%%%%%%%%%%%%%%%%%
\subsection{Time domain analysis of steady state}
For a single resonator mode, the solution to the Lyapunov equation leads to the following expressions for the momentum and position variances in steady state:
\begin{subequations}
\begin{align}
\label{Lyap.1}
\braket{p^2} &= \frac{\gamma \left(\bar{n} + 1/2 +\mathcal{C}/2\right)}{\gamma+\Gamma\cos(\omega \tau)},\\
\braket{q^2} &= \frac{\omega}{\omega+\Gamma \sin(\omega \tau)}\braket{p^2}.
\end{align}
\end{subequations}
\indent We have made use of the optomechanical cooperativity defined here as $\mathcal{C}=G^2/(\kappa \gamma)$ \cite{Aspelmeyer2014cavity}. This term brings an extra heating contribution owing to the back-action of the continuous monitoring of the field quadrature. As opposed to the cavity self-cooling scheme \cite{Aspelmeyer2014cavity}, where one aims at large cooperativities, here we aim to keep this term small. This is easily achieved by making the cavity lossy, i.e. $\kappa\gg G$.\\
\indent Notice that, generally, as pointed out also previously for the single mode feedback cooling without time delay~\cite{genes2008ground}, the equipartition theorem does not hold and therefore the damped state is not in thermal equilibrium. However, we define an approximate final occupancy quantity that assumes the following analytical expression
\begin{equation}
\label{Lyap.2}
n_\text{eff}(\tau) = \frac{1}{2}\frac{\gamma \left(\bar{n} + 1/2 +\mathcal{C}/2\right)}{\gamma+\Gamma\cos(\omega \tau)} \left[1 + \frac{\omega}{\omega+\Gamma \sin(\omega \tau)} \right].
\end{equation}
\indent In the limit of zero time delay, this reduces to the expected result $n_\text{eff}=\gamma(\bar{n} + 1/2 +\mathcal{C}/2)/(\gamma+\Gamma)$ (as in Ref.~\cite{genes2008ground}), where the effective damping rate is $\gamma+\Gamma$. With non-zero time delay the rate $\gamma+\Gamma\cos(\omega \tau)$ is reduced and eventually can become negative when the feedback acts completely out of phase and an instability can occur. If the delay is small such that $\omega \tau\ll 1$, the effect is minimal as the damping rate is still almost optimal $\Gamma \cos(\omega \tau) \approx \Gamma$. For values of $\omega \tau$ close to $\pi/2$ first inefficient cooling and then an instability will occur as it is presented in Fig.~\ref{fig:singlemode}b, c. In such a case, a good choice is to further delay the feedback response setting $\tau_\text{opt}=n \times 2\pi/\omega$. This is exemplified in Fig.~\ref{fig:singlemode}d, e, where a delay of $\tau = 4\pi/\omega$ results in a low final occupancy that is close to
\begin{equation}
\label{Lyap.2}
n_\text{eff} = \frac{\gamma \left(\bar{n} + 1/2 +\mathcal{C}/2\right)}{\gamma+\Gamma},
\end{equation}
for $\tau < \Gamma^{-1}$. For larger delay additional non-Lorentzian features in the power spectrum of the position quadrature appear (as described analytically in the next subsection); these features brought in by the feedback delay are not captured by the approximation and regions of insufficient cooling emerge, as can be seen from Fig.~\ref{fig:singlemode}f. This can be explained by the results in Fig.~\ref{fig:singlemode}e where we can see that for increasing delay $\tau$ the position spectrum deviates strongly from a Lorentzian form by additional superposed oscillations that follow $\sim \exp(i\Omega\tau)$, where since we are in Fourier space $\tau$ determines the period of the oscillations. Thereby, the larger $\tau$ is the smaller is the period of the oscillations. If the period is close (see Fig.~\ref{fig:singlemode}e inset) or  coincides with the resonance condition we witness a high occupation even for $\tau$ being a multiple of $2\pi/\omega$.\\

\subsection{Fourier analysis of damping rates}
\label{FaDecay}
Applying a Fourier transform to the coupled set of integro-differential equations allows one to provide an exact analysis of the steady state even in the non-Markovian regime (see Appendix for details). In the frequency domain one can then compute the variances of the position and momentum as
\begin{subequations}
\begin{align}
\braket{q^2} &= \int^{\infty}_{-\infty}\frac{d\Omega}{2\pi}S_{q}(\Omega),\\
\braket{p^2} &= \int^{\infty}_{-\infty}\frac{d\Omega}{2\pi}\frac{\Omega^2}{\omega^2}S_{q}(\Omega).
\end{align}
\end{subequations}
Here, we refer to the term $S_{q}(\Omega)$ as the position spectrum of the resonator.
The integration goes over the whole power spectrum of the noise which can be split into four contributions
\begin{eqnarray}
\nonumber
S_{q}(\Omega) &=& |\chi^{\text{cd}}_{\text{eff}}(\Omega)|^2 \left[ S_{\text{th}}(\Omega) + S_{\text{rp}}(\Omega) + S_{\text{fb}}(\Omega) + S_{\text{fb,rp}}(\Omega)\right],
\end{eqnarray}
stemming from the radiation pressure force $S_{\text{rp}}(\Omega) = G^2\kappa/(\kappa^2+ \Omega^2)$, the feedback back-action noise $S_{\text{fb}}(\Omega) = |g^{(0)}(\Omega)|^2/(4\kappa\eta)$ as well as its interference $S_{\text{fb,rp}}(\Omega)$ and most importantly and dominantly from the thermal fluctuations $S_{\text{th}}(\Omega) \approx \gamma(2\bar{n}+1)$ (for $\bar{n}\gg 1$). The effective susceptibility appearing above describes the modified response of the position quadrature to the external noise and takes a quasi-Lorentzian form
\begin{align}
\chi^{\text{cd}}_{\text{eff}} = \frac{\omega}{\left[\left(\omega^2_{\text{eff}}(\Omega) - \Omega^2\right) - i\Omega\gamma_{\text{eff}}(\Omega) \right]}.
\end{align}
The poles are shifted from the original position $\pm\omega$ to the effective frequency $\pm\omega_{\text{eff}}$
\begin{align}
\omega^2_{\text{eff}}(\Omega)=\omega^2 + \Omega\delta\tilde{\omega}(\Omega)\cos(\Omega\tau) + \Omega\tilde{\Gamma}(\Omega)\sin(\Omega\tau),
\end{align}
while the frequency dependent effective damping rate is also modified from $\gamma$ to
\begin{align}
\gamma_{\text{eff}}(\Omega) &=\gamma + \tilde{\Gamma}(\Omega)\cos(\Omega\tau) - \delta \tilde{\omega}(\Omega)\sin(\Omega\tau).
\end{align}
Notice that in the absence of feedback the susceptibility is simply that of a damped harmonic oscillator with damping rate $\gamma$ and resonance frequency $\omega$. The feedback for zero delay time adds a damping rate $\tilde{\Gamma}(\Omega)$ and a shifted resonance frequency $\sqrt{\omega^2+\Omega\delta\tilde{\omega}(\Omega)}$. Both quantities are then strongly dependent on the time delay as evidenced in Fig.~\ref{fig:singlemode}.\\
\indent As long as the effective mechanical susceptibility is close to a Lorentzian the integration of the spectra above becomes trivial as the only considerable contribution comes from the spectrally flat thermal power spectrum (see Fig.~\ref{fig:singlemode}c,e). We will use the fact that the integral $\textstyle 1/(2\pi) \int_{-\infty}^{\infty}d\Omega|\chi^{\text{cd}}_{\text{eff}}|^2 \approx 1/[2\gamma_{\text{eff}}(\omega)]$ for reasonably small delay times where we can set $\tilde{\Gamma}(\Omega) \approx \tilde{\Gamma}(\omega) = \Gamma$ and $\delta \tilde{\omega}(\Omega) \approx \delta \tilde{\omega}(\omega) = \delta \omega$ and we can approximate
\begin{subequations}
\begin{align}
n_\text{eff}(\tau) &\approx \frac{1}{2}\left( \bar{n} + \frac{1}{2}\right)\frac{\gamma}{\gamma_{\text{eff}}(\omega)}\left(1 + \frac{\omega^2}{\omega_{\text{eff}}^2(\omega)}\right) \\
\label{SMode.1}
&= \frac{1}{2}\frac{\gamma (\bar{n} + 1/2)}{\gamma + \Gamma\cos(\omega\tau) - \delta \omega\sin(\omega\tau)} \\\nonumber
&\times \left(1 + \frac{\omega}{\omega + \delta\omega\cos(\omega\tau) + \Gamma\sin(\omega\tau)}\right),
\end{align}
\end{subequations}
matching the expression obtained in case of the wFFLC approximation.
In case of the wFFLC we can witness a regime of instability for $\{\tau \; | \gamma + \Gamma\cos(\omega\tau) - \delta \omega \sin(\omega\tau) \leq 0 \}$ as is shown in Fig.~\ref{fig:singlemode}b which will converge to the frequency intervals $n \times \pi/2 \leq \tau \leq n\times 3\pi/2$ for $n \in \mathbb{N}$ and for $\omfb, \kappa \gg \omega_j$ and thereby converge to the sFFLC.\\

\section{Simultaneous cooling of two adjacent modes}
\label{two}
\begin{figure*}[t]
	\centering
	\includegraphics[width=1.95\columnwidth]{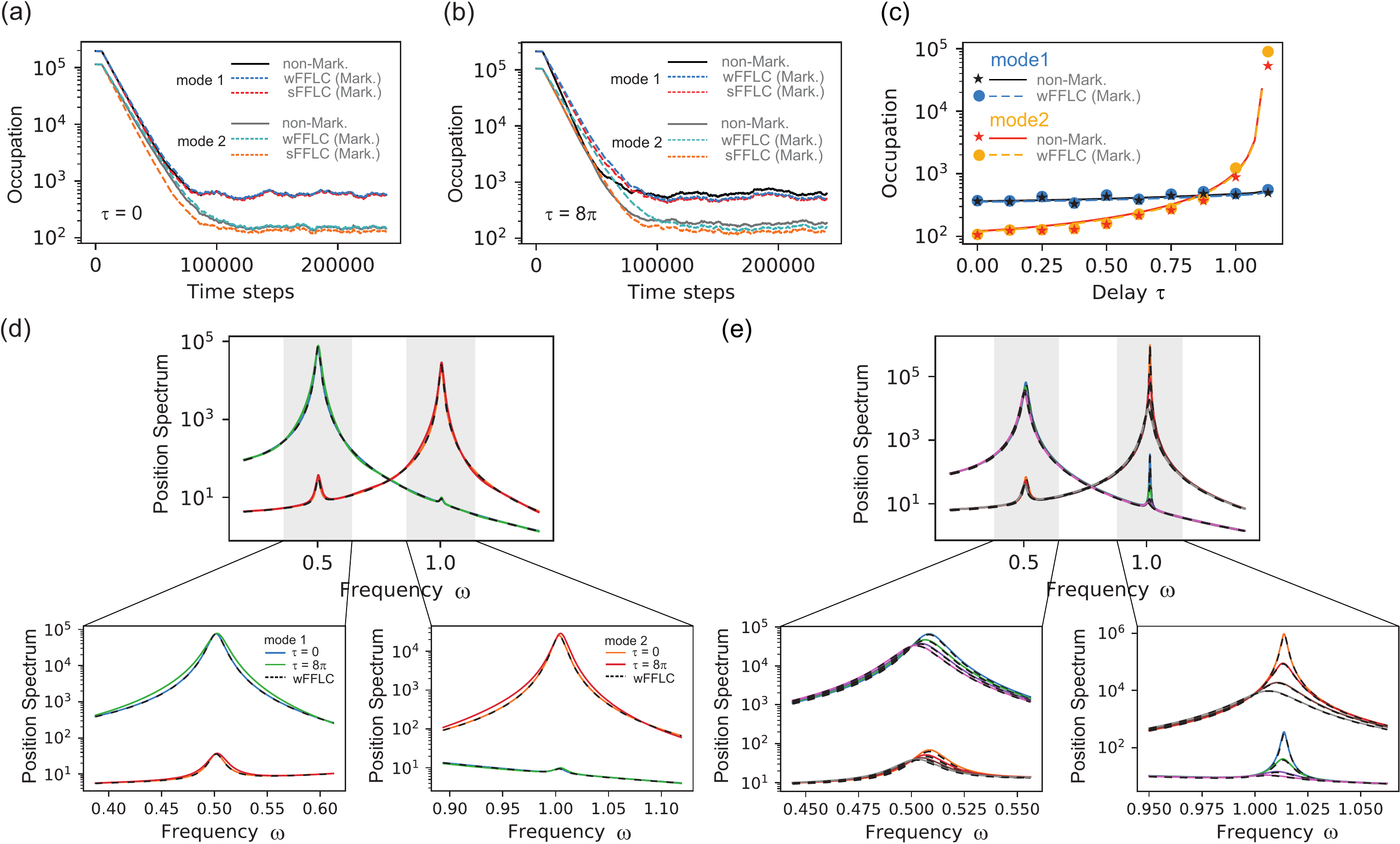}
	\caption{\textit{Simultaneous cooling of two adjacent modes with delay.} (a) (b) Evolution of occupancy for both modes for $\tau = 0$ and $\tau = 8\pi$. The solid lines give the exact solutions while the wFFLC and sFFLC approximations are represented by the dashed lines. (c) Final occupancy represented by stars (exact) dots (wFFLC) are obtained from a time-domain simulation while the solid (exact) and dashed (wFFLC) lines are derived via Fourier transform. Here, the time delay is in units of $\omega_2^{-1}$. (d) Position spectra in Fourier space for $\tau = 0$ and $\tau = 8\pi$. The mismatch in the Lorentz shape for $\tau = 8\pi$ explains the worsened agreement between the wFFLC and the exact treatment seen in (b). (e) Position spectrum in Fourier space for $\tau = (0, 1, 2, 3)\times (\pi/8)$. Here, solid lines follow the exact expression while dashed lines are given for the wFFLC approximation. We have used $\omega = 0.5, 1$, $g_{\text{cd}} = 0.6$, $G = 0.2, 0.1$, $\kappa = 4$, $\omega_{\text{fb}} = 4.5$, $\bar{n} = (2, 1)\times 10^{5}$ and $\gamma = (4, 2)\times 10^{-5}$ for a) b) and (d) while we have used the parameters $\omega = (\omega_1, \omega_2) = (0.5, 1)$, $g_{\text{cd}} = 0.6$, $G = 0.3, 0.2$, $\kappa = 4$, $\omega_{\text{fb}} = 4.5$, $\bar{n} = (2, 1)\times 10^{5}$ and $\gamma = (4, 3)\times 10^{-5}$ for (c) (e).}
	\label{fig:Doublemode}
\end{figure*}
Let us assume two adjacent resonator modes undergoing simultaneous cold-damping: analytical expressions for the final occupancies can still be obtained from the Lyapunov equation. Under the sFFLC approximation one can express the occupancies as
\begin{widetext}
\begin{eqnarray}
\label{TwoModes.1}
\nonumber
n_{i,\text{eff}}(\tau) &\approx & \frac{1}{2}\frac{(\bar{n}_i+1/2)\gamma_i + \frac{G_{i}^{2}}{2\kappa}}{\Gamma_{ii}c_i}\left(1 +  \frac{\omega_i}{(\omega_i + \Gamma_{ii}s_i)} \right) + \left(\frac{\Gamma_{ij}}{4\Gamma_{ii}}\right)\left(\frac{s_j \omega_j}{c_i(\omega^2_j-\omega^2_i)}\Lambda_{ij} - \frac{c_j}{c_i}X_{ij}\right)\left(1 +  \frac{\omega_i}{(\omega_i + \Gamma_{ii}s_i)} \right) \\
& & -\frac{\Gamma_{ij}s_j}{4(\omega_i + \Gamma_{ii}s_i)}Z_{ij} + \frac{\Gamma_{ij} c_j\omega_i}{4(\omega_i + \Gamma_{ii}s_i)(\omega^2_i-\omega^2_j)}\Lambda_{ij},
\end{eqnarray}
\end{widetext}
with $j \neq i$. The expressions for the off diagonal covariance terms $X_{ij}$ and $Z_{ij}$ as well as the $\Lambda_{ij}$ terms are more cumbersome and are therefore relegated to Appendix~\ref{B}. From Eq.~\eqref{TwoModes.1} we see that next to the term expressing the single mode solution, additional terms describing mode to mode coupling emerge, which describe mutual heating effects. For independent modes (such that $|\omega_1 - \omega_2| \rightarrow \infty$) the mutual heating vanishes.
Also notice that as the two modes are subjected to the same damping channel, correlated damping occurs leading to momentum-momentum correlations in the off-diagonal elements. Numerical results involving time domain and Fourier-transform solutions are displayed in Fig.~\ref{fig:Doublemode}.\\
\indent Such effects are particularly evident in the degenerate case with $\omega_1 = \omega_2$. For $G_1 = G_2$ and $g^{(1)}_{\text{cd}} = g^{(2)}_{\text{cd}}$ we define the collective bright $\mathcal{Q}_1 = (q_1 + q_2)/\sqrt{2}$ and dark mode $\mathcal{Q}_2 = (q_1 - q_2)/\sqrt{2}$  and express them in the Fourier space
\begin{subequations}
\begin{align}
\mathcal{Q}_1(\Omega) &= \frac{\omega}{(\omega_{\text{B,eff}}^2(\Omega)-\Omega^2) + i\Omega\gamma_{\text{B,eff}}(\Omega)}\left(\frac{\zeta_1(\Omega) + \zeta_2(\Omega)}{\sqrt{2}}\right) \\
\mathcal{Q}_2(\Omega) &= \frac{\omega}{(\omega^2-\Omega^2) + i\Omega\gamma}\left(\frac{\zeta_1(\Omega) - \zeta_2(\Omega)}{\sqrt{2}}\right).
\end{align}
\end{subequations}
The susceptibility of the bright mode is modified by the frequency dependent resonance $\omega_{\text{B,eff}}^2 = \omega^2 + 2(\Omega \delta \tilde{\omega}(\Omega)\cos(\Omega\tau) + \Omega\tilde{\Gamma}(\Omega)\sin(\Omega\tau))$ and damping rate $\gamma_{\text{B,eff}} = \gamma + 2(\tilde{\Gamma}(\Omega)\cos(\Omega\tau) - \delta \tilde{\omega}(\Omega)\sin(\Omega\tau))$. This results in a damping rate twice as large as compared to the single mode solution found in Eq.~\eqref{SMode.1} for the bright mode. Instead, the dark mode is fully decoupled from the feedback. We can also analytically list the final phonon occupancies
\begin{subequations}
\begin{align}
\nonumber
n_{\text{B,eff}}(\tau) &\approx  \frac{1}{2}\frac{\gamma (\bar{n} + 1/2)}{\gamma + 2(\Gamma \cos(\omega\tau) - \delta \omega \sin(\omega\tau))} \\
&  \times \left(1 + \frac{\omega}{\omega + 2(\delta \omega \cos(\omega\tau) + \Gamma \sin(\omega\tau))} \right)\\
n_{\text{D,eff}} &= \bar{n},
\end{align}
\end{subequations}
showing again the decoupling of the dark mode from the feedback loop and that the achievable temperature for the bright mode is half of that of an individual mode.\\

\section{Multimode cooling:}
\label{multi}
\begin{figure}[t]
	\centering
	\includegraphics[width=0.85\columnwidth]{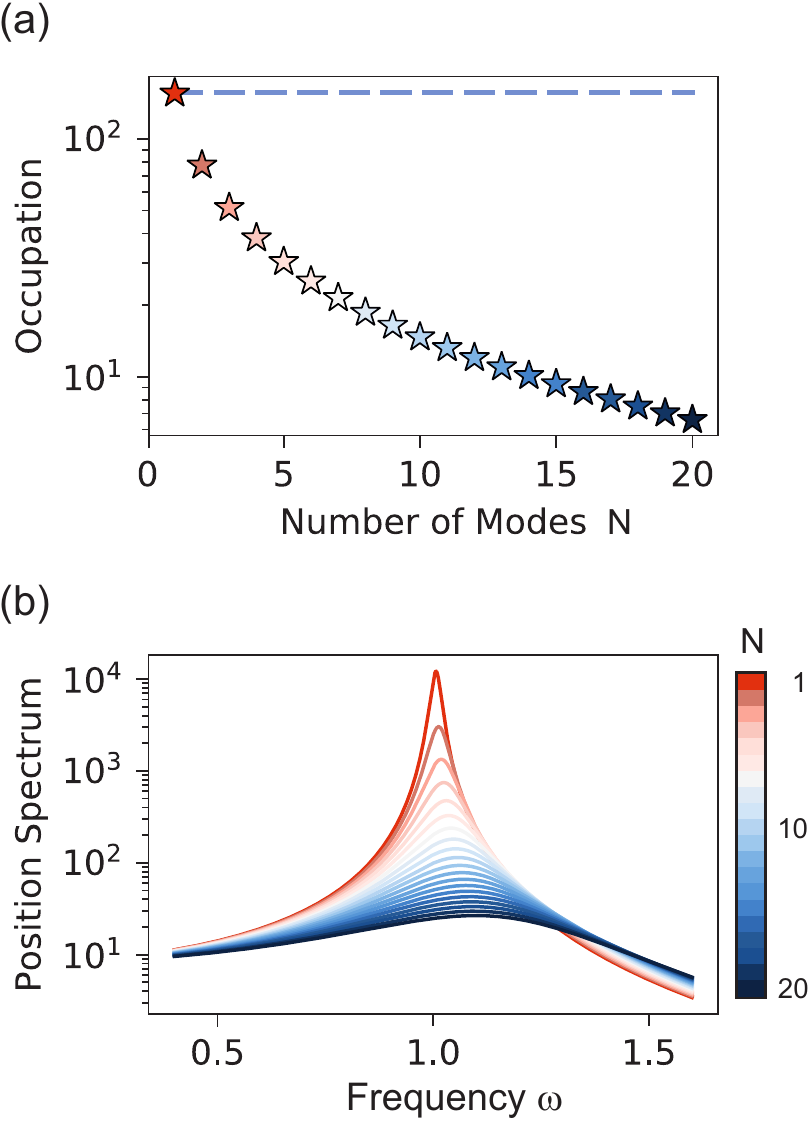}
	\caption{\textit{Multi-mode cooling for frequency degenerate modes} (a) The final occupation of the bright mode (stars) as a function of mode number is presented and compared to the final occupation of a single mode (dashed line) at $\tau = 0$. We consider the case of equal coupling to the cavity mode and equal coupling to the feedback force. (b) The corresponding position spectra are presented and show an increasing width and decreasing magnitude of the spectra with increasing mode number (color bar). We have used the parameters $\omega = 1$, $g_{\text{cd}} = 0.6$, $\kappa = 4$, $\omega_{\text{fb}} = 4.5$, $\bar{n}_1 = 1\times 10^{5}$, $G = 0.2$ and $\gamma = 4\times 10^{-5}$.}
	\label{fig:CollectiveBright}
\end{figure}
The collective basis of one bright and $N-1$ dark modes is particularly useful for the case of many resonances. Starting from the equations of motion in Fourier space Eq.~\eqref{FouEq.2} we can derive an expression for the bright mode
\begin{widetext}
\begin{equation}
\label{Multi.1}
\mathcal{Q}_1(\Omega) = \sum_j \frac{\omega_j}{\left[\left(\omega^2_{\text{j,eff}}(\Omega) - \Omega^2\right) + i\Omega\gamma_{\text{j,eff}}(\Omega) + \sum_{k \neq j}\frac{\left[\left(\omega^2_{\text{k,eff}}(\Omega) - \omega^2_k\right) + i\Omega(\gamma_{\text{k,eff}}(\Omega)-\gamma_k)\right]\left[(\omega^2_j - \Omega^2)+i\Omega \gamma_j\right]}{\left[(\omega^2_k - \Omega^2)+i\Omega \gamma_k\right]} \right]}\frac{G_j\zeta_j(\Omega)}{\sqrt{\sum_k G^2_k}},
\end{equation}
\end{widetext}
which solely depends on the noise terms for input. From this result for the bright mode $\mathcal{Q}_1$ the expressions for the $N-1$ dark modes $\mathcal{Q}_k$ can be easily obtained from the relations (details in Appendix~\ref{C})
\begin{widetext}
\begin{eqnarray}
\label{Multi.2}
\sum_j \frac{\alpha_{kj} \left[\left(\omega^2_{\text{j,eff}}(\Omega) - \omega^2_j\right) + i\Omega(\gamma_{\text{j,eff}}(\Omega)-\gamma_j) \right]}{\alpha_{1j}\left[(\omega^2_j - \Omega^2)+i\Omega \gamma_j\right]}\mathcal{Q}_1 + \mathcal{Q}_k &=& \sum_j \frac{\omega_j}{\left[(\omega^2_j - \Omega^2)+i\Omega \gamma_j\right]}\alpha_{kj}\zeta_j(\Omega).
\end{eqnarray}
\end{widetext}
\indent A simple solution can be found in the case of $N$ identical resonator modes with $\omega_j = \omega$ and $\gamma_j = \gamma$. Here, the term for the bright mode given in Eq.~\eqref{Multi.1} becomes
\begin{eqnarray}
\label{Multi.3}
\nonumber
\mathcal{Q}_1 &=& \frac{\omega}{\left[\left(\omega^2_{\text{B,eff}}(\Omega) - \Omega^2\right) + i\Omega\gamma_{\text{B,eff}}(\Omega) \right]}\sum_j\frac{G_j\zeta_j(\Omega)}{\sqrt{\sum_j G^2_j}},\\
\end{eqnarray}
where the effective frequency of the bright mode follows $\omega^2_{\text{B,eff}}(\Omega) = \omega^2 + \sum_j\Omega\delta\tilde{\omega}_j(\Omega)\cos(\Omega\tau) + \Omega\tilde{\Gamma}_j(\Omega)\sin(\Omega\tau)$ with an effective decay rate of $\gamma_{\text{B,eff}}(\Omega) =\gamma + \sum_j\tilde{\Gamma}_j(\Omega)\cos(\Omega\tau) - \delta \tilde{\omega}_j(\Omega)\sin(\Omega\tau)$.
This can even further be simplified in the case of identical coupling to the cavity mode $G_j = G$ and identical coupling to the feedback force $g^{(j)}_{\text{cd}} = g_{\text{cd}}$. Here, we obtain for the bright and dark modes the solutions
\begin{subequations}
\label{Multi.4}
\begin{align}
\mathcal{Q}_1 &= \frac{\omega}{\left[\left(\omega^2_{\text{B,eff}}(\Omega) - \Omega^2\right) + i\Omega\gamma_{\text{B,eff}}(\Omega) \right]}\sum_j\frac{\zeta_j(\Omega)}{\sqrt{N}},\\
\mathcal{Q}_k &= \frac{\omega}{\left[\left(\omega^2 - \Omega^2\right) + i\Omega\gamma \right]}\sum_j \alpha_{kj} \zeta_j(\Omega)
\end{align}
\end{subequations}
where $\omega^2_{\text{B,eff}}(\Omega) = \omega^2 + N(\Omega\delta\tilde{\omega}(\Omega)\cos(\Omega\tau) + \Omega\tilde{\Gamma}(\Omega)\sin(\Omega\tau))$ and $\gamma_{\text{B,eff}}(\Omega) =\gamma + N\left(\tilde{\Gamma}(\Omega)\cos(\Omega\tau) - \delta \tilde{\omega}(\Omega)\sin(\Omega\tau)\right)$. For zero time delay the bright mode damping rate $\gamma_{\text{B,eff}}(\Omega) =\gamma + N\tilde{\Gamma}(\Omega)$ is $N$-times larger than in the case of an individual resonator mode, which can be seen from the position spectra in Fig.~\ref{fig:CollectiveBright}b.\\
\indent The expressions in Eq.~\eqref{Multi.4} show $N$-uncoupled resonator modes. Here, each mode can be treated independently and following the procedure introduced in section~\ref{FaDecay} and here exemplified for a delay of $\tau = 0$ we obtain for the occupation number of the bright and dark modes
\begin{subequations}
\begin{align}
\frac{1}{2}\left(\braket{\mathcal{Q}_1^2} + \braket{\mathcal{P}_1^2}\right) &\approx \frac{1}{2}\frac{\gamma}{\gamma + N\Gamma}\left(\bar{n} + \frac{1}{2}\right)\left(1 + \frac{\omega}{\omega + N\delta \omega}\right)\\
\frac{1}{2}\left(\braket{Q_k^2} + \braket{P_k^2}\right) &= \left(\bar{n} + \frac{1}{2}\right),
\end{align}
\end{subequations}
showing that only the bright mode experiences cooling, but this mode can reach a much lower occupation number in comparison to the single mode case which is growing with the number $N$ of identical modes that are addressed (see Fig.~\ref{fig:CollectiveBright}a). Here, we have ignored the contributions from feedback and radiation pressure noise, which show an effect when the thermal noise term is small and increase with mode number. This will be addressed in Sec.~\ref{Groundstate} where we identify the residual occupancy stemming from radiation pressure readout noise as well as from feedback noise. In the case that couplings and frequencies vary, the bright mode which is addressed by the feedback mechanism and cooled directly couples with the dark modes and cools them indirectly, resulting in a sympathetic cooling process of the collective modes. More details of the derivation are given in Appendix~\ref{C}.\\

\section{Adjustable time delay}
\begin{figure}[t]
	\centering
	\includegraphics[width=0.90\columnwidth]{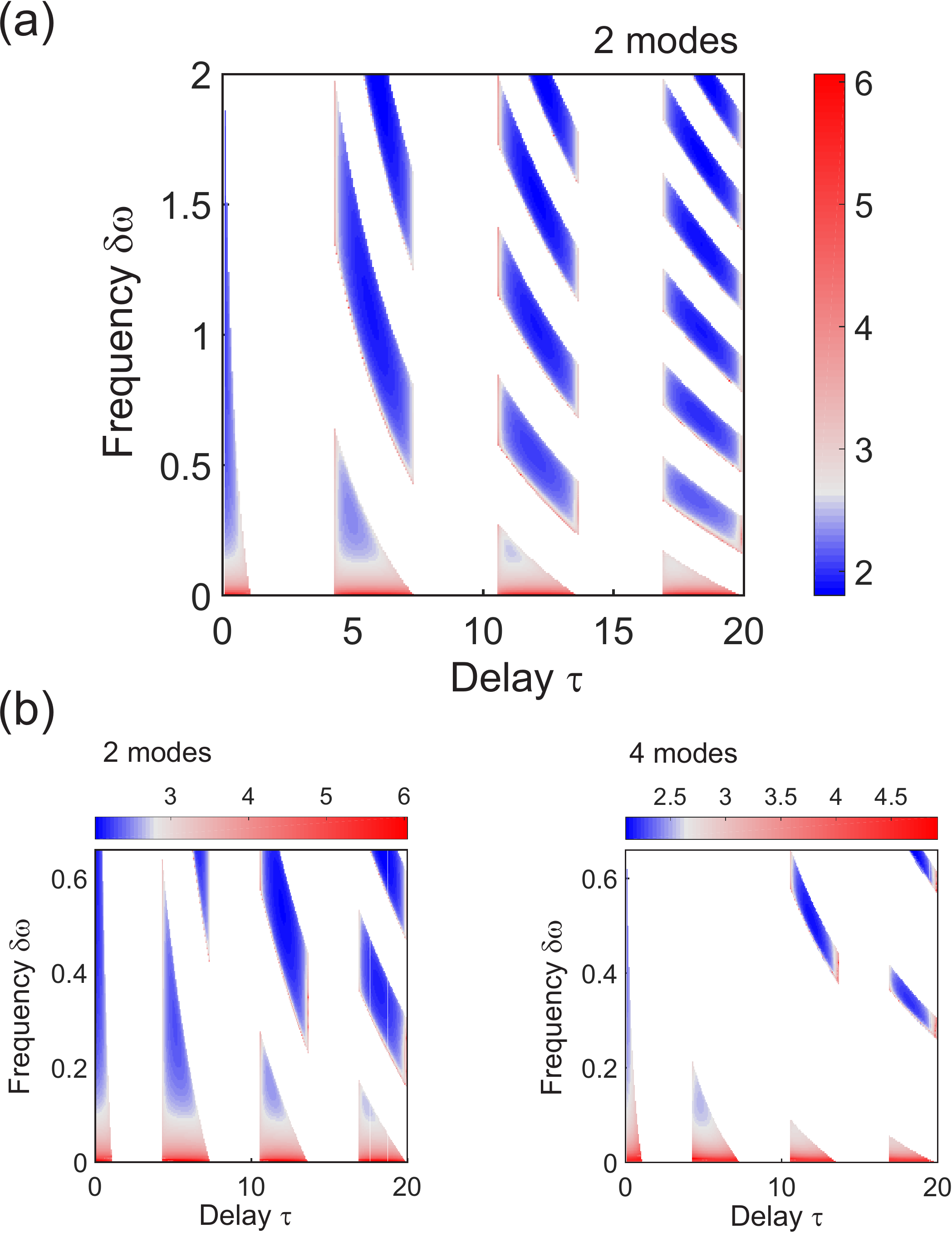}
	\caption{\textit{Multi-mode cooling with delay.} (a) The sum of the final occupations of two modes for different delays and frequency differences between the two oscillator modes. The color bars give the final occupation in logarithmic scale. Here, regions of instability where either one or both modes are unstable are represented by white empty regions. (b) The results for two- and four modes following a linear dispersion with $\delta \omega$ being the frequency difference between two neighboring modes are compared indicating an increase of the area of instability for increasing mode number. We have used the parameters $\omega_1 = 1$ with $\omega_{k+1} = \omega_1 + k\delta \omega$, $g_{\text{cd}} = 0.6$, $\kappa = 4$, $\omega_{\text{fb}} = 4.5$, $\bar{n}_1 = 1\times 10^{5}$ with $\bar{n}_k = \bar{n}_1/\omega_k$, where in (a) we chose $G = 0.3, 0.2$ and $\gamma = (4, 6)\times 10^{-5}$, while in (b) we have used $G_k = 0.2 + (k-1)\times 0.1$ and $\gamma_k = (4 + 2(k-1))\times 10^{-5}$. The time delays are in units of $\omega_{1}^{-1}$.}
	\label{fig:TwoEnergy}
\end{figure}
As in the single mode case, a strategy can be devised to improve the efficiency of cold-damping with time delay by counter-intuitively \textit{further delaying} the action of the  feedback loop. To this end we fix the condition $\cos \omega_j \tau \approx 1$ by providing an additional time $\tau_{\text{add}}$ such that $\tau = \tau_{\text{inh}} + \tau_{\text{add}}$. The condition springs from the analytical results with Markovian dynamics under the wFFLC or sFFLC approximations (and with a regime of validity implying that $\tau < ( \text{max} (\Gamma_j))^{-1}$ is fulfilled for all modes). This demands that $\omega_j\tau \approx 2\pi n_j$ with $n_j\in \mathbb{N}$. In the case that all frequencies are multiples of a common frequency $\omega$ this can be obtained by bringing the total delay to fulfill $\tau = 2\pi/\omega$. In the general case this is quite a difficult matter and for an increasing number of modes can result in additional delay times $\tau_{\text{add}}$ exceeding the time regime where the wFFLC and sFFLC approximations are valid, which even can be seen for a single mode presented in Fig.~\ref{fig:singlemode}f for very large delay times. From color-maps as presented in Fig.~\ref{fig:TwoEnergy} showing the total energy of the system it is possible to find regions with efficient cooling and avoid instabilities. From Fig.~\ref{fig:TwoEnergy}b showing the total energy for two or four modes we also see that that the region where one or many modes are unstable increases for increasing mode number. This results from the fact that many more constraints differing between each mode have to be fulfilled simultaneously. Nevertheless, the condition $\min \tau$ with $\omega_j\tau \approx 2\pi n_j$ can be a good guide to find approximate solutions. Additionally one can employ various nonlinear optimization schemes to obtain a minimum with suitable parameters.\\

\section{Ground-state cooling. Residual occupancy}
\label{Groundstate}
In the previous sections we have mainly focused on cases where the thermal noise $S_{\text{th}}$ dominates over the feedback and radiation pressure noise terms. This is true when the initial occupancy is large and the resulting feedback cooling rate is not strong enough to lead to close to the ground state final occupancy. Let us assume instead that $\gamma\bar{n}\ll\Gamma$ and the thermal noise can be almost completely attenuated via the cold-damping cooling scheme. In such a case the final residual occupancy is given by the extra heating terms stemming from the cold damping back action as well as from radiation pressure readout noise.\\
\subsection{Single mode ground-state cooling}
\indent For a single mode, the full expression of the noise term in Fourier space is given by
\begin{eqnarray}
\nonumber
\label{Eq.Ground.1}
S(\Omega) &=& S_{\text{th}}(\Omega)+ S_{\text{fb}}(\Omega)+S_{\text{rp}}(\Omega) +S_{\text{fb-rp}}(\Omega),
\end{eqnarray}
where the terms in Eq.~\eqref{Eq.Ground.1} represent the thermal noise $S_{\text{th}}(\Omega) \approx \gamma (2\bar{n} + 1)$, feedback noise $S_{\text{fb}}(\Omega) = \Omega^2\omfb^2 g^2_{\text{cd}}/(4\kappa\eta(\omfb^2+\Omega^2))$, radiation pressure noise $S_{\text{rp}}(\Omega) = \kappa G^2/(\kappa^2+\Omega^2)$ and the interference term between the feedback and radiation pressure noise given by $S_{\text{fb-rp}}(\Omega) = - \Omega(\gamma_{\text{eff}}(\Omega)-\gamma)/\omega$. From the noise spectrum we can obtain the effective population by convoluting it with the proper effective susceptibility
\begin{eqnarray}
n_{\text{eff}} &=& \int^{\infty}_{-\infty}\frac{d\Omega}{4\pi}|\chi^{\text{cd}}_{\text{eff}}|^2 S(\Omega)\left( 1 + \frac{\Omega^2}{\omega^2}\right) - \frac{1}{2} \\\nonumber
&=& \frac{\gamma}{\gamma_{\text{eff}}(\omega)}\left(\bar{n} + \frac{1}{2}\right)\left(1 + \frac{\omega^2}{\omega^2_{\text{eff}}(\omega)} \right) + n_{\text{res}}.
\end{eqnarray}
where $n_{\text{res}}$ describes the residual occupancy resulting from  the feedback and radiation pressure noise terms which form a fundamental lower bound for ground-state cooling. The residual noise amounts to
\begin{widetext}
\begin{eqnarray}
\nonumber
n_{\text{res}} &=& \frac{G^2}{4\omega^2_{\text{eff}}\gamma_{\text{eff}}}\left[\kappa - \frac{(\kappa^2-\omega^2)(\kappa+\gamma_{\text{eff}})(\omega^2_{\text{eff}} + \kappa^2 -\kappa \gamma_{\text{eff}})}{(\omega^2_{\text{eff}} + \kappa^2)^2-\gamma^2_{\text{eff}}\kappa^2}\right]\\
& & + \frac{\omfb^2 g^2_{\text{cd}}}{16\kappa\eta \omega^2_{\text{eff}}\gamma_{\text{eff}}}\left[\omega^2 + \frac{(\omega^2_{\text{eff}} + \omfb^2)(\omega^4_{\text{eff}} - \omega^2\omfb^2)+\omfb(\omfb^2 - \omega^2)\omega^2_{\text{eff}}\gamma_{\text{eff}}}{(\omega^2_{\text{eff}} + \omfb^2)^2-\gamma^2_{\text{eff}}\omfb^2} \right] - \frac{1}{2}.
\end{eqnarray}
\end{widetext}
The resulting behavior is plotted in Fig.~\ref{fig:GoundState}a as a function of delay $\tau$. Notice that the expression above is valid for any $\tau$ as the effective resonance frequency and damping rate are both including the time delay dependence. However, simple analytical results can be identified in the regime where $\kappa, \omfb \gg \omega$ which we have named the sFFLC approximation. In such a case we can approximate
\begin{eqnarray}
\nonumber
n_{\text{res}} &\approx& \frac{1}{\kappa\gamma_{\text{eff}}}\left(1 + \frac{\omega^2}{\omega^2_{\text{eff}}} \right)\left[G^2 + \frac{g^2_{\text{cd}}\omega^2_{\text{eff}}}{4\eta} \right] - \frac{1}{2} + \frac{\omfb g^2_{\text{cd}}}{16\kappa\eta}\\\nonumber
&=& \frac{1}{2\kappa\Gamma}\left(\frac{g_{\text{cd}}\omega}{2\sqrt{\eta}} - G\right)^2 + \left(\frac{g^2_{\text{cd}}\omega^2}{4\eta} - G^2\right)\frac{(\kappa + \omfb)}{4\kappa^2\omfb}\\
& & + \frac{1-\sqrt{\eta}}{2\sqrt{\eta}} + \frac{\omfb g^2_{\text{cd}}}{16\kappa\eta}
\end{eqnarray}
where we have used the approximate expression for the cooling rate $\Gamma=\omega g_{\text{cd}} G/\kappa$ and the delay $\tau = 0$. For unit efficiency $\eta=1$ the minimal residual occupancy is approximately given by $\omfb g^2_{\text{cd}}/(16\kappa)$ under the condition $\omega g_{\text{cd}}=2G$.
%For $\eta<1$ we have a modified optimal condition $\omega g_{\text{cd}}=2G\sqrt{\eta}$ and the minimal residual occupancy is $(1-\sqrt{\eta})/(2\sqrt{\eta}).$
%%%%%%%%%%%%%%%%%%%%%%%%%%%%%%%%%%
%%%%%%%%%%%%%%%%%%%%%%%%%%%%%%%%%%
\subsection{Collective bright mode}
\begin{figure}[t]
	\centering
	\includegraphics[width=0.75\columnwidth]{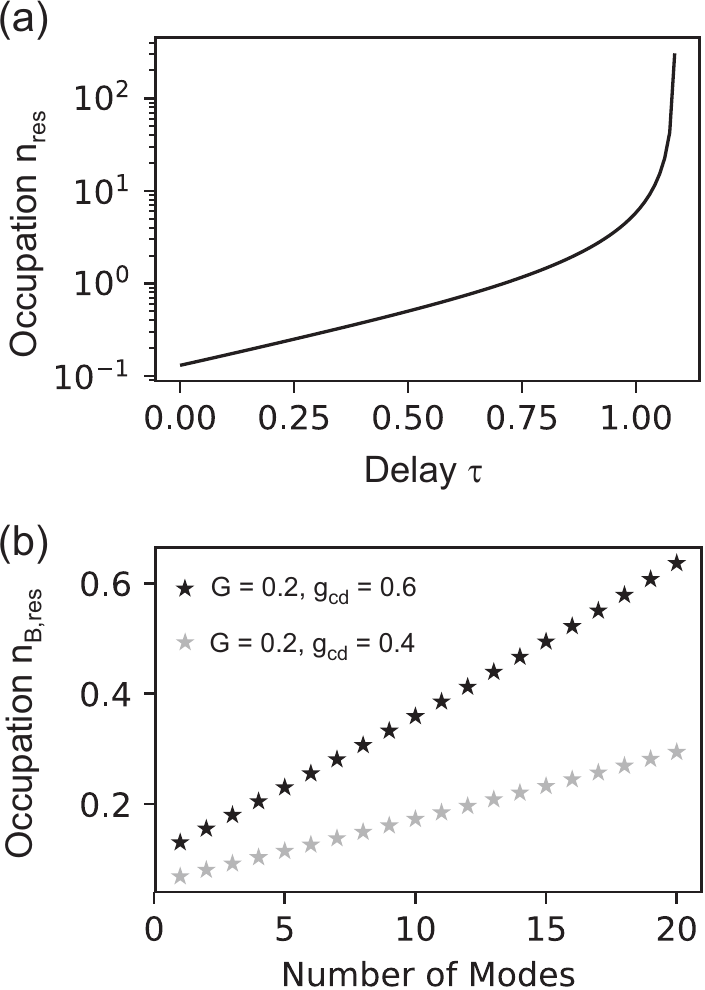}
	\caption{\textit{Residual occupancy} (a) The final residual occupancy as a function of delay $\tau$ in units of $\omega^{-1}$ is presented. Here, we have used the parameters $\omega_1 = 1$, $g_{\text{cd}} = 0.6$, $G = 0.2$, $\kappa = 4$ and $\omega_{\text{fb}} = 4.5$. (b) The final residual occupancy for the bright mode as a function of the mode number for the parameters used in Fig.~\ref{fig:CollectiveBright} with $\tau = 0$ is given by the black stars. The condition $\omega g_{\text{cd}}=2G$ for $\eta = 1$ minimizing the heating due to the residual noise is fulfilled for $g_{\text{cd}} = 0.4$, $G = 0.2$ which is represented by the gray stars.}
	\label{fig:GoundState}
\end{figure}

%%%%%%%%%%%%%%%%%%%%%%%%%%%%%%%%%%
%%%%%%%%%%%%%%%%%%%%%%%%%%%%%%%%%%
Investigating the behavior of the bright mode in the case of the $N$ fully degenerate resonator modes, we obtain the full noise term in Fourier space given by
\begin{eqnarray}
\nonumber
\label{Eq.Ground.2}
S_{\text{B,eff}}(\Omega) &=& S_{\text{th}}(\Omega)+ N\left(S_{\text{fb}}(\Omega)+S_{\text{rp}}(\Omega) +S_{\text{fb-rp}}(\Omega)\right)
\end{eqnarray}
with full derivation presented in Appendix~\ref{C}. While the thermal noise is independent on the number of modes, the residual noise terms are N-times larger. We show in Fig.~\ref{fig:GoundState}b the contribution of the residual noise terms on the final occupation as a function of the number of modes for $\tau = 0$ by numerically integrating
\begin{eqnarray}
\nonumber
n_{\text{B,res}} &=& \int^{\infty}_{-\infty}\frac{d\Omega}{4\pi}|\chi^{\text{cd}}_{\text{B,eff}}|^2 (S_{\text{B,eff}}(\Omega) - S_{\text{th}}(\Omega))\left( 1 + \frac{\Omega^2}{\omega^2}\right)\\
& & - \frac{1}{2}.
\end{eqnarray}
The modified susceptibility function is characterized by the increased damping rate $N\Gamma$ which shows the previously derived conclusion that the thermal occupancy of the bright mode decreases with $N^{-1}$. However, with increasing $N$, the Lorentzian profile of the mechanical effective susceptibility is slightly shifted and its damping goes away from the simple $N\Gamma$ scaling which means that the residual occupancy acquires a slight $N$ dependence which originates mostly from the feedback noise term. The full result is presented in the Appendix~\ref{C}. In the sFFLC approximation we obtain the expression
\begin{eqnarray}
\nonumber
n_{\text{B,res}} &\approx& \frac{1}{2\kappa\Gamma}\left(\frac{g_{\text{cd}}\omega}{2\sqrt{\eta}} - G\right)^2 + \frac{1-\sqrt{\eta}}{2\sqrt{\eta}} \\\nonumber
& & + N\left[\left(\frac{g^2_{\text{cd}}\omega^2}{4\eta} - G^2\right)\frac{(\kappa + \omfb)}{4\kappa^2\omfb} +  \frac{\omfb g^2_{\text{cd}}}{16\kappa\eta} \right],\\
\end{eqnarray}
which shows a linear dependence for the number of modes.
This dependence is shown in Fig.~\ref{fig:GoundState}b as a linear one with a slope strongly dependent on the ratio of $G/g_{\text{cd}}$. The previously derived condition for optimal residual occupancy in the single mode case holds here as well as the slope and magnitude is minimal for $\omega g_{\text{cd}}=2G\sqrt{\eta}$.
\\
\section{Conclusions}
We have analytically and numerically shown that efficient simultaneous cold-damping of many mechanical resonances is achievable as long as frequency degeneracy is avoided. Furthermore, detrimental effects stemming from the feedback's intrinsically delayed response can be mitigated by introducing an additional variable delay $\tau_{\text{add}}$ that can be adjusted to optimize the cooling efficiency. For example, for a sequence of frequencies which are multiples of a common frequency $\omega$ efficient cooling is obtained again for a total delay of $\tau = \tau_{\text{inh}} + \tau_{\text{add}} = 2\pi /\omega$. Another approach to solve this problem could be the implementation of a machine learning scheme either to find preferable settings for the delay or by using a machine learning procedure to provide for the full feedback mechanism~\cite{Sommer2020Prospects}. In the latter case the feedback would adjust the phase to accommodate for the delayed signal and minimize the final temperature.\\
\indent A main aspect of our treatment is the transformation to a collective basis. In particular for a number of quasi-degenerate modes, a bright/dark mode analysis shows that the damping can be $N$ times faster while the occupancy is $N$ times lower for a suitable identified collective bright mode (as illustrated in Fig.~\ref{fig:CollectiveBright}a)). This is a remarkable result in itself as it shows that collective optomechanics can be employed to provide more efficient cooling of an engineered collective mode. For applications aiming instead at better sensing capabilities for wider frequency intervals, the alternative, as also indicated in Ref.~\cite{Sommer2019Partial} is to engineer mechanical resonators with a dispersion relation close to linear such that mutual heating is inhibited.\\
\indent In terms of methods used, the time domain treatment has been very successful in the case of zero delay~\cite{Sommer2019Partial} allowing for fully analytical results for all mode occupancies. The complication of non-zero time delay can be dealt with more efficiently in the Fourier space where final occupancies for an arbitrary number of modes and a large variety of cases can be obtained exactly. Moreover, the effort is computationally less costly for numerical simulations compared to a brute force approach that involves solving the full set of stochastic differential equations.\\
While our calculations have assumed a particular choice of electronic feedback relevant to experiments in optomechanics with mirrors, microtoroids and levitated nano-particles, the formalism we have used is of much more general validity. In particular, the Fourier domain analysis is general and it simply requires the specific form of feedback in the final estimation of the modified mechanical susceptibility. While there is a wide variety of implementation based on feedback, optimization via machine learning techniques~\cite{Sommer2020Prospects} might provide an answer to which one could provide an optimal cooling efficiency in optomechanics.

\section{Acknowledgments}
We acknowledge financial support from the Max Planck Society and from the German Federal Ministry of Education and Research, co-funded by the European Commission (project RouTe), project number 13N14839 within the research program "Photonik Forschung Deutschland". We acknowledge initial discussions with David Vitali which have lead to identifying the feedback delay time problem as essential in cold-damping optomechanics.

\bibliography{DelayColdDamping}

\newpage
\onecolumngrid

\newpage
\appendix

%%%%%%%%%%%%%%%%%%%%%%%%%%%%%%%%%%%%%%%%%%%%%%%%%%%%%%%%%%%%%%%%%%%%%
%%%%%%%%%%%%%%%%%%%%%%%%%%%%%%%%%%%%%%%%%%%%%%%%%%%%%%%%%%%%%%%%%%%%%
\section{Multimode cold damping: Markovian versus non-Markovian regimes}
\label{A}
%%%%%%%%%%%%%%%%%%%%%%%%%%%%%%%%%%%%%%%%%%%%%%%%%%%%%%%%%%%%%%%%%%%%%
%%%%%%%%%%%%%%%%%%%%%%%%%%%%%%%%%%%%%%%%%%%%%%%%%%%%%%%%%%%%%%%%%%%%%

The equations of motion in the quantum mechanical treatment for cold damping with many resonator modes are given by
\begin{subequations}
\label{App.Full1}
\begin{align}
 \dot{q}_j &= \omega_j   p_j,\\
 \dot{p}_j &= -\omega_j   q_j - \gamma_j   p_j + G_j   x + \xi_j - \int_{-\infty}^{\infty}ds g^{(\tau)}_j(t-s)  y^{\text{est}}(s),\\
 \dot{x} &= -\kappa   x + \sqrt{2\kappa}x^{\text{in}}, \\
 \dot{y} &= -\kappa   y + \textstyle \sum_{j=1}^{N} G_j  q_j + \sqrt{2\kappa}y^{\text{in}}.
\end{align}
\end{subequations}
Here, we have kept the effective cavity detuning at $\Delta = 0$. Further on, we eliminate the cavity field quadratures by formally integrating their equations of motion to obtain:
\begin{subequations}
\begin{align}
x(t) &= \sqrt{2\kappa} \int^{t}_{-\infty} ds e^{-\kappa(t-s)}x^{\text{in}}(s),\\
y(t) &= \int^{t}_{-\infty}ds e^{-\kappa (t-s)}\sum_{j=1}^{N}G_j q_{j}(s)+\sqrt{2\kappa}\int^{t}_{-\infty}ds e^{-\kappa (t-s)} y^{\text{in}}(s).
\end{align}
\end{subequations}
With the intra-cavity phase quadrature expressed as $y^{\text{est}} = y - (y^{\text{in}} + \sqrt{\eta^{-1}-1}y^{v})/\sqrt{2\kappa}$ we can reduce the set of equations in Eq.~\eqref{App.Full1} by two.\\

\subsection*{Non-Markovian dynamics}
Terms proportional to the displacements $q_j$ as well as noise terms stemming from the cavity input noise $y^{\text{in}}(s)$ and noise from the vacuum filled port $y^v$ are introduced by $y^{\text{est}}(s)$. We can first work out the terms coming from $y$ by calculating:
\begin{eqnarray}
\label{App.Eq1}
\nonumber
(g^{(\tau)}_j \ast y) &=& g^{(j)}_{\text{cd}} \omfb  \int_{-\infty}^{\infty}ds e^{-\omfb (t-s-\tau)}\delta(t-s-\tau) y(s) - g^{(j)}_{\text{cd}}\omfb ^{2}\int_{-\infty}^{\infty}ds \theta(t-s-\tau)e^{-\omfb (t-s-\tau)} y(s) \\
&=& \int_{-\infty}^{t-\tau} ds \frac{\kappa e^{-\kappa(t-s-\tau)} - \omfb e^{-\omfb (t-s-\tau)}}{(\kappa - \omfb )} \left[\textstyle \sum_{k=1}^{N} g^{(j)}_{\text{cd}}\omfb G_k q_{k}(s) + \sqrt{2\kappa}g^{(j)}_{\text{cd}}\omfb y^{\text{in}}(s)\right].
\end{eqnarray}
We apply integration by parts to obtain a dependence with respect to $p_j$, where we notice that the convolution above contains a derivative of the following function
\begin{equation}
h_{\tau}(t-s)=\frac{e^{-\kappa(t-s-\tau)}-e^{-\omfb (t-s-\tau)}}{\omfb -\kappa}.
\end{equation}
With the relation $\dot{q}_j = \omega_j p_j$ we obtain
\begin{align}
\label{App.Eq2}
(g^{(\tau)}_j \ast y) = \sum_{k=1}^{N} g^{(j)}_{\text{cd}}\omfb G_k \omega_k \int_{-\infty}^{t-\tau}ds h_{\tau}(t-s) p_k(s)- \sqrt{2\kappa}g^{(j)}_{\text{cd}}\omfb  \int_{-\infty}^{t-\tau} ds \partial_sh_{\tau}(t-s) y^{\text{in}}(s).
\end{align}
Following these steps, we can now write in simplified notation the reduced set of equations of motion for the $2N$ resonator modes quadratures
\begin{subequations}
\begin{align}
\label{App.Full4}
\dot{q}_j &= \omega_j p_j, \\\nonumber
\dot{p}_j &= -\omega_j q_j - \int^{\infty}_{-\infty}ds\left(\gamma_j\delta(t-s) + g^{(j)}_{\text{cd}}\omfb G_j\omega_j \theta(t-s-\tau)h_{\tau}(t-s)\right) p_j(s) - \sum_{k \neq j} g^{(j)}_{\text{cd}}\omfb G_k\omega_k\int^{t-\tau}_{-\infty} ds  h_{\tau}(t-s) p_k(s)\\
& + \xi_j+\xi_\text{fb}+ \xi_\text{vac}+ \xi_\text{rp}.
\end{align}
\end{subequations}
The three sources of noise are owed to the direct feedback action, the feedback filtered vacuum action in the loss port and to the intra-cavity radiation pressure effect:
\begin{subequations}
\begin{align}
\label{App.Full5}
\xi_\text{fb} &= - \frac{g^{(j)}_{\text{cd}}\omfb }{\sqrt{2\kappa}}\int^{\infty}_{-\infty}ds \phi^{(\tau)}_1(t-s)y^{\text{in}}(s), \\
\xi_\text{vac} &= - \frac{g^{(j)}_{\text{cd}}\omfb }{\sqrt{2\kappa}}\sqrt{\eta^{-1}-1}\int^{\infty}_{-\infty} ds \phi^{(\tau)}_2(t-s)y^v(s),\\
\xi_\text{rp} &= \sqrt{2\kappa}G_j\int^{\infty}_{-\infty}ds \phi_3(t-s)x^{\text{in}}(s),
\end{align}
\end{subequations}
with the following definitions for the convolution kernels
\begin{subequations}
\begin{align}
\label{App.Full5}
\phi^{(\tau)}_{1}(t) &= \theta(t-\tau)(\omfb (\omfb +\kappa)e^{-\omfb (t-\tau)}-2\kappa^2 e^{-\kappa (t-\tau)})/(\omfb -\kappa) - \delta(t-\tau), \\
\phi^{(\tau)}_{2}(t) &= \theta(t-\tau)\omfb e^{-\omfb (t-\tau)} -\delta(t-\tau),\\
\phi_{3}(t) &= \theta(t)e^{-\kappa t}.
\end{align}
\end{subequations}
\\

\subsection*{The fast-feedback-lossy-cavity (FFLC) approximations: Markovian dynamics}

In the limit of a lossy cavity and  relatively fast feedback with $\kappa, \omfb  \geq \omega_j$ for all $j\in \{1,\dots,n\}$ where we can approximate $p_j(s) \approx A_j\cos(\omega_j s + \phi_j)$ and $q_j(s) \approx A_j\sin(\omega_j s + \phi_j)$ we obtain an estimate to first order given by
\begin{eqnarray}
\nonumber
\int^{t-\tau}_{-\infty}ds h_{\tau}(t-s) p_j(s) &\approx &  A_j\int^{t}_{-\infty}ds h_{\tau}(t-s)\cos(\omega_j s + \phi_j)\\\nonumber
&=& \left(\frac{\kappa}{\kappa^2 + \omega_{j}^2}- \frac{\omfb}{\omfb^2 + \omega_{j}^2}\right)\frac{p_{j}(t-\tau)}{\omfb - \kappa} + \left(\frac{\omega_j}{\kappa^2 + \omega_{j}^2} - \frac{\omega_j}{\omfb^2 + \omega_{j}^2} \right)\frac{q_{j}(t-\tau)}{\omfb - \kappa} \\\nonumber
&=& \left[ \frac{\left(1-\frac{\omega_k^2}{\kappa\omega_{\text{fb}}} \right)}{\kappa\omfb\left(1+\frac{\omega_k^2}{\kappa^2} \right)\left(1+\frac{\omega_k^2}{\omega_{\text{fb}}^2} \right)} \right]\left(p_{j}(t)\cos(\omega_j \tau) + q_j(t) \sin(\omega_j \tau) \right) \\
&+& \left[ \frac{\omega_{j}\left(\omfb + \kappa \right)}{\kappa^2 \omfb^2 \left(1+\frac{\omega_k^2}{\kappa^2} \right)\left(1+\frac{\omega_k^2}{\omega_{\text{fb}}^2} \right)} \right]\left(q_{j}(t)\cos(\omega_j \tau) - p_{j}(t)\sin(\omega_j \tau) \right)
\end{eqnarray}
and we end up with a set of coupled linear differential equations for the mechanical mode quadratures:
\begin{subequations}
\begin{align}
\label{App.Fullof5}
\dot{q}_j &= \omega_j p_j,\\\nonumber
\dot{p}_j &= -\left(\omega_j + \delta \omega_{jj}\cos(\omega_j\tau) + \Gamma_{jj}\sin(\omega_{j}\tau)\right)q_j - \left(\gamma_j + \Gamma_{jj}\cos(\omega_{j}\tau) - \delta \omega_{jj}\sin(\omega_{j}\tau) \right) p_j,\\
&- \sum_{k\neq j} \left( \Gamma_{jk}\sin(\omega_{k}\tau) + \delta \omega_{jk}\cos(\omega_{k}\tau) \right)q_k - \sum_{k\neq j} \left( \Gamma_{jk}\cos(\omega_{k}\tau) - \delta \omega_{jk}\sin(\omega_{k}\tau) \right)p_k + \xi_j+\xi_\text{fb}+ \xi_\text{vac}+ \xi_\text{rp},
\end{align}
\end{subequations}
We refer to this as the weak fast-feedback-lossy-cavity (wFFLC) approximation.
Here, the rate terms are defined as
\begin{equation}
\Gamma_{jk} = \frac{g_{\text{cd}}^{(j)}G_k\omega_k \omega_{\text{fb}} \left(\kappa\omega_{\text{fb}}-\omega_k^2 \right)}{\left(\kappa^2+\omega_k^2 \right)\left(\omega_{\text{fb}}^2 +\omega_k^2\right)}
\end{equation}
and the frequency shift terms are defined as
\begin{equation}
\delta \omega_{jk} = \frac{g_{\text{cd}}^{(j)}G_k\omega_{\text{fb}}\omega_k^2 \left(\omega_{\text{fb}} + \kappa \right)}{\left(\kappa^2+\omega_k^2 \right)\left(\omega_{\text{fb}}^2 +\omega_k^2 \right)}.
\end{equation}
\\
In the case that $\kappa, \omfb  \gg \omega_j$ the rate terms converge to $\Gamma_{jk} = g_{\text{cd}}^{(j)}G_k\omega_k/\kappa$ and $\delta \omega_{jk} = 0$ and we obtain the set of coupled differential equations for the mechanical mode quadratures:
\begin{subequations}
\begin{align}
\label{App.Fullof6}
\dot{q}_j &= \omega_j p_j,\\
\dot{p}_j &= -\omega_j q_j - \left(\gamma_j + \Gamma_{jj}\cos(\omega_{j}\tau)\right) p_j - \sum_{k\neq j} \Gamma_{jk}\sin(\omega_{k}\tau) q_k - \sum_{k\neq j} \Gamma_{jk}\cos(\omega_{k}\tau) p_k + \xi_j+\xi_\text{fb}+ \xi_\text{vac}+ \xi_\text{rp}.
\end{align}
\end{subequations}
We refer to this approximation for the dynamics formed by this conditions as the strong fast-feedback-lossy-cavity (sFFLC) approximation.
\\
In case we have $N$ equal resonator modes with the same frequency $\omega$ and decay rate $\gamma$ it is favorable to write the equations of motions in the collective basis where $\mathcal{Q}_1 = \sum_k \alpha_{1k} q_k =\left(\sqrt{\sum_l G^2_l}\right)^{-1}\sum_k G_k q_k$ describes the bright mode while the $N-1$ dark modes are given by $\mathcal{Q}_j = \sum_k \alpha_{jk} q_k$ with $\sum_k \alpha_{ik}\alpha_{jk} = \delta_{ij}$ and can be obtained via a Gram-Schmidt procedure. Starting from the wFFLC the equation of motion for the bright mode results in
\begin{subequations}
\label{App.Fullof7}
\begin{align}
\dot{\mathcal{Q}}_1 &= \omega \mathcal{P}_1 , \\
\dot{\mathcal{P}}_1 &= -\left(\omega + \sum_k \delta \omega_{kk}\cos(\omega\tau) + \Gamma_{kk}\sin(\omega\tau) \right)\mathcal{Q}_1 - \left(\gamma + \sum_k \Gamma_{kk}\cos(\omega\tau) - \delta \omega_{kk}\sin(\omega\tau) \right)\mathcal{P}_1 + \sum_k\alpha_{1k} \zeta_k,
\end{align}
\end{subequations}
where all noise terms have been gathered into a single term $\zeta_j = \xi_j+\xi_\text{fb}+ \xi_\text{vac}+ \xi_\text{rp}$. The $N-1$ dark modes are given by
\begin{subequations}
\begin{align}
\dot{\mathcal{Q}}_j &= \omega \mathcal{P}_j ,\\\nonumber
\dot{\mathcal{P}}_j &= -\omega \mathcal{Q}_j - \gamma \mathcal{P}_j - \left( \sum_k \frac{\alpha_{jk}}{\alpha_{1k}}(\delta \omega_{kk}\cos(\omega\tau) + \Gamma_{kk}\sin(\omega\tau))\right)\mathcal{Q}_1 - \left( \sum_k \frac{\alpha_{jk}}{\alpha_{1k}}(\Gamma_{kk}\cos(\omega\tau) - \delta \omega_{kk}\sin(\omega\tau))\right)\mathcal{P}_1\\
& + \sum_k\alpha_{jk} \zeta_k .
\end{align}
\end{subequations}
By solving Eq.~\eqref{App.Fullof7} in the case the system approaches steady state we obtain
\begin{subequations}
\begin{align}
\nonumber
\mathcal{Q}_1(t) &= \int_{-\infty}^{t} ds \frac{\omega}{\sqrt{\omega_{\text{B}}\omega - \left(\frac{\Gamma_{\text{B}}}{2}\right)^2}} e^{-\frac{\Gamma_{\text{B}}}{2}(t-s)}\sin\left(\sqrt{\omega_{\text{B}}\omega - \left(\Gamma_{\text{B}}/2\right)^2}(t-s)\right) \sum_j \alpha_{1j}\zeta_j(s) \\
&= \int_{-\infty}^{t} ds \Theta_q(t-s) \sum_j \alpha_{1j}\zeta_j(s)\\\nonumber
\mathcal{P}_1(t) &= \int_{-\infty}^{t} ds \frac{1}{\sqrt{\omega_{\text{B}}\omega - \left(\frac{\Gamma_{\text{B}}}{2}\right)^2}}\left(\sqrt{\omega_{\text{B}}\omega - \left(\Gamma_{\text{B}}/2\right)^2} e^{-\frac{\Gamma_{\text{B}}}{2}(t-s)}\cos\left(\sqrt{\omega_{\text{B}}\omega - \left(\Gamma_{\text{B}}/2\right)^2}(t-s)\right) \right. \\\nonumber
& \left. - \frac{\Gamma_{\text{B}}}{2}e^{-\frac{\Gamma_{\text{B}}}{2}(t-s)}\sin\left(\sqrt{\omega_{\text{B}}\omega - \left(\Gamma_{\text{B}}/2\right)^2}(t-s)\right)\right)  \sum_j \alpha_{1j}\zeta_j(s)\\
&= \int_{-\infty}^{t} ds \Theta_p(t-s) \sum_j \alpha_{1j}\zeta_j(s),
\end{align}
\end{subequations}
where $\omega_{\text{B}} = \omega + \sum_k \delta \omega_{kk}\cos(\omega\tau) + \Gamma_{kk}\sin(\omega\tau)$ and $\Gamma_{\text{B}} = \gamma + \sum_k \Gamma_{kk}\cos(\omega\tau) - \delta \omega_{kk}\sin(\omega\tau)$. This allows us to reduce the expressions for the dark states to
\begin{subequations}
\begin{align}
\dot{\mathcal{Q}}_j &= \omega \mathcal{P}_j , \\
\dot{\mathcal{P}}_j &= -\omega \mathcal{Q}_j - \gamma \mathcal{P}_j + \Xi_j ,
\end{align}
\end{subequations}
where the modified noise terms are given by
\begin{eqnarray}
\nonumber
\Xi_l &=& \left( \sum_k \frac{\alpha_{lk}}{\alpha_{1k}}(\delta \omega_{kk}\cos(\omega\tau) + \Gamma_{kk}\sin(\omega\tau))\right)\int_{-\infty}^{t} ds \Theta_q(t-s)\sum_j \alpha_{1j}\zeta_j(s) \\
& & - \left( \sum_k \frac{\alpha_{lk}}{\alpha_{1k}}(\Gamma_{kk}\cos(\omega\tau) - \delta \omega_{kk}\sin(\omega\tau))\right)\int_{-\infty}^{t} ds \Theta_p(t-s)\sum_j \alpha_{1j}\zeta_j(s) + \sum_j \alpha_{lj}\zeta_j(t).
\end{eqnarray}
\\

\subsection*{Solving the Lyapunov equation}
\label{LyapunovSolv}
The set of differential equations presented in Eq.~\eqref{App.Full5} can be cast into the form
\begin{equation}
\label{App.Lya1}
\dot{\bold{v}} = M \bold{v} + \bold{n}_{\text{in}}
\end{equation}
where we define $\bold{v} = (q_1, p_1, \dots q_N, p_N)^{\top}$ and $ \bold{n}_{\text{in}} = (0, \zeta_1, \dots , 0, \zeta_N)$. The general formal solution of this set of equations is given by
\begin{equation}
\bold{v}(t) = e^{M(t-t_0)}\bold{v}(t_0) + \int^{t}_{t_0} ds e^{M(t-s)}\bold{n}_{\text{in}}(s)
\end{equation}
which allows us to obtain the correlation matrix of the resonator system
\begin{equation}
\label{App.Lya2}
V = \langle \bold{v}(t) \bold{v}^{\top}(t) \rangle = \int^{t}_{t_0}ds \int^{t}_{t_0}ds' e^{M(t-s)}\langle \bold{n}_{\text{in}}(s) \bold{n}_{\text{in}}^{\top}(s')\rangle e^{M^{\top}(t-s')}.
\end{equation}
Here, we have ignored the transient solution which will decay strongly at large time scales $t$. Regarding the noise correlation term $\langle \bold{n}_{\text{in}}(s)\bold{n}_{\text{in}}^{\top}(s')\rangle$ we can obtain $\langle n_{\text{in},i}(s)n_{\text{in},j}(s') \rangle \neq 0$ only for components where $i$ and $j$ are both even numbers. For these components we obtain the expressions
\begin{subequations}
\begin{align}
\label{App.Lya3}
\nonumber
\langle n_{\text{in},2i}(s)n_{\text{in},2j}(s') \rangle &= \langle \zeta_{i}(s)\zeta_{j}(s')\rangle \\\nonumber
&= \langle \xi_i(s)\xi_j(s')\rangle + \frac{g^{(i)}_{\text{cd}}g^{(j)}_{\text{cd}}\omfb ^{2}}{2\kappa}\left[ \langle (\phi_1^{(\tau)} \ast y^{\text{in}})(s)(\phi_1^{(\tau)} \ast y^{\text{in}})(s') \rangle + (\eta^{-1} -1) \langle (\phi_2^{(\tau)} \ast y^{v})(s)(\phi_2^{(\tau)} \ast y^{v})(s') \rangle\right]\\\nonumber
& + 2\kappa G_i G_j \langle (\phi_3 \ast x^{\text{in}})(s)(\phi_3 \ast x^{\text{in}})(s')\rangle - g^{(i)}_{\text{cd}}\omega_{\text{fb}}G_j \langle (\phi_1^{(\tau)} \ast y^{\text{in}})(s)(\phi_3 \ast x^{\text{in}})(s') \rangle  \\\nonumber
& - g^{(j)}_{\text{cd}}\omega_{\text{fb}}G_i \langle (\phi_3 \ast x^{\text{in}})(s)(\phi_1^{(\tau)} \ast y^{\text{in}})(s') \rangle \\\nonumber
&= (2\bar{n}_{i}+1)\gamma_{i}\delta_{ij}\delta(s-s') + \frac{g^{(i)}_{\text{cd}}g^{(j)}_{\text{cd}}\omfb ^{2}}{4\kappa\eta}\left(\delta(s-s') - \frac{\omfb }{2}e^{-\omfb |s-s'|} \right) + \frac{G_i G_j}{\kappa}\frac{\kappa}{2}e^{-\kappa|s-s'|}\\\nonumber
& + ig^{(i)}_{\text{cd}}G_j \omega_{\text{fb}} \theta((s-\tau)-s')\left(\frac{\omega_{\text{fb}}e^{-\omega_{\text{fb}}((s-\tau)-s')}-\kappa e^{-\kappa((s-\tau)-s')}}{2(\omega_{\text{fb}}-\kappa)} \right) \\
& - ig^{(j)}_{\text{cd}}G_i \omega_{\text{fb}} \theta((s'-\tau)-s) \left(\frac{\omega_{\text{fb}}e^{-\omega_{\text{fb}}((s'-\tau)-s)}-\kappa e^{-\kappa((s'-\tau)-s)}}{2(\omega_{\text{fb}}-\kappa)} \right),
\end{align}
\end{subequations}
for $i,j \in \{1,\dots,N\}$. Here, the delay $\tau$ only emerges in the cross terms of the $x^{\text{in}}$, $y^{\text{in}}$ correlations. For $\omfb , \kappa \gg \omega_j, \Gamma_j$ describing the regime of the sFFLC, we can approximate $\delta(t) \approx (\omfb /2)e^{-\omfb  |t|}$ as well as $\delta(t) \approx (\kappa/2)e^{-\kappa |t|}$ resulting in
\begin{equation}
\label{App.Lya4}
\langle \zeta_{i}(s)\zeta_{j}(s')\rangle \approx \left( (2\bar{n}_{i}+1)\gamma_{i}\delta_{ij} + \frac{G_iG_j}{\kappa}\right)\delta(s-s'),
\end{equation}
that exhibits no dependence on delay $\tau$. For $\delta$-correlated noise we can simplify the correlation matrix to
\begin{equation}
\label{App.Lya5}
V = \int^{t}_{t_0} ds e^{M(t-s)}\mathcal{D}_{\text{in}}e^{M^{\top}(t-s)},
\end{equation}
where $\mathcal{D}_{\text{in},2i,2j} =  (2\bar{n}_{i}+1)\gamma_{i}\delta_{ij} + G_i G_j/\kappa$ for even index numbers and is zero otherwise.
The Lyapunov equation for the $N$-oscillator system which determines the steady solution of the correlation matrix can be derived using integration by parts for
\begin{eqnarray}
\label{App.Eq3}
\nonumber
MV + VM^{\top} &=& \int^{t}_{t_0}ds Me^{M(t-s)}\mathcal{D}_{\text{in}}e^{M^{\top}(t-s)} + VM^{\top} \\\nonumber
&=& -e^{M(t-s)}\mathcal{D}_{\text{in}}e^{M^{\top}(t-s)} \Big|^{t}_{t_0} -VM^{\top} + VM^{\top} \\
&=& -\mathcal{D}_{\text{in}}.
\end{eqnarray}
Evaluating the individual components from the Lyapunov equation we obtain the set of equations in the sFFLC
\begin{subequations}
\label{App.Eq4a}
\begin{align}
\label{App.Eq4aa}
Y_{ii} &= 0, \\
\omega_j Y_{ij} +\omega_i Y_{ji} &= 0,\\
(\gamma_i + \Gamma_{ii}c_i) X_{ii} + \sum_{j\neq i}\Gamma_{ij}s_{j} Y_{ji} + \sum_{j\neq i}\Gamma_{ij}c_j X_{ij} - (2\bar{n}_i + 1)\gamma_i - \frac{G_i^2}{\kappa} &= 0, \\
\omega_i X_{ii} - \left(\omega_i  + \Gamma_{ii}s_{i}\right)Z_{ii} - \sum_{j \neq i} \Gamma_{ij}s_{j} Z_{ij} - \sum_{j \neq i} \Gamma_{ij}c_j Y_{ij} &= 0, \\
\omega_i X_{ij} - \left(\omega_j + \Gamma_{jj}s_j \right)Z_{ij} - \left(\gamma_j + \Gamma_{jj}c_j \right)Y_{ij} - \sum_{k\neq j} \Gamma_{jk}s_k Z_{ik} - \sum_{k\neq j} \Gamma_{jk}c_k Y_{ik}  &= 0, \\
\omega_j X_{ij} - \left(\omega_i + \Gamma_{ii}s_i \right)Z_{ij} - \left(\gamma_i + \Gamma_{ii}c_i \right)Y_{ij} - \sum_{k\neq i} \Gamma_{ik}s_k Z_{jk} - \sum_{k\neq i} \Gamma_{ik}c_k Y_{jk}  &= 0, \\
\label{App.Eq4aaa1}
-\frac{\left(\omega_i^2 - \omega_j^2 \right)}{\omega_i} Y_{ij} -\Gamma_{ji}c_i X_{ii}-\Gamma_{ij}c_j X_{jj} - \sum_{k \neq j} \Gamma_{ik}s_k Y_{kj} - \sum_{k \neq i} \Gamma_{jk}s_k Y_{ki}  - \sum_{k \neq j} \Gamma_{ik}c_k X_{jk} - \sum_{k \neq i} \Gamma_{jk}c_k X_{ik}  + \frac{2G_{i}G_{j}}{\kappa}  &= 0,
\end{align}
\end{subequations}
with $X_{ij} = \langle  p_i p_j + p_j p_i \rangle$, $Y_{ij} = \langle q_i p_j + p_j q_i  \rangle$ and $Z_{ij} = \langle q_i q_j + q_j q_i \rangle$. Additionally we define $c_i = \cos(\omega_j\tau)$ and $s_i = \sin(\omega_j\tau)$ for simplification. In the case that $\gamma_j \ll (g^{(j)}_{\text{cd}}G_{j}\omega_{j})/\kappa$ where $\Gamma_{jj} \approx (g^{(j)}_{\text{cd}}G_{j}\omega_{j})/\kappa$ and $\Gamma_{ij} = (g^{(i)}_{\text{cd}}/g^{(j)}_{\text{cd}})\Gamma_{jj}$, we can simplify the expression in Eq.~\eqref{App.Eq4a1} to acquire the relation
\begin{equation}
\label{App.Eq4aa}
-\frac{\left(\omega_i^2 - \omega_j^2 \right)}{\omega_i} Y_{ij} - \left(\frac{g^{(j)}_{\text{cd}}}{g^{(i)}_{\text{cd}}}\right)(2\bar{n}_i + 1)\gamma_i - \left(\frac{g^{(i)}_{\text{cd}}}{g^{(j)}_{\text{cd}}}\right)(2\bar{n}_j + 1)\gamma_j - \frac{\left(g^{(j)}_{\text{cd}}G_i - g^{(i)}_{\text{cd}}G_j \right)^2}{\kappa g^{(i)}_{\text{cd}}g_{cd,j}} = 0,
\end{equation}
where we have defined
\begin{equation}
\label{App.Eq4a}
\Lambda_{ij} := \left(\left(\frac{g^{(j)}_{\text{cd}}}{g^{(i)}_{\text{cd}}}\right)(2\bar{n}_i + 1)\gamma_i + \left(\frac{g^{(i)}_{\text{cd}}}{g^{(j)}_{\text{cd}}}\right)(2\bar{n}_j + 1)\gamma_j + \frac{(g^{(j)}_{\text{cd}}G_i - g^{(i)}_{\text{cd}}G_j)^2}{\kappa g^{(i)}_{\text{cd}}g^{(j)}_{\text{cd}}}\right).
\end{equation}
\\
For $\tau = 0$ we can get, with respect to the approximations introduced above, exact solutions for the final energies of each mode as has been reported in \cite{Sommer2019Partial} and is presented below
\begin{subequations}
\begin{align}
\label{App.Eq6}
\nonumber
\frac{1}{2}\left(\langle p_i^2\rangle + \langle q_i^2\rangle\right) &= \left(\bar{n}_i + \frac{1}{2} \right)\frac{\gamma_i}{\Gamma_{ii}} + \frac{G_i^2}{2\Gamma_{ii}\kappa} \\
&+ \sum_{j \neq i}\left[ \frac{\Gamma_{ij}}{2\Gamma_{ii}}\left\{\frac{\left(\omega_i^2\Gamma_{jj} + \omega_j^2\Gamma_{ii} \right)\Lambda_{ij}}{\left(\omega_i^2 -\omega_j^2 \right)^2} + \sum_{k \neq i,j} \frac{1}{\left( \omega_i^2 -\omega_j^2\right)}\left(\frac{\omega_i^2 \Gamma_{jk} \Lambda_{ik}}{\left(\omega_i^2 -\omega_k^2 \right)} - \frac{\omega_j^2 \Gamma_{ik}\Lambda_{jk}}{\left(\omega_j^2 - \omega_k^2 \right)} \right) \right\} + \frac{\Gamma_{ij} \Lambda_{ij}}{4\left(\omega_i^2 - \omega_j^2 \right)} \right] \\\nonumber
&\approx \bar{n}_i\frac{\gamma_i}{\Gamma_{ii}} + \frac{G_i^2}{2\Gamma_{ii}\kappa} + \sum_{j \neq i} \left[ \frac{\Gamma_{ij}}{\Gamma_{ii}}\left\{ \frac{\left(\omega_i^2\Gamma_{jj} + \omega_j^2\Gamma_{ii} \right)\left(\left(g^{(j)}_{\text{cd}}\right)^{2}\bar{n}_i\gamma_i + \left(g^{(i)}_{\text{cd}}\right)^{2}\bar{n}_j \gamma_j\right)}{(g^{(i)}_{\text{cd}}g^{(j)}_{\text{cd}})\left(\omega_i^2 -\omega_j^2 \right)^2} \right. \right. \\\nonumber
&+ \left. \sum_{k \neq i,j} \frac{1}{\left( \omega_i^2 -\omega_j^2\right)}\left(\frac{\omega_i^2\Gamma_{jk}\left(\left( g^{(k)}_{\text{cd}}\right)^{2}\bar{n}_i\gamma_i + \left(g^{(i)}_{\text{cd}}\right)^{2}\bar{n}_k\gamma_k\right)}{(g^{(i)}_{\text{cd}}g^{(k)}_{\text{cd}})\left(\omega_i^2 -\omega_k^2 \right)} - \frac{\omega_j^2 \Gamma_{ik} \left( \left(g^{(k)}_{\text{cd}}\right)^{2}\bar{n}_j\gamma_j + \left(g^{(j)}_{\text{cd}}\right)^{2}\bar{n}_k\gamma_k \right)}{(g^{(j)}_{\text{cd}}g^{(k)}_{\text{cd}})\left(\omega_j^2 - \omega_k^2 \right)} \right) \right\} \\
& + \left. \frac{\Gamma_{ij}\left(\left(g^{(j)}_{\text{cd}}\right)^{2}\bar{n}_i \gamma_i + \left(g^{(i)}_{\text{cd}}\right)^{2}\bar{n}_j \gamma_j\right)}{2(g^{(i)}_{\text{cd}}g^{(j)}_{\text{cd}})\left(\omega_i^2 - \omega_j^2 \right)} \right].
\end{align}
\end{subequations}
For $\tau > 0$ it is possible to obtain analytic solutions of Eqs.~\eqref{App.Eq4a} for a single and two oscillator modes.\\

The equations derived from the Lyapunov equation in case of the wFFLC, where we have to work in the regime $k_{\text{B}}T \gg \hbar \omega_j$ where the thermal noise dominates since here the non delta-like noise correlation terms can be ignored, are stated below
\begin{subequations}
\label{App.Eq4a}
\begin{align}
\label{App.Eq4aa}
Y_{ii} &= 0, \\
\omega_j Y_{ij} +\omega_i Y_{ji} &= 0,\\
(\gamma_i + \Delta \Gamma_{ii}(\tau)) X_{ii} + \sum_{j\neq i}\Delta \omega_{ij}(\tau) Y_{ji} + \sum_{j\neq i}\Delta \Gamma_{ij}(\tau) X_{ij} - (2\bar{n}_i + 1)\gamma_i &= 0, \\
\omega_i X_{ii} - \left(\omega_i  + \Delta \omega_{ii}(\tau)\right)Z_{ii} - \sum_{j \neq i} \Delta \omega_{ij}(\tau) Z_{ij} - \sum_{j \neq i} \Delta \Gamma_{ij}(\tau) Y_{ij} &= 0, \\
\omega_i X_{ij} - \left(\omega_j + \Delta \omega_{jj}(\tau) \right)Z_{ij} - \left(\gamma_j + \Delta \Gamma_{jj}(\tau) \right)Y_{ij} - \sum_{k\neq j} \Delta \omega_{jk}(\tau) Z_{ik} - \sum_{k\neq j} \Delta \Gamma_{jk}(\tau) Y_{ik}  &= 0, \\
\omega_j X_{ij} - \left(\omega_i + \Delta \omega_{ii}(\tau) \right)Z_{ij} - \left(\gamma_i + \Delta \Gamma_{ii}(\tau) \right)Y_{ij} - \sum_{k\neq i} \Delta \omega_{ik}(\tau) Z_{jk} - \sum_{k\neq i} \Delta \Gamma_{ik}(\tau) Y_{jk}  &= 0, \\
\label{App.Eq4aaa1}
-\frac{\left(\omega_i^2 - \omega_j^2 \right)}{\omega_i} Y_{ij} - \sum_{k \neq j} \Delta \omega_{ik}(\tau) Y_{kj} - \sum_{k \neq i} \Delta \omega_{jk}(\tau) Y_{ki}  - \sum_{k} \Delta \Gamma_{ik}(\tau) X_{jk} - \sum_{k} \Delta \Gamma_{jk}(\tau) X_{ik} &= 0,
\end{align}
\end{subequations}
with $\Delta \Gamma_{ij}(\tau) = \Gamma_{ij}c_j - \delta\omega_{ij}s_j$ and $\Delta \omega_{ij}(\tau) = \Gamma_{ij}s_j + \delta\omega_{ij}c_j$.
\\

%%%%%%%%%%%%%%%%%%%%%%%%%%%%%%%%%%%%%%%%%%%%%%%%%%%%%%%%%%%%%%%%%%%%%
%%%%%%%%%%%%%%%%%%%%%%%%%%%%%%%%%%%%%%%%%%%%%%%%%%%%%%%%%%%%%%%%%%%%%
\section{Cooling of two adjacent modes}
\label{B}
%%%%%%%%%%%%%%%%%%%%%%%%%%%%%%%%%%%%%%%%%%%%%%%%%%%%%%%%%%%%%%%%%%%%%
%%%%%%%%%%%%%%%%%%%%%%%%%%%%%%%%%%%%%%%%%%%%%%%%%%%%%%%%%%%%%%%%%%%%%

In the sFFLC and under the approximation $\gamma \ll \Gamma_{jj}$ carried out in the drift matrix, we can find analytic solutions for two modes. First we express the diagonal elements as

\begin{subequations}
\begin{align}
\label{DCDEq.7}
X_{ii} &= \frac{(2\bar{n}_i+1)\gamma_i + \frac{G_{i}^{2}}{\kappa}}{\Gamma_{ii}c_i} + \left(\frac{\Gamma_{ij}}{\Gamma_{ii}}\right) \frac{s_j \omega_j}{c_i(\omega^2_j-\omega^2_i)}\Lambda_{ij} - \left(\frac{\Gamma_{ij}}{\Gamma_{ii}}\right)\frac{c_j}{c_i}X_{ij}, \\
Z_{ii} &= \frac{\omega_i}{(\omega_i + \Gamma_{ii}s_i)}X_{ii} -\frac{\Gamma_{ij}s_j}{(\omega_i + \Gamma_{ii}s_i)}Z_{ij} + \frac{\Gamma_{ij} c_j\omega_i}{(\omega_i + \Gamma_{ii}s_i)(\omega^2_i-\omega^2_j)}\Lambda_{ij}
\end{align}
\end{subequations}
This results in
\begin{eqnarray}
\nonumber
n_{i,\text{eff}}(\tau) &=& \frac{1}{4}(X_{ii} + Z_{ii})\\\nonumber
&=& \left( \frac{(\bar{n}_i+1/2)\gamma_i + \frac{G_{i}^{2}}{2\kappa}}{2\Gamma_{ii}c_i} + \left(\frac{\Gamma_{ij}}{4\Gamma_{ii}}\right) \frac{s_j \omega_j}{c_i(\omega^2_j-\omega^2_i)}\Lambda_{ij} - \left(\frac{\Gamma_{ij}}{4\Gamma_{ii}}\right)\frac{c_j}{c_i}X_{ij}\right)\left(1 +  \frac{\omega_i}{(\omega_i + \Gamma_{ii}s_i)} \right) \\
& & -\frac{\Gamma_{ij}s_j}{4(\omega_i + \Gamma_{ii}s_i)}Z_{ij} + \frac{\Gamma_{ij} c_j\omega_i}{4(\omega_i + \Gamma_{ii}s_i)(\omega^2_i-\omega^2_j)}\Lambda_{ij},
\end{eqnarray}
for the final occupation of each mode.
The off diagonal elements $X_{12}$ and $Z_{12}$ are given by

\begin{subequations}
\begin{align}
\nonumber
X_{12} &= \frac{\left(\frac{\Gamma_{21}}{\Gamma_{11}}\right)\omega_1 t_1\left((2\bar{n}_1+1)\gamma_1 + \frac{G_{1}^{2}}{\kappa} \right) - \left(\frac{\Gamma_{12}}{\Gamma_{22}}\right)\omega_{2} t_2 \left((2\bar{n}_2+1)\gamma_2 + \frac{G_{2}^{2}}{\kappa} \right)}{\left[\omega^2_1 - \omega^2_2 + \omega_1\Gamma_{11}s_1 - \omega_2\Gamma_{22}s_2 + \omega_1 \Gamma_{22}t_1c_2 - \omega_2 \Gamma_{11} t_2c_1 \right]} \\
&- \frac{\Lambda_{12}}{\left(\omega^2_1-\omega^2_2 \right)}\frac{\left[\Gamma_{22}t_1s_2\omega_1\omega_2 + \Gamma_{11}t_2s_1\omega_1\omega_2 + \Gamma_{22}c_2\omega^2_1 + \Gamma_{11}c_1\omega^2_2\right]}{\left[\omega^2_1 - \omega^2_2 + \omega_1\Gamma_{11}s_1 - \omega_2\Gamma_{22}s_2 + \omega_1 \Gamma_{22}t_1c_2 - \omega_2 \Gamma_{11} t_2c_1 \right]} \\\nonumber
Z_{12} &= \frac{K_{22}\left(\frac{\Gamma_{21}}{\Gamma_{11}}\right)\omega_1 t_1\left((2\bar{n}_1+1)\gamma_1 + \frac{G_{1}^{2}}{\kappa} \right) - K_{11}\left(\frac{\Gamma_{12}}{\Gamma_{22}}\right)\omega_{2} t_2 \left((2\bar{n}_2+1)\gamma_2 + \frac{G_{2}^{2}}{\kappa} \right)}{(\omega_1\omega_2+\omega_1\Gamma_{22}s_2 + \omega_2\Gamma_{11}s_1)\left[\omega^2_1 - \omega^2_2 + \omega_1\Gamma_{11}s_1 - \omega_2\Gamma_{22}s_2 + \omega_1 \Gamma_{22}t_1c_2 - \omega_2 \Gamma_{11} t_2c_1 \right]} \\
&- \frac{\Lambda_{12}}{\left(\omega^2_1-\omega^2_2 \right)}\frac{\left[K_{22}(\Gamma_{22}t_1s_2\omega_1\omega_2 + \Gamma_{22}c_2\omega^2_1) + K_{11}(\Gamma_{11}t_2s_1\omega_1\omega_2 + \Gamma_{11}c_1\omega^2_2)\right]}{(\omega_1\omega_2+\omega_1\Gamma_{22}s_2 + \omega_2\Gamma_{11}s_1)\left[\omega^2_1 - \omega^2_2 + \omega_1\Gamma_{11}s_1 - \omega_2\Gamma_{22}s_2 + \omega_1 \Gamma_{22}t_1c_2 - \omega_2 \Gamma_{11} t_2c_1 \right]},
\end{align}
\end{subequations}
where $K_{22} = [(\omega_2+\Gamma_{22}s_2)\omega_2 + \omega_2\Gamma_{11}t_2c_1]$ and $K_{11} = [(\omega_1+\Gamma_{11}s_1)\omega_1 + \omega_1\Gamma_{22}t_1c_2]$ and $t_j = \tan(\omega_j\tau)$.\\

\section{Analysis of cooling rates in the Fourier domain}
\label{C}
Starting with the equations of motion from Eqs.~\eqref{App.Full1} a Fourier transformation defined as $O(\Omega) = (1/\sqrt{2\pi})\int^{\infty}_{-\infty}dt e^{-i\Omega t}O(t)$ will result in the set of coupled linear equations
\begin{subequations}
\begin{align}
\label{App.Four2a}
i\Omega q_j(\Omega) &= \omega_j p_j(\Omega),\\
\label{App.Four2b}
i\Omega p_j(\Omega) &= -\omega_j q_j(\Omega) - \gamma_j  p_j(\Omega) +G_j x(\Omega) - g^{(\tau)}_j(\Omega) y^{\text{est}}(\Omega) + \xi_j(\Omega),\\
\label{App.Four2c}
i\Omega x(\Omega) &= -\kappa x(\Omega) + \sqrt{2\kappa} x^{\text{in}}(\Omega), \\
\label{App.Four2d}
i\Omega y(\Omega) &= -\kappa y(\Omega) + \textstyle \sum_{j=1}^{N} G_jq_j(\Omega) + \sqrt{2\kappa}y^{\text{in}}(\Omega).
\end{align}
\end{subequations}
Using $y^{\text{est}}(\Omega) = y(\Omega) - \left(y^{\text{in}}(\Omega) + \sqrt{(1/\eta)-1}y^{v}(\Omega) \right)/\sqrt{2\kappa}$ and Eq.~\eqref{App.Four2a}, Eq.~\eqref{App.Four2c}, Eq.~\eqref{App.Four2d} we can rewrite Eq.~\eqref{App.Four2b} to be
\begin{equation}
\label{App.Four2e}
i\left[\left(\Omega^2 - \omega^2_j \right) - i\gamma_j\Omega - \frac{g^{(0)}_j(\Omega) e^{-i\Omega\tau}\omega_jG_j}{(i\Omega + \kappa)} \right]\frac{1}{\Omega}p_{j}(\Omega) - \sum_{k\neq j} \frac{ig^{(0)}_j(\Omega) G_k \omega_k e^{-i\Omega\tau}}{\Omega(i\Omega + \kappa)} p_k(\Omega) = \zeta_j(\Omega),
\end{equation}
where the driving noise term is given by
\begin{eqnarray}
\label{App.Four3}
\zeta_j(\Omega) &=& \xi_j(\Omega) + \frac{g^{(0)}_j(\Omega)(i\Omega-\kappa)e^{-i\Omega\tau}}{\sqrt{2\kappa}(i\Omega+\kappa)}y^{\text{in}}(\Omega) + \frac{\sqrt{\eta^{-1}-1} g^{(0)}_j(\Omega)e^{-i\Omega\tau}}{\sqrt{2\kappa}}y^{v}(\Omega) + \frac{\sqrt{2\kappa}G_j}{(i\Omega+\kappa)} x^{\text{in}}(\Omega).
\end{eqnarray}
For high temperatures ($k_{\text{B}}T \gg \hbar \omega_j$) this can be approximated by $\zeta_j(\Omega) \approx \xi_j(\Omega)$.\\

\noindent Using Eq.~\eqref{App.Four2a} we can rewrite Eq.~\eqref{App.Four2e} with respect to the position which is expressed by
\begin{subequations}
\label{App.Four4a}
\begin{align}
\left[\left( \omega^2_j - \Omega^2 \right) + i\gamma_j\Omega + \frac{g^{(0)}_j(\Omega) e^{-i\Omega\tau}\omega_jG_j}{(i\Omega + \kappa)} \right]\frac{1}{\omega_j}q_{j}(\Omega) + \sum_{k\neq j} \frac{g^{(0)}_j(\Omega) G_k \omega_k e^{-i\Omega\tau}}{\omega_k (i\Omega + \kappa)} q_k(\Omega) &= \zeta_j(\Omega),\\
\sum_{k=1}^{N}(\chi^{-1})_{jk}(\Omega)q_k(\Omega) &= \zeta_j(\Omega),
\end{align}
\end{subequations}
where $\pmb{\chi}(\Omega)$ describes the effective susceptibility matrix and we obtain $\pmb{q}(\Omega) = \pmb{\chi}(\Omega)\pmb{\zeta}(\Omega)$. For the momentum we can use the relation $\pmb{p}(\Omega) = i\Omega \pmb{\hat{\omega}}^{-1} \pmb{q}(\Omega)$ where $\pmb{\hat{\omega}}_{ij} = \omega_{i}\delta_{ij}$. We can calculate the oscillator position variance from the Fourier transform by
\begin{eqnarray}
\nonumber
\braket{ \pmb{q}(t) \pmb{q}^{\top}(t) } &=& \frac{1}{2\pi} \int^{\infty}_{-\infty}\int^{\infty}_{-\infty} d\Omega d\Omega' e^{i(\Omega + \Omega')t}\braket{\pmb{q}(\Omega)\pmb{q}^{\top}(\Omega')},\\
&=& \frac{1}{2\pi} \int^{\infty}_{-\infty}\int^{\infty}_{-\infty} d\Omega d\Omega' e^{i(\Omega + \Omega')t}\pmb{\chi}(\Omega)\braket{\pmb{\zeta}(\Omega)\pmb{\zeta}^{\top}(\Omega')}\pmb{\chi}^{\top}(\Omega').
\end{eqnarray}
Here, we obtain for
\begin{eqnarray}
\nonumber
\braket{\zeta_j(\Omega)\zeta_k (\Omega')}  &=& \left[ \frac{g^{(0)}_j(\Omega)g^{(0)}_k(\Omega')(i\Omega-\kappa)(i\Omega'-\kappa)e^{-i(\Omega+\Omega')\tau}}{4\kappa (\kappa + i\Omega)(\kappa + i\Omega')} + \frac{\left(\eta^{-1}-1\right)g^{(0)}_j(\Omega)g^{(0)}_k(\Omega')e^{-i(\Omega+\Omega')\tau}}{4\kappa} + \frac{\kappa G_j G_k}{(\kappa + i\Omega)(\kappa + i\Omega')} \right. \\
& & \left. -\frac{ig^{(0)}_j(\Omega)(i\Omega-\kappa)G_k e^{-i\Omega\tau}}{2(\kappa + i\Omega)(\kappa + i\Omega')} +\frac{ig^{(0)}_k(\Omega')(i\Omega'-\kappa)G_j e^{-i\Omega'\tau}}{2(\kappa + i\Omega)(\kappa + i\Omega')} + \frac{\gamma_j \Omega}{\omega_j}\coth\left(\frac{\hbar \Omega}{2k_{\text{B}}T} \right)\delta_{jk}\right]\delta(\Omega + \Omega'),
\end{eqnarray}
following a delta distribution in Fourier space which is typical for colored noise and
which allows us to perform one integration resulting in
\begin{equation}
\braket{ \pmb{q}(t) \pmb{q}^{\top}(t) } = \frac{1}{2\pi} \int^{\infty}_{-\infty} d\Omega \pmb{\chi}(\Omega)\pmb{S}(\Omega)\pmb{\chi}^{\dagger}(\Omega),
\end{equation}
where $\pmb{\chi}^{-1}(-\Omega) = (\pmb{\chi}^{-1})^{*}(\Omega)$ and the full noise spectrum is given by
\begin{eqnarray}
\nonumber
S_{jk}(\Omega) &=& \frac{\gamma_j \Omega}{\omega_j}\coth\left(\frac{\hbar \Omega}{2k_{\text{B}}T} \right)\delta_{jk} + \frac{g^{(0)}_j(\Omega)g^{(0)}_k(\Omega)^{*}}{4\kappa\eta}  + \frac{\kappa G_jG_k}{\left(\Omega^2 + \kappa^2 \right)} + \frac{ig^{(0)}_j(\Omega)G_k e^{-i\Omega\tau}}{2(\kappa + i\Omega)} - \frac{ig^{(0)}_k(\Omega)^{*}G_j e^{i\Omega\tau}}{2(\kappa-i\Omega)} \\
&\approx & \gamma_j (2\bar{n}_j + 1)\delta_{jk} + \frac{g^{(0)}_j(\Omega)g^{(0)}_k(\Omega)^{*}}{4\kappa\eta}  + \frac{\kappa G_jG_k}{\left(\Omega^2 + \kappa^2 \right)} + \frac{ig^{(0)}_j(\Omega)G_k e^{-i\Omega\tau}}{2(\kappa + i\Omega)} - \frac{ig^{(0)}_k(\Omega)^{*}G_j e^{i\Omega\tau}}{2(\kappa-i\Omega)},
\end{eqnarray}
where the approximation is valid for resonably high temperatures meaning $k_B T \gg \hbar \omega_j$ . Here, we have kept the feedback response function $g^{(0)}_j(\Omega)$ general. In the case where $g^{(0)}_j(\Omega) = i\Omega\omfb g^{(j)}_{\text{cd}}/(\omfb + i\Omega)$ the noise spectrum follows the expression
\begin{eqnarray}
\label{App.Noise1}
\nonumber
S_{jk}(\Omega) &\approx & \gamma_j (2\bar{n}_j + 1)\delta_{jk} + \frac{\Omega^2\omfb^2 g^{(j)}_{\text{cd}}g^{(k)}_{\text{cd}}}{4\kappa\eta(\omfb^2 + \Omega^2)} + \frac{\kappa G_j G_k}{(\Omega^2 + \kappa^2)} - \frac{\Omega}{2}\left( \left(\frac{\tilde{\Gamma}_{jk}(\Omega)}{\omega_k} + \frac{\tilde{\Gamma}_{kj}(\Omega)}{\omega_j} \right)\cos(\Omega\tau) \right. \\\nonumber
& & \left. - \left(\frac{\delta\tilde{\omega}_{jk}(\Omega)}{\omega_k} + \frac{\delta\tilde{\omega}_{kj}(\Omega)}{\omega_j} \right)\sin(\Omega\tau) -i\left[\left(\frac{\delta\tilde{\omega}_{jk}(\Omega)}{\omega_k} - \frac{\delta\tilde{\omega}_{kj}(\Omega)}{\omega_j} \right)\cos(\Omega\tau) + \left(\frac{\tilde{\Gamma}_{jk}(\Omega)}{\omega_k} - \frac{\tilde{\Gamma}_{kj}(\Omega)}{\omega_j} \right)\sin(\Omega\tau) \right] \right),\\
\end{eqnarray}
with
\begin{subequations}
\begin{align}
\delta \tilde{\omega}_{jk}(\Omega) &= \frac{g^{(j)}_{\text{cd}}G_k\omega_k\omega_{\text{fb}}\Omega \left(\omega_{\text{fb}} + \kappa \right)}{\left(\kappa^2+\Omega^2 \right)\left(\omega_{\text{fb}}^2+\Omega^2 \right)} \\
\tilde{\Gamma}_{jk}(\Omega) &= \frac{g^{(j)}_{\text{cd}}G_k\omega_k \omega_{\text{fb}} \left(\kappa\omega_{\text{fb}}-\Omega^2 \right)}{\left(\kappa^2+\Omega^2 \right)\left(\omega_{\text{fb}}^2+\Omega^2 \right)}.
\end{align}
\end{subequations}
For the oscillator momentum variance we obtain
\begin{equation}
\braket{ \pmb{p}(t) \pmb{p}^{\top}(t) } = \frac{1}{2\pi} \int^{\infty}_{-\infty} d\Omega \Omega^2  \pmb{\hat{\omega}}^{-1}\pmb{\chi}(\Omega)\pmb{S}(\Omega)\pmb{\chi}^{\dagger}(\Omega)\pmb{\hat{\omega}}^{-1},
\end{equation}
which allows us to express the oscillator-energy matrix by
\begin{equation}
\frac{1}{2}\left(\braket{ \pmb{q}(t) \pmb{q}^{\top}(t) } + \braket{ \pmb{p}(t) \pmb{p}^{\top}(t) } \right) = \frac{1}{2\pi} \int^{\infty}_{-\infty} d\Omega \frac{1}{2}\left[\pmb{\chi}(\Omega)\pmb{S}(\Omega)\pmb{\chi}^{\dagger}(\Omega) + \Omega^2  \pmb{\hat{\omega}}^{-1}\pmb{\chi}(\Omega)\pmb{S}(\Omega)\pmb{\chi}^{\dagger}(\Omega)\pmb{\hat{\omega}}^{-1} \right].
\end{equation}
The energies presented as occupations of the individual modes are located on the diagonal of the matrix and are given by
\begin{equation}
\left(n_{\text{eff}}\right)_j = \frac{1}{2\pi}\int^{\infty}_{\infty}d\Omega \frac{1}{2}\left[ \left(\pmb{\chi}(\Omega)\pmb{S}(\Omega)\pmb{\chi}^{\dagger}(\Omega)\right)_{jj}\left(1 + \frac{\Omega^2}{\omega^{2}_{j}} \right) \right].
\end{equation}
For high temperatures where we can approximate $S_{jk}(\Omega) \approx \gamma_j (2\bar{n}_j + 1)\delta_{jk}$ meaning that the thermal noise is much larger than the feedback and radiation pressure noise, we can simplify the expression above to
\begin{eqnarray}
\label{App.Four3a}
\nonumber
\left(n_{\text{eff}}\right)_j &= \frac{1}{2\pi}\int^{\infty}_{\infty}d\Omega \frac{1}{2}\left[\sum_{k=1}^{N} |\chi_{jk}(\Omega)|^{2}\gamma_{k}(2\bar{n}_{k} + 1) \left(1 + \frac{\Omega^2}{\omega^{2}_{j}} \right) \right]\\
&= \sum_{k=1}^{N} \frac{1}{2\pi}\int^{\infty}_{\infty}d\Omega \frac{1}{2}\left[ \frac{|\text{adj}(\pmb{\chi}^{-1}(\Omega))_{jk}|^{2}\gamma_{k}(2\bar{n}_{k} + 1) }{|\det(\pmb{\chi}^{-1}(\Omega))|^{2}}\left(1 + \frac{\Omega^2}{\omega^{2}_{j}} \right) \right],
\end{eqnarray}
where we have used that the susceptibility matrix can be derived by $\chi_{jk} = \det(\pmb{\chi}^{-1})^{-1} (\text{adj}(\pmb{\chi}^{-1}))_{jk}$ from its inverse which is expressed in Eqs.~\eqref{App.Four3a}.\\

\subsection*{Single mode}
\label{Ea}
For a single mode we obtain
\begin{equation}
q(\Omega)  = \frac{\omega}{\left[\omega^2 - \Omega^2 + i\gamma\Omega + \frac{g^{(0)}(\Omega) e^{-i\Omega\tau}G\omega}{(i\Omega + \kappa)} \right]}\zeta(\Omega)= \chi^{\text{cd}}_{\text{eff}}(\Omega) \zeta(\Omega),
\end{equation}
and $p(\Omega)=i\Omega q(\Omega)/\omega $. The effective susceptibility takes a quasi-Lorentzian form
\begin{align}
\chi^{\text{cd}}_{\text{eff}} &= \frac{\omega}{\left[\left(\omega^2_{\text{eff}}(\Omega) - \Omega^2\right) - i\Omega\gamma_{\text{eff}}(\Omega) \right]},
\end{align}
where we have the effective resonance and damping rates are frequency and time delay dependent
\begin{subequations}
\begin{align}
\omega^2_{\text{eff}}(\Omega)&=\omega^2 + \Omega\delta\tilde{\omega}(\Omega)\cos(\Omega\tau) + \Omega\tilde{\Gamma}(\Omega)\sin(\Omega\tau),\\
\gamma_{\text{eff}}(\Omega) &=\gamma + \tilde{\Gamma}(\Omega)\cos(\Omega\tau) - \delta \tilde{\omega}(\Omega)\sin(\Omega\tau).
\end{align}
\end{subequations}

From the susceptibility we obtain the position fluctuation spectrum via
\begin{eqnarray}
S_{q}(\Omega) &=& |\chi^{\text{cd}}_{\text{eff}}(\Omega)|^2 \left[ S_{\text{th}}(\Omega) + S_{\text{rp}}(\Omega) + S_{\text{fb}}(\Omega) + S_{\text{fb-rp}}(\Omega)\right],
\end{eqnarray}
with the radiation pressure noise term $S_{\text{rp}}(\Omega) = G^2\kappa/(\kappa^2+ \Omega^2)$, the feedback noise $S_{\text{fb}}(\Omega) = |g^{(0)}(\Omega)|^2/(4\kappa\eta)$, the interference between the feedback and the radiation pressure noise term $S_{\text{fb-rp}}(\Omega) = ig^{(0)}(\Omega)G e^{-i\Omega\tau}/(2(\kappa + i\Omega)) - ig^{(0)}(\Omega)^{*}G e^{i\Omega\tau}/(2(\kappa-i\Omega))$ and the thermal noise $S_{\text{th}}(\Omega) \approx \gamma(2\bar{n}+1)$ for high temperature which dominates the noise spectrum. From integration of the fluctuation spectrum we obtain the position and momentum variances at steady state
\begin{subequations}
\begin{align}
\braket{q^2} &= \int^{\infty}_{-\infty}\frac{d\omega}{2\pi}S_{q}(\Omega),\\
\braket{p^2} &= \int^{\infty}_{-\infty}\frac{d\Omega}{2\pi}\frac{\Omega^2}{\omega^2}S_{q}(\Omega).
\end{align}
\end{subequations}
For example we obtain the expression
\begin{equation}
\braket{p^2} = \gamma\left( \bar{n}+ \frac{1}{2}\right)\int^{\infty}_{-\infty}d\Omega \frac{\Omega^2/\pi}{\left(\omega_{\text{eff}}(\Omega)^2 - \Omega^2 \right)^2 + \Omega^2\gamma_{\text{eff}}(\Omega)^2}\approx \gamma\left( \bar{n}+ \frac{1}{2}\right)\int^{\infty}_{0}d\Omega \frac{2/\pi}{4\left(\Omega - \omega_{\text{eff}}(\Omega) \right)^2 + \gamma_{\text{eff}}(\Omega)^2},
\end{equation}
where we have used $\left(\omega_{\text{eff}}(\Omega)^2 - \Omega^2 \right) \approx 2\Omega\left(\Omega -\omega_{\text{eff}}(\Omega) \right)$ given for the near resonance approximation. Approximating $\omega_{\text{eff}}(\Omega) = \omega_{\text{eff}}(\omega)$ and $\gamma_{\text{eff}}(\Omega) = \gamma_{\text{eff}}(\omega)$ we can perform the integration and we obtain
\begin{eqnarray}
\label{App.Five00}
\nonumber
\frac{1}{2}\left(\braket{q^2} + \braket{p^2}\right) &=& \frac{1}{2}\left( \bar{n} + \frac{1}{2}\right)\frac{\gamma}{\gamma_{\text{eff}}(\omega)}\left(1 + \frac{\omega^2}{\omega_{\text{eff}}(\omega)^2}\right)\\
&=& \frac{1}{2}\frac{\gamma (\bar{n} + 1/2)}{\gamma + \tilde{\Gamma}(\omega)\cos(\omega\tau) - \delta \tilde{\omega}(\omega)\sin(\omega\tau)}\left(1 + \frac{\omega}{\omega + \delta\tilde{\omega}(\omega)\cos(\omega\tau) +\tilde{\Gamma}(\omega)\sin(\omega\tau)}\right)
\end{eqnarray}
forming the same result that we have got for the wFFLC-Markovian-approximation.\\

First we want to investigate the behavior for a delay induced phase shift $\omega\tau = 2\pi m$. A first glimpse of the delay dependence can be obtained by looking at $|\chi^{\text{cd}}_{\text{eff}}(\Omega)|^2$. In the case that $\tau = 0$ we get the expression
\begin{eqnarray}
\label{App.Five0}
|\chi^{\text{cd}}_{\text{eff}}(\Omega)|^2 &=& \frac{\omega^2}{\left[\left(\omega^2 + \Omega\delta\tilde{\Omega}(\Omega) - \Omega^2\right)^2 + \Omega^2\left(\gamma + \tilde{\Gamma}(\Omega)\right)^2 \right]},
\end{eqnarray}
while for multiples of $2\pi$ with respect to the phase given by $\tau = 2\pi m/\omega$ we obtain
\begin{eqnarray}
|\chi^{\text{cd}}_{\text{eff}}(\Omega)|^2 &=& \frac{\omega^2}{\left[\left(\omega^2 + \Omega\delta\tilde{\Omega}(\Omega)\cos(\alpha_{n}) + \Omega\tilde{\Gamma}(\Omega)\sin(\alpha_{n})  - \Omega^2\right)^2 + \Omega^2\left(\gamma + \tilde{\Gamma}(\Omega)\cos(\alpha_{n}) - \delta \tilde{\omega}(\Omega)\sin(\alpha_{n}) \right)^2 \right]},
\end{eqnarray}
where $\alpha_m = 2\pi m (\Omega/\omega)$. In the approximation where we set $\omega_{\text{eff}}(\omega)$ and $\gamma_{\text{eff}}(\omega)$ the phase is given by $\alpha_m \approx 2\pi m$ this collapses back to the expression given in Eq.~\eqref{App.Five0}. This shows that the approximation employed to calculate the final occupancy in Eq.~\ref{App.Five00} is only valid for relatively small delay times $\tau$.\\

\subsection*{Two modes}
For two modes we revisit the general equations Eq.~\eqref{App.Four3a} for $N$-modes which by using the nomenclature from the previous section can be cast into the form
\begin{subequations}
\label{App.Five1}
\begin{align}
\frac{(\omega^2_{j,\text{eff}}(\Omega) - \Omega^2 ) + i\Omega\gamma_{j,\text{eff}}(\Omega)}{\omega_j}q_{j}(\Omega) + \sum_{k\neq j} \left(\frac{g^{(j)}_{\text{cd}}}{g^{(k)}_{\text{cd}}}\right)\frac{(\omega^2_{k,\text{eff}}(\Omega)-\omega^2_k) + i\Omega(\gamma_{k,\text{eff}}(\Omega)-\gamma_k)}{\omega_k} q_k(\Omega) &= \zeta_j(\Omega),\\
\sum_{k=1}^{N}(\chi^{-1})_{jk}(\Omega)q_k(\Omega) &= \zeta_j(\Omega),
\end{align}
\end{subequations}
where we follow the corresponding definitions for $\omega^2_{j,\text{eff}}(\Omega)$ and $\gamma_{j,\text{eff}}(\Omega)$ from the previous section. Since the susceptibility is given by $\pmb{\chi} = \det(\pmb{\chi}^{-1})^{-1}\text{adj}(\pmb{\chi}^{-1})$ we obtain for two modes
\begin{equation}
\pmb{\chi}(\Omega) = \frac{1}{\det(\pmb{\chi}^{-1})}\left( \begin{array}{cc} \frac{(\omega^2_{2,\text{eff}}(\Omega) - \Omega^2 ) + i\Omega\gamma_{2,\text{eff}}(\Omega)}{\omega_2} & -\left(\frac{g^{(1)}_{\text{cd}}}{g^{(2)}_{\text{cd}}}\right)\frac{(\omega^2_{2,\text{eff}}(\Omega)-\omega^2_2) + i\Omega(\gamma_{2,\text{eff}}(\Omega)-\gamma_2)}{\omega_2} \\ -\left(\frac{g^{(2)}_{\text{cd}}}{g^{(1)}_{\text{cd}}}\right)\frac{(\omega^2_{1,\text{eff}}(\Omega)-\omega^2_1) + i\Omega(\gamma_{1,\text{eff}}(\Omega)-\gamma_1)}{\omega_1} & \frac{(\omega^2_{1,\text{eff}}(\Omega) - \Omega^2 ) + i\Omega\gamma_{1,\text{eff}}(\Omega)}{\omega_1} \end{array} \right),
\end{equation}
where
\begin{eqnarray}
\nonumber
\det(\pmb{\chi}^{-1}) &=& \frac{1}{\omega_1\omega_2}\left[\left((\omega^2_{1,\text{eff}}(\Omega) - \Omega^2 ) + i\Omega\gamma_{1,\text{eff}}(\Omega)\right)\left((\omega^2_{2,\text{eff}}(\Omega) - \Omega^2 ) + i\Omega\gamma_{2,\text{eff}}(\Omega) \right) \right. \\
& & \left. -\left( (\omega^2_{1,\text{eff}}(\Omega)-\omega^2_1) + i\Omega(\gamma_{1,\text{eff}}(\Omega)-\gamma_1)\right)\left( (\omega^2_{2,\text{eff}}(\Omega)-\omega^2_2) + i\Omega(\gamma_{2,\text{eff}}(\Omega)-\gamma_2) \right) \right].
\end{eqnarray}
To investigate Eq.~\eqref{App.Four4a} which gives the result for the final occupation of each mode for high thermal noise we evaluate the terms
\begin{subequations}
\begin{align}
\nonumber
(\pmb{\chi}(\Omega)\pmb{S}(\Omega)\pmb{\chi}^{\dagger}(\Omega))_{11} &= \frac{\omega^{2}_{1}}{\omega^2_1\omega^2_2|\det(\pmb{\chi}^{-1}(\Omega))|^2}(\gamma_1(2\bar{n}_1+1)\left[(\omega^2_{2,\text{eff}}(\Omega) - \Omega^2 )^2 + \Omega^2\gamma_{2,\text{eff}}(\Omega)^2 \right]
\\
& + \gamma_2(2\bar{n}_2+1)\left(\frac{g^{(1)}_{\text{cd}}}{g^{(2)}_{\text{cd}}} \right)^2\left[ (\omega^2_{2,\text{eff}}(\Omega) - \omega_2^2 )^2 + \Omega^2(\gamma_{2,\text{eff}}(\Omega)-\gamma_2)^2 \right] ) \\\nonumber
(\pmb{\chi}(\Omega)\pmb{S}(\Omega)\pmb{\chi}^{\dagger}(\Omega))_{22} &= \frac{\omega^{2}_{2}}{\omega^2_1\omega^2_2|\det(\pmb{\chi}^{-1}(\Omega))|^2}(\gamma_2(2\bar{n}_2+1)\left[(\omega^2_{1,\text{eff}}(\Omega) - \Omega^2 )^2 + \Omega^2\gamma_{1,\text{eff}}(\Omega)^2 \right]
\\
& + \gamma_1(2\bar{n}_1+1)\left(\frac{g^{(2)}_{\text{cd}}}{g^{(1)}_{\text{cd}}} \right)^2\left[ (\omega^2_{1,\text{eff}}(\Omega) - \omega_1^2 )^2 + \Omega^2(\gamma_{1,\text{eff}}(\Omega)-\gamma_1)^2 \right]),
\end{align}
\end{subequations}
with
\begin{eqnarray}
\nonumber
|\det(\pmb{\chi}^{-1}(\Omega)|^{2} &= \frac{1}{\omega^2_1\omega^2_2}\left\{\left[(\omega^2_{1,\text{eff}}(\Omega) - \Omega^2 )^2 + \Omega^2\gamma_{1,\text{eff}}(\Omega)^2 \right]\left[(\omega^2_{2,\text{eff}}(\Omega) - \Omega^2 )^2 + \Omega^2\gamma_{2,\text{eff}}(\Omega)^2 \right] \right. \\\nonumber
& \left. + \left[(\omega^2_{1,\text{eff}}(\Omega) - \omega^2_1)^2 + \Omega^2(\gamma_{1,\text{eff}}(\Omega)-\gamma_1)^2 \right]\left[(\omega^2_{2,\text{eff}}(\Omega) - \omega^2_2 )^2 + \Omega^2(\gamma_{2,\text{eff}}(\Omega)-\gamma_2)^2 \right] \right. \\\nonumber
& \left. + 2\left\{\left[(\omega^2_{1,\text{eff}}(\Omega) - \Omega^2) (\omega^2_{1,\text{eff}}(\Omega) - \omega^2_1) + \Omega^2\gamma_{1,\text{eff}}(\Omega)\left(\gamma_{1,\text{eff}}(\Omega)-\gamma_1 \right) \right] \right. \right. \\\nonumber
& \left. \left. \times \left[(\omega^2_{2,\text{eff}}(\Omega) - \Omega^2) (\omega^2_{2,\text{eff}}(\Omega) - \omega^2_2) + \Omega^2\gamma_{2,\text{eff}}(\Omega)\left(\gamma_{2,\text{eff}}(\Omega)-\gamma_2 \right) \right] \right. \right. \\
& \left. \left. - \Omega^2\left[\gamma_{1,\text{eff}}(\Omega)(\Omega^2-\omega^2_1) + \gamma_1(\omega^2_{1,\text{eff}}(\Omega)-\Omega^2) \right]\left[\gamma_{2,\text{eff}}(\Omega)(\Omega^2-\omega^2_2) + \gamma_2(\omega^2_{2,\text{eff}}(\Omega)-\Omega^2) \right] \right\} \right\}.
\end{eqnarray}
These results can be reshaped into a more convenient form given by
\begin{subequations}
\label{App.Five5}
\begin{align}
\nonumber
(\pmb{\chi}(\Omega)\pmb{S}(\Omega)\pmb{\chi}^{\dagger}(\Omega))_{11} &= \frac{\omega^2_1 \gamma_1(2\bar{n}_1 + 1)}{\left[(\omega^2_{1,\text{eff}}(\Omega) - \Omega^2 )^2 + \Omega^2\gamma_{1,\text{eff}}(\Omega)^2 \right]\left[1 - f(\Omega) \right]} \\
& + \left(\frac{g^{(1)}_{\text{cd}}}{g^{(2)}_{\text{cd}}} \right)^2 \frac{\left[ (\omega^2_{2,\text{eff}}(\Omega) - \omega_2^2 )^2 + \Omega^2(\gamma_{2,\text{eff}}(\Omega)-\gamma_2)^2 \right]}{\left[(\omega^2_{1,\text{eff}}(\Omega) - \Omega^2 )^2 + \Omega^2\gamma_{1,\text{eff}}(\Omega)^2 \right]}\frac{\omega^2_1 \gamma_2 (2\bar{n}_2 + 1)}{\left[(\omega^2_{2,\text{eff}}(\Omega) - \Omega^2 )^2 + \Omega^2\gamma_{2,\text{eff}}(\Omega)^2 \right][1 - f(\Omega)]} \\\nonumber
(\pmb{\chi}(\Omega)\pmb{S}(\Omega)\pmb{\chi}^{\dagger}(\Omega))_{22} &= \frac{\omega^2_2 \gamma_2(2\bar{n}_2 + 1)}{\left[(\omega^2_{2,\text{eff}}(\Omega) - \Omega^2 )^2 + \Omega^2\gamma_{2,\text{eff}}(\Omega)^2 \right]\left[1 - f(\Omega) \right]} \\
& + \left(\frac{g^{(2)}_{\text{cd}}}{g^{(1)}_{\text{cd}}} \right)^2 \frac{\left[ (\omega^2_{1,\text{eff}}(\Omega) - \omega_1^2 )^2 + \Omega^2(\gamma_{1,\text{eff}}(\Omega)-\gamma_1)^2 \right]}{\left[(\omega^2_{2,\text{eff}}(\Omega) - \Omega^2 )^2 + \Omega^2\gamma_{2,\text{eff}}(\Omega)^2 \right]}\frac{\omega^2_2 \gamma_1 (2\bar{n}_1 + 1)}{\left[(\omega^2_{1,\text{eff}}(\Omega) - \Omega^2 )^2 + \Omega^2\gamma_{1,\text{eff}}(\Omega)^2 \right][1 - f(\Omega)]}
\end{align}
\end{subequations}
where the function $f(\Omega)$ is expressed by
\begin{eqnarray}
\nonumber
f(\Omega) &=& \frac{\left[ (\omega^2_{1,\text{eff}}(\Omega) - \omega_1^2 )^2 + \Omega^2(\gamma_{1,\text{eff}}(\Omega)-\gamma_1)^2 \right]\left[ (\omega^2_{2,\text{eff}}(\Omega) - \omega_2^2 )^2 + \Omega^2(\gamma_{2,\text{eff}}(\Omega)-\gamma_2)^2 \right]}{\left[(\omega^2_{1,\text{eff}}(\Omega) - \Omega^2 )^2 + \Omega^2\gamma_{1,\text{eff}}(\Omega)^2 \right]\left[(\omega^2_{2,\text{eff}}(\Omega) - \Omega^2 )^2 + \Omega^2\gamma_{2,\text{eff}}(\Omega)^2 \right]} \\
& & - 2\Re \left\{ \frac{\left[ (\omega^2_{1,\text{eff}}(\Omega) - \omega_1^2 ) + i\Omega(\gamma_{1,\text{eff}}(\Omega)-\gamma_1) \right]\left[ (\omega^2_{2,\text{eff}}(\Omega) - \omega_2^2 ) + i\Omega(\gamma_{2,\text{eff}}(\Omega)-\gamma_2) \right]}{\left[(\omega^2_{1,\text{eff}}(\Omega) - \Omega^2 ) + i\Omega\gamma_{1,\text{eff}}(\Omega) \right]\left[(\omega^2_{2,\text{eff}}(\Omega) - \Omega^2 ) + i\Omega\gamma_{2,\text{eff}}(\Omega) \right]}  \right\}.
\end{eqnarray}
Without loss of generality, in the case that $\omega_1 \ll \omega_2$ we see from Eq.~\eqref{App.Five5} that all terms harboring products of the resonance terms $(\omega^2_{1,\text{eff}}(\Omega) - \Omega^2 ) + i\Omega\gamma_{1,\text{eff}}(\Omega)$ and $(\omega^2_{2,\text{eff}}(\Omega) - \Omega^2 ) + i\Omega\gamma_{2,\text{eff}}(\Omega)$ in the denominator become very small in comparison to terms with single resonance terms in the denominator and can be neglected and since also $f(\Omega) \rightarrow 0$ we obtain the limit of independent solutions for each mode matching with the single mode solutions.
\\
A simple solution can be obtained in the case when we have two identical oscillators with identical coupling.
Here we can uncouple the mode for the center of mass oscillation from the mode describing the relative motion.
Here, we obtain
\begin{subequations}
\label{App.Five7}
\begin{align}
Q(\Omega) &= \frac{\omega}{(\omega_{\text{B,eff}}^2(\Omega)-\Omega^2) + i\Omega\gamma_{\text{B,eff}}(\Omega)}\left(\frac{\zeta_1(\Omega) + \zeta_2(\Omega)}{\sqrt{2}}\right) \\
\delta q(\Omega) &= \frac{\omega}{(\omega^2-\Omega^2) + i\Omega\gamma}\left(\frac{\zeta_1(\Omega) - \zeta_2(\Omega)}{\sqrt{2}}\right),
\end{align}
\end{subequations}
where we have $\omega_{\text{B,eff}}^2 = \omega^2 + 2\Omega \delta \tilde{\omega}(\Omega)\cos(\Omega\tau) + 2\Omega\tilde{\Gamma}(\Omega)\sin(\Omega\tau)$ and $\gamma_{\text{B,eff}} = \gamma + 2\tilde{\Gamma}(\Omega)\cos(\Omega\tau) - 2\delta \tilde{\omega}(\Omega)\sin(\Omega\tau)$.
In the case that $k_{\text{B}} T \gg \hbar \omega$ where we can ignore the feedback and radiation pressure noise terms we obtain $\braket{(\zeta_1(\Omega) \pm \zeta_2(\Omega))(\zeta_1(\Omega) \pm \zeta_2(\Omega))}/2 = \braket{\zeta_1(\Omega)\zeta_1(\Omega)}/2 + \braket{\zeta_2(\Omega)\zeta_2(\Omega)}/2$ in both cases resulting in the position spectra
\begin{subequations}
\begin{align}
S_{Q}(\Omega) &= \frac{\omega^2\gamma(2\bar{n}+1)}{(\omega_{\text{B,eff}}^2(\Omega)-\Omega^2)^2 + \Omega^2\gamma^2_{\text{B,eff}}(\Omega)} \\
S_{\delta q}(\Omega) &= \frac{\omega^2\gamma(2\bar{n}+1)}{(\omega^2-\Omega^2)^2 + \Omega^2\gamma^2}.
\end{align}
\end{subequations}
For the mode carrying the relative motion we can obtain the occupation by integration of
\begin{equation}
\frac{1}{2}\left(\braket{\delta q^2} + \braket{\delta p^2}\right) = \frac{1}{2}\int^{\infty}_{-\infty}\frac{d\omega}{2\pi}S_{\delta q}(\Omega)\left(1 + \frac{\Omega^2}{\omega^2} \right) = \left( \bar{n} + \frac{1}{2} \right),
\end{equation}
showing the occupation of an oscillator mode that is completely unaffected by the feedback.
For the center of mass mode we have to use the approximation introduced above which results in
\begin{eqnarray}
\nonumber
\frac{1}{2}\left(\braket{Q^2} + \braket{P^2}\right) &=& \frac{1}{2}\int^{\infty}_{-\infty}\frac{d\omega}{2\pi}S_{Q}(\Omega)\left(1 + \frac{\Omega^2}{\omega^2} \right)\\
&\approx & \frac{1}{2}\frac{\gamma (\bar{n} + 1/2)}{\gamma + 2(\tilde{\Gamma}(\omega)\cos(\omega\tau) - \delta \tilde{\omega}(\omega)\sin(\omega\tau))}\left(1 + \frac{\omega}{\omega + 2(\delta\tilde{\omega}(\omega)\cos(\omega\tau) +\tilde{\Gamma}(\omega)\sin(\omega\tau))}\right).
\end{eqnarray}
\\

\subsection*{Many modes}
For $N$-resonator modes it is far more difficult to obtain simple analytic solutions for arbitrary delay times $\tau$. Nevertheless for the collective basis in Fourier domain we can obtain analytic expressions for the position spectrum of each collective mode that upon integration can deliver steady-state final occupations of the collective modes and following retransformation we can obtain the final occupation of each individual mode. By starting from the Fourier domain we obtain the equations
\begin{subequations}
\begin{align}
\frac{(\omega^2_{\text{j,eff}}(\Omega) - \Omega^2 ) + i\Omega\gamma_{\text{j,eff}}(\Omega)}{\omega_j}q_{j}(\Omega) + \sum_{k\neq j} \left(\frac{g_{\text{cd}}^{(j)}}{g_{\text{cd}}^{(k)}}\right) \frac{(\omega^2_{\text{k,eff}}(\Omega)-\omega^2_k) + i\Omega(\gamma_{\text{k,eff}}(\Omega)-\gamma_k)}{\omega_k} q_k(\Omega) &= \zeta_j(\Omega) \\
\frac{(\omega^2_j - \Omega^2 ) + i\Omega\gamma_j}{\omega_j}q_{j}(\Omega) + \sum_{k} \left(\frac{g_{\text{cd}}^{(j)}}{g_{\text{cd}}^{(k)}}\right) \frac{(\omega^2_{\text{k,eff}}(\Omega)-\omega^2_k) + i\Omega(\gamma_{\text{k,eff}}(\Omega)-\gamma_k)}{\omega_k} q_k(\Omega) &= \zeta_j(\Omega)\\
\frac{(\omega^2_j - \Omega^2 ) + i\Omega\gamma_j}{\omega_j}q_{j}(\Omega) + g_{\text{cd}}^{(j)}\Omega\left(\delta\bar{\omega}(\Omega)c + \bar{\Gamma}(\Omega)s + i(\bar{\Gamma}(\Omega)s - \delta \bar{\omega}(\Omega)s) \right) \sum_{k} G_k q_k(\Omega) &= \zeta_j(\Omega)\\
\label{App.Multi2}
\frac{(\omega^2_j - \Omega^2 ) + i\Omega\gamma_j}{\omega_j}q_{j}(\Omega) + g_{\text{cd}}^{(j)}\Omega\left(\delta\bar{\omega}(\Omega)c + \bar{\Gamma}(\Omega)s + i(\bar{\Gamma}(\Omega)s - \delta \bar{\omega}(\Omega)s) \right) \sqrt{\sum_l G^2_l}\mathcal{Q}_1(\Omega) &= \zeta_j(\Omega)
\end{align}
\end{subequations}
where $Q_1(\Omega) = \left(\sqrt{\sum_l G^2_l}\right)^{-1}\sum_{k} G_k q_k(\Omega)$ is the position quadrature of the bright mode that is directly addressed by the feedback mechanism and the we have defined the terms $\delta \bar{\omega} = \omfb \Omega (\kappa + \omfb)/((\kappa^2 + \Omega^2)(\omfb^2 + \Omega^2)) $, $\bar{\Gamma}(\Omega) = \omfb(\kappa\omfb - \Omega^2)/((\kappa^2 + \Omega^2)(\omfb^2 + \Omega^2))$ and $c = \cos(\Omega\tau)$, $s = \sin(\Omega\tau)$. The $N-1$ additional collective dark modes of the resonator can be obtained from a Gram-Schmidt procedure. In general we obtain $Q_j(\Omega) = \sum_k \alpha_{jk} q_k(\Omega)$ which follows the rule $\sum_j \alpha^{*}_{kj} \alpha_{k'j} = \delta_{kk'}$. By forming a weigthed sum with the weights $G_j\omega_j/\left(\sqrt{\sum_l G^2_l}((\omega^2_j - \Omega^2 ) + i\Omega\gamma_j)\right)$ over Eq.~\eqref{App.Multi2} we obtain
\begin{equation}
\mathcal{Q}_1(\Omega) + \sum_j \frac{(\omega^2_{\text{j,eff}}(\Omega)-\omega^2_j) + i\Omega(\gamma_{\text{j,eff}}(\Omega)-\gamma_j)}{((\omega^2_j - \Omega^2 ) + i\Omega\gamma_j)}\mathcal{Q}_1(\Omega) = \sum_j \frac{\omega_j}{((\omega^2_j - \Omega^2 ) + i\Omega\gamma_j)}\frac{G_j\zeta_j(\Omega)}{\sqrt{\sum_l G^2_l}},
\end{equation}
resulting in
\begin{equation}
\label{App.Multi3}
\mathcal{Q}_1(\Omega) = \sum_j \frac{\omega_j}{\left[\left(\omega^2_{\text{j,eff}}(\Omega) - \Omega^2\right) + i\Omega\gamma_{\text{j,eff}}(\Omega) + \sum_{k \neq j}\frac{\left[\left(\omega^2_{\text{k,eff}}(\Omega) - \omega^2_k\right) + i\Omega(\gamma_{\text{k,eff}}(\Omega)-\gamma_k)\right]\left[(\omega^2_j - \Omega^2)+i\Omega \gamma_j\right]}{\left[(\omega^2_k - \Omega^2)+i\Omega \gamma_k\right]} \right]}\frac{G_j\zeta_j(\Omega)}{\sqrt{\sum_k G^2_k}}.
\end{equation}
By summing over the weights $\alpha_{kj}\omega_{j}/((\omega^2_j - \Omega^2 ) + i\Omega\gamma_j)$ we obtain the relation
\begin{equation}
\mathcal{Q}_k(\Omega) + \sum_j \frac{\alpha_{kj}\left[\left(\omega^2_{\text{j,eff}}(\Omega) - \omega^2_j\right) + i\Omega(\gamma_{\text{j,eff}}(\Omega)-\gamma_j) \right]}{\alpha_{1j}\left[(\omega^2_j - \Omega^2)+i\Omega \gamma_j\right]}\mathcal{Q}_1(\Omega)  = \sum_j \frac{\omega_j}{\left[(\omega^2_j - \Omega^2)+i\Omega \gamma_j\right]}\alpha_{kj}\zeta_j(\Omega),
\end{equation}
which allows us to obtain the solutions for the dark modes $\mathcal{Q}_k$ by injecting the solution for the bright mode from Eq.~\eqref{App.Multi3}.
In the case that we have $N$ degenerate modes of frequency $\omega$ and natural decay rate $\gamma$ the solution for the bright mode can be simplified to
\begin{equation}
\mathcal{Q}_1 = \frac{\omega}{\left[\left(\omega^2_{\text{B,eff}}(\Omega) - \Omega^2\right) + i\Omega\gamma_{\text{B,eff}}(\Omega) \right]}\sum_j\frac{G_j\zeta_j(\Omega)}{\sqrt{\sum_j G^2_j}},
\end{equation}
with $\omega^2_{\text{B,eff}}(\Omega) = \omega^2 + \sum_j\Omega\delta\tilde{\omega}_j(\Omega)\cos(\Omega\tau) + \Omega\tilde{\Gamma}_j(\Omega)\sin(\Omega\tau)$ and with an effective decay rate of $\gamma_{\text{B,eff}}(\Omega) =\gamma + \sum_j\tilde{\Gamma}_j(\Omega)\cos(\Omega\tau) - \delta \tilde{\omega}_j(\Omega)\sin(\Omega\tau)$.\\
In case that the coupling coefficients to the cavity mode $G_k = G$ and the coupling coefficients to the feedback force $g^{(k)}_{\text{cd}}$ are the same for each mode we obtain
\begin{subequations}
\label{App.Multi4}
\begin{align}
\mathcal{Q}_1 &= \frac{\omega}{\left[\left(\omega^2_{\text{B,eff}}(\Omega) - \Omega^2\right) + i\Omega\gamma_{\text{B,eff}}(\Omega) \right]}\sum_j\frac{\zeta_j(\Omega)}{\sqrt{N}},\\
\mathcal{Q}_k &= \frac{\omega}{\left[\left(\omega^2 - \Omega^2\right) + i\Omega\gamma \right]}\sum_j \alpha_{kj} \zeta_j(\Omega)
\end{align}
\end{subequations}
with $\omega^2_{\text{B,eff}}(\Omega) =  \omega^2 + N\Omega \delta \tilde{\omega}(\Omega)\cos(\Omega\tau) + N\Omega\tilde{\Gamma}(\Omega)\sin(\Omega\tau)$ and $\gamma_{\text{B,eff}} = \gamma + N\tilde{\Gamma}(\Omega)\cos(\Omega\tau) - N\delta \tilde{\omega}(\Omega)\sin(\Omega\tau)$. For the collective bright mode this results in
\begin{eqnarray}
\nonumber
\frac{1}{2}\left(\braket{Q_1^2} + \braket{P_1^2}\right) &=& \int^{\infty}_{-\infty}\frac{d\omega}{4\pi}\left(1 + \frac{\Omega^2}{\omega^2} \right)\frac{\omega^2}{\left[\left(\omega^2_{\text{B,eff}}(\Omega) - \Omega^2\right)^2 + \Omega^2\gamma_{\text{B,eff}}(\Omega)^2 \right]}\frac{1}{N}\sum_j\sum_k S_{jk}(\Omega) \\\nonumber
&=& \int^{\infty}_{-\infty}\frac{d\omega}{4\pi}\left(1 + \frac{\Omega^2}{\omega^2} \right)\frac{\omega^2}{\left[\left(\omega^2_{\text{B,eff}}(\Omega) - \Omega^2\right)^2 + \Omega^2\gamma_{\text{B,eff}}(\Omega)^2 \right]}\left(\gamma(2\bar{n}+1) + N\left(\frac{\Omega^2\omfb^2 g^{2}_{\text{cd}}}{4\kappa\eta(\omfb^2 + \Omega^2)} \right. \right. \\\nonumber
& & \left. \left. + \frac{\kappa G^2}{(\Omega^2 + \kappa^2)}  - \frac{\Omega}{\omega}(\gamma_{\text{B,eff}}(\Omega)-\gamma)  \right) \right). \\
\end{eqnarray}
To obtain analytical results for the occupation of the bright mode considering all noise terms we can use the approximation
\begin{equation}
\frac{\omega^2}{\left[\left(\omega^2_{\text{B,eff}}(\Omega) - \Omega^2\right)^2 + \Omega^2\gamma_{\text{B,eff}}(\Omega)^2 \right]} \approx \frac{\omega^2}{\left[\left(\omega^2_{\text{B,eff}}(\omega) - \Omega^2\right)^2 + \Omega^2\gamma_{\text{B,eff}}(\omega)^2 \right]}
\end{equation}
for the susceptibility function which gives us the expression
\begin{eqnarray}
\nonumber
\frac{1}{2}\left(\braket{Q_1^2} + \braket{P_1^2}\right) &\approx& \int^{\infty}_{-\infty}\frac{d\omega}{4\pi}\left(1 + \frac{\Omega^2}{\omega^2} \right)\frac{\omega^2}{\left[\left(\omega^2_{\text{B,eff}}(\omega) - \Omega^2\right)^2 + \Omega^2\gamma_{\text{B,eff}}(\omega)^2 \right]}\left(\gamma(2\bar{n}+1) + N\left(\frac{\Omega^2\omfb^2 g^{2}_{\text{cd}}}{4\kappa\eta(\omfb^2 + \Omega^2)} \right. \right. \\\nonumber
& & \left. \left. + \frac{\kappa G^2}{(\Omega^2 + \kappa^2)}  - \frac{\Omega}{\omega}(\gamma_{\text{B,eff}}(\Omega)-\gamma)  \right) \right) \\\nonumber
&=& \frac{1}{2}\frac{\gamma}{\gamma_{\text{B,eff}}}\left(\bar{n} + \frac{1}{2} \right)\left(1 + \frac{\omega^2}{\omega_{\text{B,eff}}} \right) + \frac{NG^2}{4\omega^2_{\text{B,eff}}\gamma_{\text{B,eff}}}\left[\kappa - \frac{(\kappa^2-\omega^2)(\kappa+\gamma_{\text{B,eff}})(\omega^2_{\text{B,eff}} + \kappa^2 -\kappa \gamma_{\text{B,eff}})}{(\omega^2_{\text{B,eff}} + \kappa^2)^2-\gamma^2_{\text{B,eff}}\kappa^2}\right]\\
& & + \frac{N\omfb^2 g^2_{\text{cd}}}{16\kappa\eta \omega^2_{\text{B,eff}}\gamma_{\text{B,eff}}}\left[\omega^2 + \frac{(\omega^2_{\text{B,eff}} + \omfb^2)(\omega^4_{\text{B,eff}} - \omega^2\omfb^2)+(\omfb^2 - \omega^2)\omfb\omega^2_{\text{B,eff}}\gamma_{\text{B,eff}}}{(\omega^2_{\text{B,eff}} + \omfb^2)^2-\gamma^2_{\text{B,eff}}\omfb^2} \right]
\end{eqnarray}
In the high temperature limit and for a sufficiently low number of modes $N$ we can ignore the feedback and radiation pressure noise terms which results in
\begin{eqnarray}
\frac{1}{2}\left(\braket{Q_1^2} + \braket{P_1^2}\right) &\approx & \frac{1}{2}\frac{\gamma (\bar{n} + 1/2)}{\gamma + N(\tilde{\Gamma}(\omega)\cos(\omega\tau) - \delta \tilde{\omega}(\omega)\sin(\omega\tau))}\left(1 + \frac{\omega}{\omega + N(\delta\tilde{\omega}(\omega)\cos(\omega\tau) +\tilde{\Gamma}(\omega)\sin(\omega\tau))}\right),
\end{eqnarray}
which for $\tau = 0$ is given by
\begin{equation}
\frac{1}{2}\left(\braket{Q_1^2} + \braket{P_1^2}\right) \approx \frac{1}{2}\frac{\gamma}{\gamma + N\Gamma}\left(\bar{n} + \frac{1}{2}\right)\left(1 + \frac{\omega}{\omega + N\delta \omega}\right).
\end{equation}
Here, the $(N-1)$ collective modes representing the relative motion all have an unmodified occupation number which is independent of $\tau$ and given by
\begin{equation}
\frac{1}{2}\left(\braket{Q_j^2} + \braket{P_j^2}\right) = \left(\bar{n} + \frac{1}{2}\right),
\end{equation}
where the index fulfills the condition $j \neq 1$. This shows that only the bright mode is accessible to cooling with an effective decay rate being $N$ times larger than the decay rate for a single mode. This opens up an avenue for single mode cooling where an $N$ times lower temperature can be reached for the collective bright mode in comparison to addressing one of the individual identical modes.
\\
In the case that $k_{\text{B}}T \sim \omega$ where the feedback damping can approach the quantum limit we have to consider the contribution from feedback and radiation pressure noise.
Neglecting the thermal noise we obtain the residual occupation solely created by the feedback and radiation pressure terms which for $\tau = 0$ can be approximated by
\begin{eqnarray}
\nonumber
n_{\text{B,res}} &=& \int^{\infty}_{0}\frac{d\omega}{2\pi}\left(1 + \frac{\Omega^2}{\omega^2} \right)\frac{\omega^2}{\left[\left(\omega^2_{\text{B,eff}}(\Omega) - \Omega^2\right)^2 + \Omega^2\gamma_{\text{B,eff}}(\Omega)^2 \right]} N\left(\frac{\Omega^2\omfb^2 g^{(j)}_{\text{cd}}g^{(k)}_{\text{cd}}}{4\kappa\eta(\omfb^2 + \Omega^2)} + \frac{\kappa G_j G_k}{(\Omega^2 + \kappa^2)} \right)  - \frac{1}{2}\\\nonumber
&\approx & \frac{NG^2}{4\omega^2_{\text{B,eff}}\gamma_{\text{B,eff}}}\left[\kappa - \frac{(\kappa^2-\omega^2)(\kappa+\gamma_{\text{B,eff}})(\omega^2_{\text{B,eff}} + \kappa^2 -\kappa \gamma_{\text{B,eff}})}{(\omega^2_{\text{B,eff}} + \kappa^2)^2-\gamma^2_{\text{B,eff}}\kappa^2}\right] \\
& & + \frac{N\omfb^2 g^2_{\text{cd}}}{16\kappa\eta \omega^2_{\text{B,eff}}\gamma_{\text{B,eff}}}\left[\omega^2 + \frac{(\omega^2_{\text{B,eff}} + \omfb^2)(\omega^4_{\text{B,eff}} - \omega^2\omfb^2)+(\omfb^2 - \omega^2)\omfb\omega^2_{\text{B,eff}}\gamma_{\text{B,eff}}}{(\omega^2_{\text{B,eff}} + \omfb^2)^2-\gamma^2_{\text{B,eff}}\omfb^2} \right] - \frac{1}{2},
\end{eqnarray}
where an additional dependence on $N$ comes from the terms $\omega_{\text{B,eff}}(\omega)^2 = \omega^2 + N\omega\delta\tilde{\omega}(\omega)$ and $\gamma_{\text{B,eff}}(\omega) = \gamma + N\tilde{\Gamma}(\omega)$.

%%%%%%%%%%%%%%%%%%%%%%%%%%%%%%%%%%%%%%%%%%%%%%%%%%%%%%%%%%%%%%%%%%%%%
%%%%%%%%%%%%%%%%%%%%%%%%%%%%%%%%%%%%%%%%%%%%%%%%%%%%%%%%%%%%%%%%%%%%%
\section{Numerical integration of Langevin equations}
\label{D}
%%%%%%%%%%%%%%%%%%%%%%%%%%%%%%%%%%%%%%%%%%%%%%%%%%%%%%%%%%%%%%%%%%%%%
%%%%%%%%%%%%%%%%%%%%%%%%%%%%%%%%%%%%%%%%%%%%%%%%%%%%%%%%%%%%%%%%%%%%%

We perform numerical Monte-Carlo simulations for the equations of motion to test the results at steady state derived by solving the Lyapunov equation or from working with the Fourier transform. Here, the initial conditions are obtained from a Boltzmann distribution representing the initial thermal state. From the differential forms of the stochastic differential equations of motion the numerical integration can be obtained. Here, we work in the high temperature regime described by $k_{\text{B}}T \gg \hbar \omega_j$ that allows us to treat the observables as classical variables where the commutation relations can be ignored. The set of differential forms is given by
\begin{subequations}
\begin{align}
\label{App.Eq8}
dq_j &= \omega_j p_j dt, \\
dp_j &= -\omega_j q_j dt - \gamma_j p_j dt - (g_j\ast y) dt + \sqrt{(2\bar{n}_j+1)\gamma_j}dW(t),\\
dy &= -\kappa y dt + \sum_{j=1}^{N} G_j q_j dt,
\end{align}
\end{subequations}
where $dq_j(t) \approx q_j(t+dt) - q_j(t)$, $dp_j(t) \approx p_j(t+dt) - p_j(t)$ and $dy(t) \approx y(t+dt) - y(t)$. Here, $dW(t)$ describes an infinitesimal Wiener increment which follows the condition $dW^2 = dt$ and guarantees that the fluctuation dissipation theorem is fulfilled \cite{Jacobs2010Stochastic}. Numerical stability for the integration is obtained by employing the Runge-Kutta fourth-order method (RK4).
\\
%%%%%%%%%%%%%%%%%%%%%%%%%%%%%%%%%%%%%%%%%%%%%%%%%%%%%%%%%%%%%%%%%%%%%%%%%%%%%%%

\section{Approximations orders}
\label{E}
Using the full equations of motion
\begin{subequations}
\begin{align}
\label{App.FullApp1}
\dot{q}_j &= \omega_j p_j, \\
\dot{p}_j &= -\omega_j q_j - \gamma_j p_j - \sum_{k} g^{(j)}_{\text{cd}}\omfb G_k\omega_k\int^{t-\tau}_{-\infty} ds  h_{\tau}(t-s) p_k(s) + \zeta_{j}
\end{align}
\end{subequations}
where $\zeta_j = \xi_j+\xi_\text{fb}+ \xi_\text{vac}+ \xi_\text{rp}$, we can obtain successive orders of approximation from integration by substitution and injecting the equations of motion into the term $\int^{t-\tau}_{-\infty} ds h_{\tau}(t-s)p_j(s)$ which for example for a single injection results in
\begin{eqnarray}
\label{App.FullApp2}
\int^{t-\tau}_{-\infty} ds h_{\tau}(t-s)p_j(s) &=& \frac{1}{\kappa\omega_{\text{fb}}}p_j(t-\tau) -\frac{1}{(\omega_{\text{fb}}-\kappa)}\int^{t-\tau}_{-\infty}ds h^{(1)}_{\tau}(t-s)\dot{p}_{j}(s) \\\nonumber
&=& \frac{1}{\kappa\omega_{\text{fb}}}p_j(t-\tau) + \frac{\omega_j (\omega_{\text{fb}} + \kappa)}{\kappa^2\omega_{\text{fb}}^2}q_{j}(t-\tau) - \frac{\omega^2_j}{(\omega_{\text{fb}}-\kappa)}\int^{t-\tau}_{-\infty}ds h^{(2)}_{\tau}(t-s)p_j(s)\\\nonumber
& & + \sum_{k} \frac{g^{(j)}_{\text{cd}}\omega_{\text{fb}}G_k\omega_k}{(\omega_{\text{fb}}-\kappa)}\int^{t-\tau}_{-\infty}ds h^{(1)}_{\tau}(t-s)\int^{s-\tau}_{-\infty}ds' h_{\tau}(s-s')p_k(s') \\
& & -\frac{1}{(\omega_{\text{fb}}-\kappa)}\int^{t-\tau}_{-\infty}ds h^{(1)}_{\tau}(t-s)\zeta_{j}(s),
\end{eqnarray}
where $h^{(l)}_{\tau}(t) = (e^{-\kappa(t-\tau)}/\kappa^{l} - e^{-\omega_{\text{fb}}(t-\tau)}/\omega_{\text{fb}}^{l})$ and where we have omitted any term proportional to $\gamma$. By repeating this procedure for the term $- \frac{\omega^2_j}{(\omega_{\text{fb}}-\kappa)}\int^{t-\tau}_{-\infty}ds h^{(2)}_{\tau}(t-s)p_j(s)$ infinitely many times we obtain for the first order approximation
\begin{subequations}
\begin{align}
\nonumber
\int^{t-\tau}_{-\infty} ds h_{\tau}(t-s)p_j(s) &= \frac{\kappa\omega_{\text{fb}} - \omega^{2}_{j}}{(\kappa^2 + \omega^2_j)(\omega_{\text{fb}}^2 + \omega^2_j)}p_{j}(t-\tau) + \frac{\omega_{j}(\omega_{\text{fb}} + \kappa)}{(\kappa^2 + \omega^2_j)(\omega_{\text{fb}}^2 + \omega^2_j)}q_{j}(t-\tau)  \\\nonumber
& + \sum_{k} \frac{g^{(j)}_{\text{cd}}\omega_{\text{fb}}G_k\omega_k}{(\omega_{\text{fb}}-\kappa)} \int^{t-\tau}_{-\infty}ds \left[\frac{\kappa e^{-\kappa(t-\tau-s)}}{(\kappa^2 + \omega^2_j)} - \frac{\omega_{\text{fb}} e^{-\omega_{\text{fb}}(t-\tau-s)}}{(\omega_{\text{fb}}^2 + \omega^2_j)}\right]\int^{s-\tau}_{-\infty}ds' h_{\tau}(s-s')p_{k}(s') \\
& - \frac{1}{(\omega_{\text{fb}} - \kappa)}\int^{t-\tau}_{-\infty}ds \left[\frac{\kappa e^{-\kappa(t-\tau-s)}}{(\kappa^2 + \omega^2_j)} - \frac{\omega_{\text{fb}} e^{-\omega_{\text{fb}}(t-\tau-s)}}{(\omega_{\text{fb}}^2 + \omega^2_j)}  \right]\zeta_j(s) \\\nonumber
& \approx \hat{\Gamma}_j p_{j}(t-\tau) + \delta \hat{\omega}_j q_{j}(t-\tau) \\
& - \frac{1}{(\omega_{\text{fb}} - \kappa)}\int^{t-\tau}_{-\infty}ds \left[\frac{\kappa e^{-\kappa(t-\tau-s)}}{(\kappa^2 + \omega^2_j)} - \frac{\omega_{\text{fb}} e^{-\omega_{\text{fb}}(t-\tau-s)}}{(\omega_{\text{fb}}^2 + \omega^2_j)}  \right]\zeta_j(s),
\end{align}
\end{subequations}
where we have defined $\hat{\Gamma}_j = (\kappa\omega_{\text{fb}} - \omega^2_j)/\left[(\kappa^2 + \omega^2_j)(\omega^2_{\text{fb}} + \omega^2_j)\right]$ and $\delta \hat{\omega}_j = \omega_j (\omega_{\text{fb}}+\kappa)/\left[(\kappa^2 + \omega^2_j)(\omega^2_{\text{fb}} + \omega^2_j)\right]$.\\
Implementing this into the equations of motion results in
\begin{subequations}
\begin{align}
\dot{q}_j(t) &= \omega_j p_j(t), \\\nonumber
\dot{p}_j(t) &= -\omega_j q_j(t) - \gamma_j p_j(t) - \sum_{k} \left[ \Gamma_{jk}p_k(t-\tau) + \delta \omega_{jk}q_k(t-\tau)  \right]+ \zeta_{j}(t) \\
& + \sum_{k} \frac{g^{(j)}\omega_{\text{fb}}G_k\omega_k}{(\omega_{\text{fb}} - \kappa)}\int^{t-\tau}_{-\infty}ds \left[\frac{\kappa e^{-\kappa(t-\tau-s)}}{(\kappa^2 + \omega^2_j)} - \frac{\omega_{\text{fb}} e^{-\omega_{\text{fb}}(t-\tau-s)}}{(\omega_{\text{fb}}^2 + \omega^2_j)}  \right]\zeta_k(s).
\end{align}
\end{subequations}
Following this strategy for the second order approximation we obtain
\begin{subequations}
\begin{align}
\nonumber
\int^{t-\tau}_{-\infty} ds h_{\tau}(t-s)p_j(s) &\approx \hat{\Gamma}_j p_{j}(t-\tau)  + \delta \hat{\omega}_j q_j(t-\tau) - \frac{1}{(\omega_{\text{fb}} - \kappa)}\int^{t-\tau}_{-\infty}ds \left[\frac{\kappa e^{-\kappa(t-\tau-s)}}{(\kappa^2 + \omega^2_j)} - \frac{\omega_{\text{fb}} e^{-\omega_{\text{fb}}(t-\tau-s)}}{(\omega_{\text{fb}}^2 + \omega^2_j)}  \right]\zeta_j(s) \\\nonumber
& + \sum_{k} \frac{g^{(j)}\omega_{\text{fb}}G_k\omega_k}{(\omega_{\text{fb}} - \kappa)}\left[\hat{\Gamma}_k\left(\frac{\kappa^2}{(\kappa^2 + \omega^2_j)(\kappa^2 + \omega^2_k)} - \frac{\omega_{\text{fb}}^2}{(\omega_{\text{fb}}^2 + \omega^2_j)(\omega_{\text{fb}}^2 + \omega^2_k)} \right) \right. \\\nonumber
& \left. - \delta \hat{\omega}_k \omega_k \left(\frac{\kappa}{(\kappa^2 + \omega^2_j)(\kappa^2 + \omega^2_k)} - \frac{\omega_{\text{fb}}}{(\omega_{\text{fb}}^2 + \omega^2_j)(\omega_{\text{fb}}^2 + \omega^2_k)} \right) \right] p_k(t-2\tau) \\\nonumber
& + \sum_{k} \frac{g^{(j)}\omega_{\text{fb}}G_k\omega_k}{(\omega_{\text{fb}} - \kappa)}\left[\delta \hat{\omega}_k\left(\frac{\kappa^2}{(\kappa^2 + \omega^2_j)(\kappa^2 + \omega^2_k)} - \frac{\omega_{\text{fb}}^2}{(\omega_{\text{fb}}^2 + \omega^2_j)(\omega_{\text{fb}}^2 + \omega^2_k)} \right) \right. \\\nonumber
& \left. + \hat{\Gamma}_k \omega_k \left(\frac{\kappa}{(\kappa^2 + \omega^2_j)(\kappa^2 + \omega^2_k)} - \frac{\omega_{\text{fb}}}{(\omega_{\text{fb}}^2 + \omega^2_j)(\omega_{\text{fb}}^2 + \omega^2_k)} \right) \right] q_k(t-2\tau) \\\nonumber
& + \sum_k \delta \hat{\omega}_k \frac{g^{(j)}\omega_{\text{fb}}G_k\omega_k}{(\omega_{\text{fb}} - \kappa)}\int^{t-\tau}_{-\infty} ds \omega_k\left[ \frac{\kappa e^{-\kappa(t-\tau-s)}}{(\kappa^2 + \omega^2_j)(\kappa^2 + \omega^2_k)} - \frac{\omega_{\text{fb}} e^{-\omega_{\text{fb}}(t-\tau-s)}}{(\omega_{\text{fb}}^2 + \omega^2_j)(\omega_{\text{fb}}^2 + \omega^2_k)}  \right]\zeta_k(s-\tau) \\\nonumber
& - \sum_k \hat{\Gamma}_k \frac{g^{(j)}\omega_{\text{fb}}G_k\omega_k}{(\omega_{\text{fb}} - \kappa)}\int^{t-\tau}_{-\infty} ds \left[ \frac{\kappa^2 e^{-\kappa(t-\tau-s)}}{(\kappa^2 + \omega^2_j)(\kappa^2 + \omega^2_k)} - \frac{\omega_{\text{fb}}^2 e^{-\omega_{\text{fb}}(t-\tau-s)}}{(\omega_{\text{fb}}^2 + \omega^2_j)(\omega_{\text{fb}}^2 + \omega^2_k)}  \right]\zeta_k(s-\tau) \\\nonumber
& - \sum_k  \frac{g^{(j)}\omega_{\text{fb}}G_k\omega_k}{(\omega_{\text{fb}} - \kappa)^2}\int^{t-\tau}_{-\infty}ds \left[ \frac{\kappa e^{-\kappa(t-\tau-s)}}{(\kappa^2 + \omega^2_j)} - \frac{\omega_{\text{fb}} e^{-\omega_{\text{fb}}(t-\tau-s)}}{(\omega_{\text{fb}}^2 + \omega^2_j)} \right] \\
& \times \int^{s-\tau}_{-\infty}ds' \left[ \frac{\kappa e^{-\kappa(s-\tau-s')}}{(\kappa^2 + \omega^2_k)} - \frac{\omega_{\text{fb}} e^{-\omega_{\text{fb}}(s-\tau-s')}}{(\omega_{\text{fb}}^2 + \omega^2_k)} \right] \zeta_k(s').
\end{align}
\end{subequations}
This derivation suggests that proceeding with this approach will result in a system of differential equations which at each time step $t$ depends additionally on a series of former timesteps located at $t - n\tau$ for all $n \in \mathbb{N}$.
\\
\end{document}